\documentclass[reprint,
 amsmath,amssymb,
 aps, prx]{revtex4-2}
\usepackage[]{graphicx}
\usepackage{bm}
\usepackage[dvipsnames]{xcolor}
\usepackage[colorlinks = true,allcolors=BlueViolet]{hyperref}
\usepackage{float}
\usepackage{mathtools}
\newcommand{\dif}{\mathrm{d}\;\!\!}
\usepackage{adjustbox}

\usepackage{tikz}
\usetikzlibrary{calc,fadings,decorations.markings,decorations.pathreplacing,arrows,decorations.pathmorphing}
\tikzset{snake it/.style={decorate, decoration=snake}}
\tikzset{apply style/.code={\tikzset{#1}}}
\usepackage[pdftex,outline]{contour}

\newcommand {\Seq}	{Schr\"odinger equation }
\newcommand {\Schr}	{Schr\"odinger }
\newcommand {\Wn} 	{Wannier }
\newcommand {\BZ} 		{Brillouin zone }

\newcommand {\WW}	{Weisskopf-Wigner }

\newcommand{\ket}[1]{\left|#1\right\rangle}
\newcommand{\bra}[1]{\left\langle#1\right|}

\newcommand{\domega}{\delta\omega}
\newcommand{\dC}{\delta C}
\newcommand{\ddC}{\delta^2 C}
\newcommand{\mero}{r}
\newcommand{\circled}[1]{{\scalebox{0.5}{\raisebox{.5pt}{\textcircled{\raisebox{-.9pt}{#1}}}}}}
\newcommand{\fun}{\nu}

\begin{document}

\title{Exact solution for the collective non-Markovian decay of two fully excited quantum emitters}
\author{Alfonso Lanuza}
\thanks{alfonso.lanuza@stonybrook.edu}
\author{Dominik Schneble}
\affiliation{Department of Physics and Astronomy, Stony Brook University, Stony Brook, NY 11794-3800, USA}

\date{\today}

\begin{abstract}
Waveguide quantum electrodynamics constitutes a modern paradigm for the interaction of light and matter, in which strong coupling, bath structure, and propagation delays can break the radiative conditions that quantum emitters typically encounter in free space. These characteristics intertwine the excitations of quantum emitters and guided radiation modes to form complex multiphoton dynamics. So far, combining the collective decay of the emitters with the non-Markovian effects induced by the modes has escaped a full solution and the detailed physics behind these systems remains unknown. Here we analyze such a collective non-Markovian decay in a minimal system of two excited emitters coupled to a one-dimensional single-band waveguide. We develop an exact solution for this system in terms of elementary functions that unveils hidden symmetries and predicts new forms of spontaneous decay. The collective non-Markovian dynamics, which are strongly dependent on the vacuum coupling and the detuning from the center of the band, show exotic features that can be characterized with a simple and readily available criterion. Our analytic methods shed light on the complexity of collective light-matter interactions and open up a pathway for understanding multiparticle open quantum systems.
\end{abstract}

\maketitle

\section{Introduction}
\begin{figure*}[t]
	\begin{center}
        \resizebox{\textwidth}{!}{\begin{tikzpicture}[x=.5pt,y=.5pt]

\def\labelScale{5}

\coordinate (Center) at (2632,1056);
\coordinate (topleft) at (0,2112);

\coordinate (a) at (95,2037);
\coordinate (b) at (2089,1737);
\coordinate (c) at (95,1357);

\node[inner sep=0pt] at (Center)
    {\includegraphics[width=2632pt]{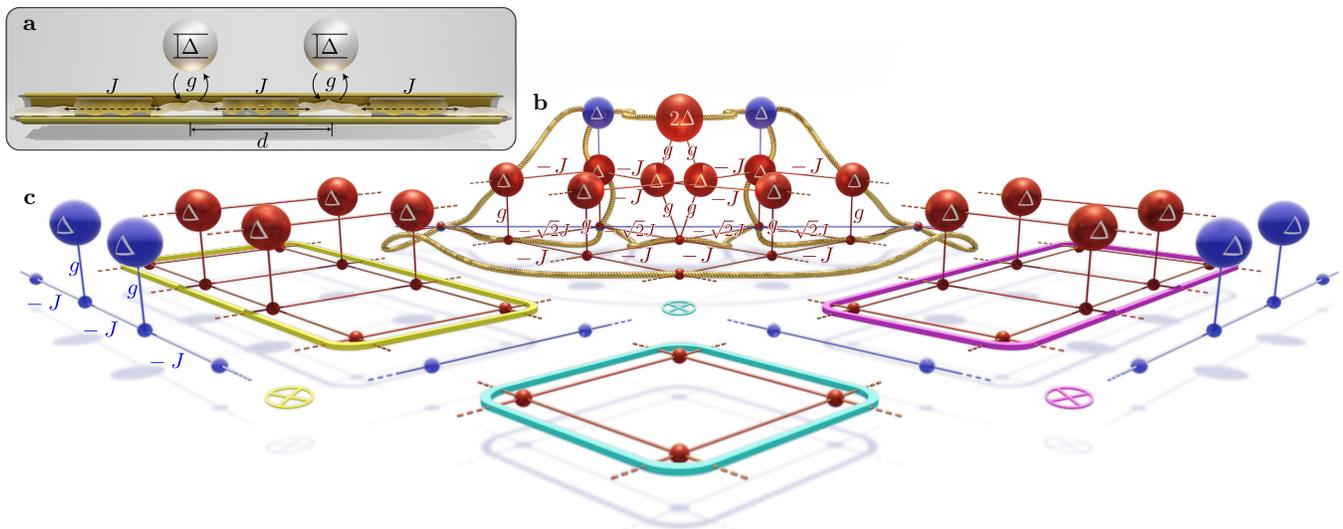}};

\node[scale=\labelScale] at (a) {\textbf{a}};
\node[scale=\labelScale] at (b) {\textbf{b}};
\node[scale=\labelScale] at (c) {\textbf{c}};

\end{tikzpicture}}
        \caption{\label{fig:2ExcitationDiagram} \textbf{The system.}\ \textbf{a} Two quantum emitters coupled to a waveguide. The strong periodic modulation of the index of refraction in the waveguide induces a single-band dispersion relation for the photons inside. \textbf{b} Adjacency graph of the matrix representation of the system Hamiltonian restricted to a single excitation (in blue) and two excitations (in red). The connection between the two is given by strings with two blue nodes at the sides and a red node at the center. For clarity, some of the strings are omitted. \textbf{c} Three asymptotic regions of the red graph that can be decoupled as tensor products between the blue graph of either the system and the waveguide (yellow and magenta regions) or twice the waveguide (cyan region).}
        \end{center}
\end{figure*}

In a pioneering paper~\cite{Dicke1954}, Dicke departed from the classical idea of emitters decaying independently. He showed that, despite the lack of photon-to-photon interactions, quantum emitters might be mutually influenced by sharing the same electromagnetic modes, thus decaying collectively. From the Markovian point of view, the field modes establish decay channels that might be super- or subradiant, but this picture is also incomplete. Given that there is only a semantic difference between a quantum emitter reabsorbing a photon in a certain mode and the mode emitting the excitation back into the quantum emitter, the reabsorption of photons can only be accounted for by placing modes and emitters at the same level \cite{Dirac1927}. Collective non-Markovian decay is the natural next step in the study of superradiance, where the ``collective" not only refers to the quantum emitters but the emitters and field modes altogether.\\

Several factors can cause and modify non-Markovian decay, such as a strong coupling between the emitters and the bath of electromagnetic modes \cite{Rzazewski1982, Bay1997}, a structured bath \cite{Bykov1975,John1990,Lambropoulos2000} with one \cite{Calajo2016} or multiple energy bands \cite{Lanuza2022}, the topology \cite{Bello2019} and dimensionality \cite{GTudela2017} of the bath, the size \cite{Guo2017,Guo2020,Soro2023} and the arrangement \cite{Lanuza2022, Facchi2019, Burgess2022} of the emitters, and the delay of radiation traveling between them \cite{Calajo2019, Sinha2020, delAngel2023, Alvarez2023, Dinc2019, Zheng2013b}. Such effects could become prominent in large quantum networks \cite{Alvarez2023, Kimble2008}; but despite rapid experimental advancements on multi-excitation super- and subradiance \cite{Sheremet2023}, their interplay with non-Markovian dynamics is an incipient area of experimentation \cite{Kim2024}.\\

Theoretical studies of collective non-Markovian effects in quantum optics rely on effective Hamiltonians~\cite{Shi2015,Shi2018,GGutierrez2021}, numerical methods \cite{Zheng2013b,Calajo2016,Cascio2019,deVega2017}, Feynman diagrams \cite{Laakso2014,Schneider2016,Roulet2016} or analytic approximations \cite{Alvarez2023}. While these approximate methods tend to be simpler and more versatile, an exact solution would benefit the field in many ways: providing checkpoints for the approximated methods, inspiring new ans\"{a}tze for related problems, unveiling hidden symmetries and new phenomena, and developing mathematical tools to approach the problem. But the solvability of non-Markovian systems beyond the single-excitation sector is unclear \cite{Calajo2016,Shi2018,Masson2022}, as an infinite number of modes makes the dimension of the Hilbert space diverge and the collective nature of the decay couples the dynamics of the individual excitations through effective interactions caused by photon blockade.\\

In this paper, we present an exact solution of collective non-Markovian decay for a minimal system featuring two adjacent quantum emitters spontaneously radiating two excitations into a 1D single-band waveguide. We develop techniques to analyze and solve this problem and emphasize the connection between the sectors with one and two excitations. The solution has a plethora of features: multiple super- and subradiant states, algebraic decay, mixed algebraic and exponential decay, fractional decay with bound states in and out of multiple continua, as well as logarithmic corrections to the algebraic decay. We also establish a simple criterion to ascertain the presence of collective non-Markovian decay.\\

The paper is structured as follows. In Section~\ref{sec:system}, we introduce the system and propose a way to visualize the two-excitation sector. In Sec.~\ref{sec:single excitation}, we review the solution to the single-excitation sector. In Sec.~\ref{sec:secular equations}, we find a convenient form to write down the equations describing the dynamical evolution of the two excitations. In Sec.~\ref{sec:symmetries}, we look at the symmetries of the system, with emphasis on an abstract symmetry that is crucial to our analysis, despite deviating from the common notion of physical symmetry. In Sec.~\ref{sec:analytic structure}, we present the full solution together with a description of its analytic structure. In Sec.~\ref{sec:spectrum}, we study the resulting spectrum of the two-excitation sector and its physical implications for collective non-Markovian decay. In Sec.~\ref{sec:bound states} we explain how the solution also captures the spatial distribution of the bound states before concluding in Sec.~\ref{sec:Conclusion}.\\

\section{The system}\label{sec:system}
We consider the problem of two identical quantum emitters (QEs) coupled by $g$ to a 1D structured waveguide (or coupled-cavity array \cite{Lombardo2014}) with period $d$ and hopping rate $J$. The dispersion relation of the waveguide features a single band (${\omega_q=-2J\cos(q d)}$, for $q\in (-\pi/d,\pi/d]$ the quasimomentum of a photon in the waveguide). The excitation energy $\hbar\Delta$ of the QEs is best interpreted as the detuning $\Delta$ from the middle of this band, as it can also have a negative value. The QEs are fixed to adjacent lattice sites of the waveguide (represented in Fig.~\ref{fig:2ExcitationDiagram}\textbf{a}).\\

Such a system is described by a \WW Hamiltonian, which in the \Wn basis reads as
\begin{equation}\label{eq:WannierH}
\begin{aligned}
    \hat{H}/\hbar=\Delta\sum_{j=1}^2 \hat{a}_j^\dagger&\hat{a}_j-J\sum_{j=-\infty}^\infty \left( \hat{b}_j^\dagger\hat{b}_{j+1}+\hat{b}_j\hat{b}_{j+1}^\dagger\right)
    \\
   & +g\sum_{j=1}^2\left(\hat{a}_j\hat{b}_j^\dagger+\hat{a}_j^\dagger\hat{b}_j\right),
\end{aligned}
\end{equation}
where $\hat{a}^\dagger_j=|1^a_{j}\rangle\bra{0}$ is a fermionic creation operator and $\hat{b}^\dagger_j=|1^b_{j}\rangle\bra{0}+\sqrt{2}|2^b_{j}\rangle\langle 1^b_{j}|+...$ is a bosonic one.\\

We will study the dynamic evolution when the two emitters start off excited. Since the Hamiltonian conserves the number of excitations, we could expand it in the sub-basis of two-excitation states
\begin{equation}
\begin{aligned}\label{eq:psi}
    \ket{\psi(t)}&=A(t)\hat{a}^\dagger_1\hat{a}^\dagger_2\ket0+\sum_{j<j'}B_{j,j'}(t)\hat{b}^\dagger_j\hat{b}^\dagger_{j'}\ket0
    \\
    +\sum_{j=-\infty}^\infty& B_{j,j}(t)\left(\tfrac{1}{\sqrt{2}}\hat{b}^\dagger_j\hat{b}^\dagger_j\right)\ket0
    +\sum_{j,j'}C_{j,j'}(t)\hat{a}^\dagger_j\hat{b}^\dagger_{j'}\ket0,
\end{aligned}
\end{equation}
where $A(0)=1$ and $B_{j,j'}(0)=C_{j,j'}(0)=0$. The factor of $1/\sqrt2$ is necessary to make the basis orthonormal.\\

Following the above approach results in a large, sparse matrix representation of the Hamiltonian. To gain insight into the underlying geometry, without necessarily computing this matrix, we introduce a graphical representation of the problem, illustrated in Fig.~\ref{fig:2ExcitationDiagram}. We start with the adjacency graph of the Hamiltonian for a single excitation (Fig.~\ref{fig:2ExcitationDiagram}\textbf{b}, in blue), which resembles the actual system (Fig.~\ref{fig:2ExcitationDiagram}\textbf{a}). Then we denote two-excitation states by drawing strings between any two points of this graph. Because of fermionic exclusion, no such string can be drawn with both ends at the same QE single-excitation basis state (represented as a blue ball with a $\Delta$). To produce the nodes of a new graph representing two-excitation states, we mark the centers of these strings (in red) with the total energy of the state, which is given by the sum of energies at the two ends of the string due to the absence of interactions between excitations (other than fermionic exclusion in the emitters).\\

Edges in the blue graph induce corresponding edges in the red graph. More specifically, every excitation experiencing coupling or hopping is represented by not only a blue edge but also several red ones, one for every two-excitation state containing the excitation. The induced transition rates are identical to the original rates, except for an additional factor of $\sqrt{2}$ when either the initial or final configuration contains two excitations in the same site (bosonic enhancement).\\

The resulting red graph is the adjacency graph of the Hamiltonian restricted to the subspace of two excitations; despite the 1D nature of the waveguide, the graph resembles that of a coupled 2D system. There are asymptotic regions that factor into tensor products of simpler graphs. Those simpler graphs are (1, in yellow and magenta) the original blue graph, and the one corresponding to an emitterless waveguide containing a single excitation, or (2, in cyan) two such emitterless waveguide graphs. Physically, this decomposition implies that the two excitations are independent and distinguishable when they are infinitely apart. These decompositions will be instrumental in finding and interpreting the general solution, which justifies treating the single-excitation case first (see Sec.~\ref{sec:single excitation}).\\

We emphasize that the bosonic/fermionic properties of the particles are engraved in the geometry of this diagram. As a result, even a classical (non-planar) circuit board \cite{Lenggenhager2022} following this graph could serve to test and simulate the dynamics of this quantum system.

\section{Single excitation case}\label{sec:single excitation}
A system of quantum emitters containing a single excitation and coupled to a band structure has previously been analyzed in \cite{Lanuza2022}. Here we apply the formalism developed in that paper to the present configuration.\\

For the rest of the paper and appendices let us take half of the bandwidth, $2J$, as the natural frequency scale ($2J=1$) and denote by $\tilde f(\omega\in\mathbb{C})$ the Wick-rotated Laplace transform of a function,  $\tilde f(\omega)=-i\mathcal{L}\lbrace f\rbrace(-i\omega)$. This particular choice will be helpful later (in Section~\ref{sec:bound states}) to characterize the form of the system's bound states. The inverse transformation is given by
\begin{equation}\label{eq:inverseLaplace}
 f(t)=-\frac{1}{2\pi i}\int_{-\infty+i0^+}^{+\infty+i0^+}\tilde f(\omega)e^{-i\omega t}\dif \omega.
\end{equation}

We note that, together with the number of excitations, the system also conserves the parity $\sigma$ of an excitation that is distributed symmetrically ($\sigma=+1$) or antisymmetrically ($\sigma=-1$) over the two emitters. Thus, one can take decoupled bases for both cases,
\begin{equation}
\begingroup
  \thinmuskip=\muexpr\thinmuskip*2/8\relax
  \medmuskip=\muexpr\medmuskip*2/8\relax  
\begin{aligned}
    \ket{\psi_\sigma(t)}
    =
    \frac{a_\sigma(t)}{\sqrt{2}}\left(\hat{a}^\dagger_1+\sigma\hat{a}^\dagger_2\right)\ket0
    +
    \sum_{j=2}^\infty \frac{b_{\sigma,j}(t)}{\sqrt{2}}\left(\hat{b}^\dagger_j+\sigma\hat{b}^\dagger_{3-j}\right)\ket0.
\end{aligned}
    \endgroup
\end{equation}
For an emission process ($a_\sigma(0)=1$ and $b_{\sigma,j}(0)=0$), the resulting transformed excitation amplitude of the emitter pair
\begin{equation}
\begin{aligned}\label{eq:ta}
    \tilde a_\sigma(\omega)=\left(\omega-\Delta-\sigma g^2+\sigma g^2\frac{\sqrt{\omega-\sigma}}{\sqrt{\omega+\sigma}} \right)^{-1}
\end{aligned}
\end{equation}
features decay to an `even edge' at frequency $-\sigma$, where the parity of the emitter state matches that of the emitted waves, and to an `odd edge' at $\sigma$, where the parity of the emitter state is incompatible with that of the emitted waves.\\

The singularities of Eqn.~\eqref{eq:ta} encode the following single-excitation decay behaviors (cf. Fig.~\ref{fig:Spectra}):\\

\textbf{\textit{i}.} There is always a bound state in the gap beyond the even edge and no bound state in the continuum. Another bound state can be found in the gap beyond the odd edge if the detuning lies within this gap or the coupling is large enough ($g^2>1-\sigma\Delta$).\\

\textbf{\textit{ii}.} The band edges are a source of algebraic decay of order 3/2. More specifically, the even edge contributes asymptotically to $a_\sigma(t)$ with
\begin{equation}
    \frac{(\sigma +i) e^{i \sigma  t}}{4 \sqrt{\pi } g^2 t^{3/2}}
\end{equation}
while the odd edge contributes with
\begin{equation}
    \frac{g^2 (\sigma -i) e^{-i \sigma  t}}{4 \sqrt{\pi } \left(g^2-1+\sigma\Delta\right)^2 t^{3/2}}.
\end{equation}
An exception to this is the incidental case that $g^2-1+\sigma\Delta=0$, where the algebraic order of the odd edge changes to $1/2$. Its influence becomes longer lived as the result of the spectral overlap between this edge and one of the bound states.\\

\textbf{\textit{iii}.} The Markovian approximation is applicable for in-band detunings, $\lvert\Delta\rvert<1$, with weak edge effects, $g^2\ll1-\lvert\Delta\rvert$. In this limit the unemitted population $\lvert a_\sigma(t)\rvert^2$ decays through a single channel with an exponential rate $(1-\sigma\Delta)\Gamma$, where $\Gamma={2 g^2}/{\sqrt{1-\Delta ^2}}$ is the  decay rate of an isolated QE. The prefactor $(1-\sigma\Delta)$ indicates that there is single-particle superradiance when $\sigma\Delta<0$ and subradiance if $\sigma\Delta>0$. No collective decay exists at  $\Delta=0$: in the band center, the symmetry between two parity sectors leads to a suppression of collective decay, in opposition to the naive idea that every system symmetry favors collective decay. These characteristics also hold for 2 excitations, see Appendix~\ref{app:Markovian limit}.\\

\section{Secular equations}\label{sec:secular equations}
The \Seq for~\eqref{eq:psi} simplifies when written in terms of the transformed field $\ket{\tilde{\psi}(\omega)}$ and then brought to a form that accommodates the asymptotic solutions of the system (see Sec.~\ref{sec:system}). For this purpose, we introduce the Bloch modes of the waveguide via $\hat{b}_q=\sum_j e^{-iqd(j-3/2)}\hat{b}_j$, where $q\in(-\pi/d,+\pi/d]$ is the quasimomentum (restricted to the first \BZ because we only consider one band). The transformed wavefunction in this picture is described by a set of amplitudes
\begin{equation}
\begin{aligned}\label{eq:tpsi}
    \ket{\tilde\psi(\omega)}&=\tilde A(\omega)\hat{a}^\dagger_1\hat{a}^\dagger_2\ket0+\sum_{p<q}\tilde B_{p,q}(\omega)\hat{b}^\dagger_p\hat{b}^\dagger_q\ket0\\
    &+\sum_q \tilde B_{q,q}(\omega)\left(\tfrac{1}{\sqrt{2}}\hat{b}^\dagger_q\hat{b}^\dagger_q\right)\ket0
    +\sum_{j,q}\tilde C_{j,q}(\omega)\hat{a}^\dagger_j\hat{b}^\dagger_q\ket0
\end{aligned}
\end{equation}
that are connected by (see Appendix \ref{app:dynamics of the transformed amplitudes})
\begin{equation}\label{eq:tildeA,BfromC}
\left\lbrace
\begin{array}{l}
   \tilde A=\frac{1}{\omega -2\Delta}+\frac{g d}{\omega -2\Delta}\int_{-\pi/d}^{\pi/d} \left(e^{-\frac{iqd}{2}} \tilde C_{2,q}+e^\frac{iqd}{2} \tilde C_{1,q}\right)\frac{\dif q}{2\pi}
   \\
    \tilde B_{p<q}=\frac{g}{\omega -\omega_p-\omega_q}\sum\limits_{j=1}^2 \left(e^{ipd\left(\frac{3}{2}-j\right)}\tilde C_{j,q}+ e^{iqd\left(\frac{3}{2}-j\right)}\tilde C_{j,p}\right)
    \\
    \tilde B_{q,q}=\frac{\sqrt{2} g}{\omega-2\omega_q}\sum\limits_{j=1}^2 e^{iqd\left(\frac{3}{2}-j\right)}\tilde C_{j,q}\ ,
\end{array}\right.
\end{equation}
and in which $\tilde C_{2,q}(\omega)=\tilde C^*_{1,q}(\omega)$ for $\omega\in \mathbb{R}$; while determining $\tilde{C}_{1,q}(\omega)$ interweaves positions and momenta, evading a simple treatment. However, it is possible to proceed by introducing the analytic function 
\begin{equation}\label{eq:CLaurent}
C(\omega,z)\coloneqq\sum_{j=-\infty}^{\infty}\tilde C_{1,j}(\omega) z^{1-j} \qquad(\text{with } z\in \mathbb{C}),
\end{equation}
which simultaneously captures the position distribution of the amplitude $\tilde{C}_{1,j}(\omega)$ in its Laurent coefficients around $z=0$, and the momentum distribution on the unit circle $z=e^{i q d}\in\mathbb{S}^1$,
\begin{equation}
    C(\omega,e^{i q d})=e^{-\frac{i q d}{2}}\tilde C_{1,q}(\omega).
\end{equation}

After (anti-)symmetrizing this function,
\begin{equation}\label{eq:Csym}
    C_\sigma(\omega,z)
    \coloneqq
    \frac{1}{2}\left(C(\omega,z)+\frac{\sigma}{z} C\left(\omega,z^{-1}\right)\right),
\end{equation}
the secular equation becomes
\begin{equation}\label{eq:secularEqn}
\begin{aligned}
C_\sigma(\omega,z)
=
g&\left(\frac{1}{z}+\sigma\right)\tilde a_\sigma(\domega)
\bigg(
\frac{1/2+g\tilde C_{1,2}(\omega)}{\omega-2\Delta}
\\&
   +
   g
    \oint\frac{z'+\sigma}{1+2\domega\  z'+z'^2}
    \frac{C_\sigma(\omega,z')\dif z'}{2\pi i}
\bigg),
\end{aligned}
\end{equation}
where the contour integral is positively oriented around $\mathbb{S}^1$ and $\domega\coloneqq\omega+(z+z^{-1})/2$ is the dimensionless counterpart of the energy left in the state after losing one of the excitations to the waveguide,  $\hbar(\omega-\omega_q)$. The reappearance of~\eqref{eq:ta} establishes the connection between one and two excitations analytically. We clarify that $\tilde C_{1,2}$ refers to the transformed field amplitude of having one excitation in emitter 1 and another in the waveguide at the position of emitter 2; this term can be treated as a constant from the perspective of solving the integral equation, although it couples the symmetry sectors $\sigma=\pm1$.\\

To the best of our knowledge, this complex integral equation has not been studied in the literature. In the next two sections, we develop the analytical tools to solve it based on two intertwined concepts: symmetries and the analytic structure of the solution.\\

\section{Symmetries}\label{sec:symmetries}
In this section, we discuss symmetries of the system. These symmetries generate a group of transformations of the double complex plane $(\omega,z)\in\mathbb{C}\times\mathbb{C}$. We highlight the following 3 generators:\\

\textbf{\textit{i.}} The inversion $(\omega, z)\to (\omega, z^{-1})$ physically represents left-right parity in the system.\\

\textbf{\textit{ii.}} The reciprocation $(\omega, z)\to (-\omega,-z)$ finds its physical origin in the symmetry of the band structure $\omega_q=-\omega_{q+\pi/d}$. This symmetry becomes explicit in parameter space under the change $\Delta\to-\Delta$ (see Fig.~\ref{fig:Waterfall}) and is broken when multiple bands are considered~\cite{Lanuza2022}. It is the one responsible for the suppression of collective effects at $\Delta=0$ (see Sec.~\ref{sec:single excitation}.\textbf{\textit{iii}} or App.~\ref{app:Markovian limit}). Signatures of this symmetry in single-excitation, single-QE bound states coupled to a single band were observed in~\cite{Stewart2020}.\\

\textbf{\textit{iii.}} The substitution $(\omega, z)\to (\omega,\zeta(\omega,z))$ is motivated by the exchange of function variable $z$ with the contributing pole of the two canceling the denominator in \eqref{eq:secularEqn}'s integrand,
\begin{equation}\label{eq:zeta}
    \zeta^{\pm 1}(\omega,z):=-\domega\pm\sqrt{\domega+1}\sqrt{\domega-1},
\end{equation}
where by `contributing' we mean that it lies within the integration contour, $\lvert \zeta(\omega,z)\rvert\leq1$. This inequality is generally strict except for $z\in \zeta^{\pm1}\left(\omega, \mathbb{S}^1\right)$, at which the integration contour crosses a pole and the integral is not well defined. In this way, $\zeta^{\pm1}\left(\omega, \mathbb{S}^1\right)$ defines two mutually-inverse branch cuts in the shape of curve segments connecting the branch points $\zeta^{\pm1}\left(\omega, 1\right)$ and $\zeta^{\pm1}\left(\omega, -1\right)$. These cuts are present not only in $C_\sigma(\omega,z)$ but also in $\tilde{a}_\sigma(\domega)$. \\

The function $\zeta$ has many other mathematical properties, such as $\zeta(\omega,z)=\zeta(\omega,z^{-1})$, $\zeta(\omega,z)=-\zeta(-\omega,-z)$, or $\zeta(\omega,\zeta(\omega,z))=z^{\pm1}$ (for $\lvert z\rvert^{\pm1}<1$), that ensure the closure of the symmetry group. From the point of view of physics however, the substitution is a rather abstract symmetry: it leaves certain properties of $C_\sigma(z)$ invariant but not $C_\sigma(z)$ itself (see Appendix~\ref{app:soft symmetry}).\\

\section{Analytic structure}\label{sec:analytic structure}
 \begin{figure}[b]
	\begin{center}
        \resizebox{0.5\textwidth}{!}{\begin{tikzpicture}[x=.5pt,y=.5pt]
\clip (-21.5,4) rectangle (572.73,504.55+14);
\def\labelScale{1}
\def\labelSCALE{1.1}

\coordinate (C4) at (1.57,423.07);
\coordinate (C2) at (79.45,167.12);

\coordinate (C3) at (280.84,313.9);
\coordinate (C1) at (282.07,135.21);

\coordinate (Phi4) at (332.03,493.67);
\coordinate (Phi2) at (323.01,292.06);

\coordinate (Phi3) at (538.19,369.57);
\coordinate (Phi1) at (491.89,206.22);

\node[inner sep=0pt] at (286.36,252.27)
    {\includegraphics[width=286.36pt]{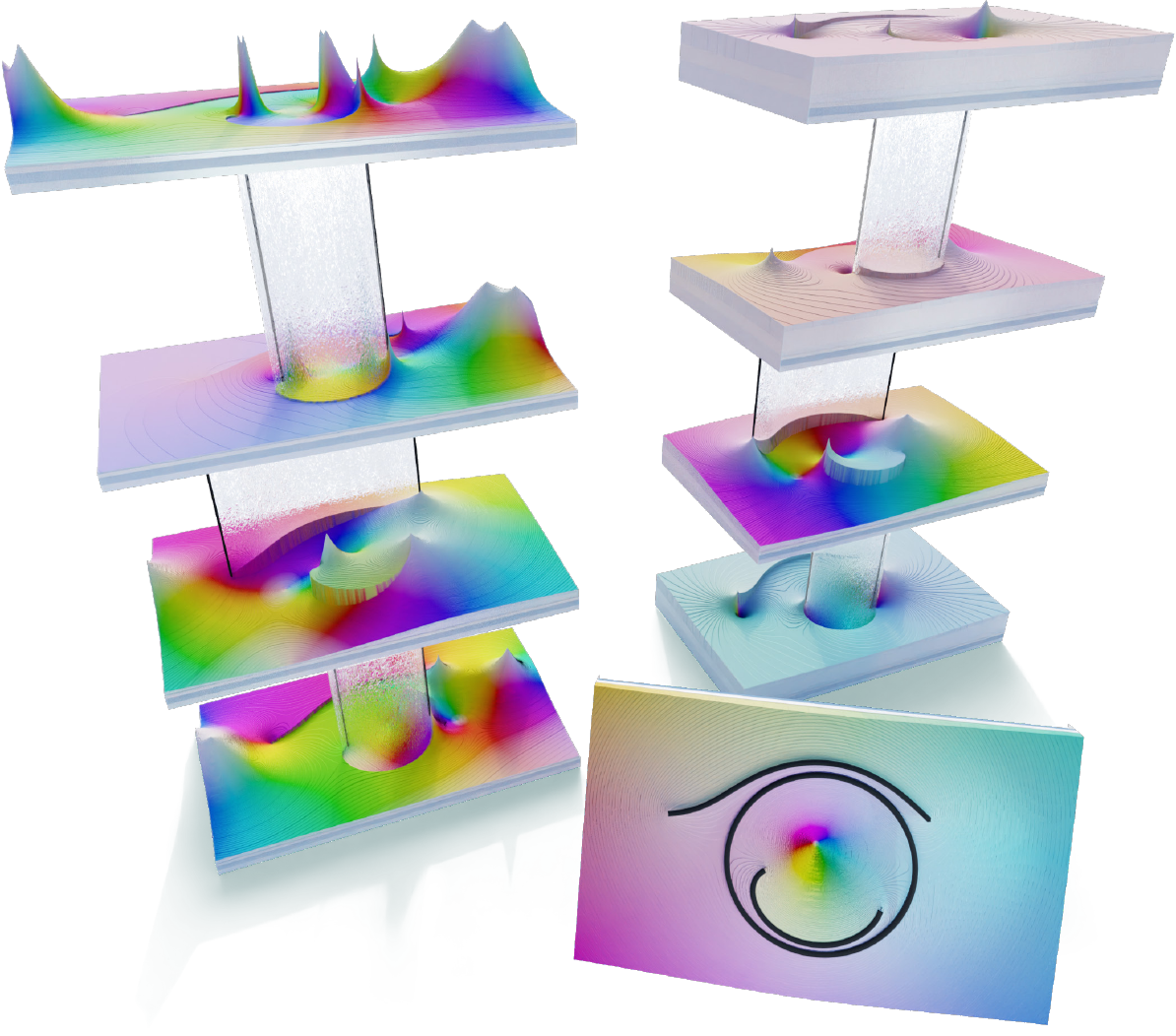}};
    
    \draw[-latex'] (C4) -- (C2) node[midway,left,sloped, rotate=90,scale=\labelScale] {$-\ddC_\sigma$};
    
    \draw[-latex'] (C3) -- (C1) node[pos=0.4,right,sloped, rotate=90,scale=\labelScale] {$\ddC_\sigma$};
    
    \draw[-latex'] (Phi4) -- (Phi2) node[midway,left,sloped, rotate=-90,scale=\labelScale] {$-2$};
    
    \draw[-latex'] (Phi3) -- (Phi1) node[midway,right,sloped, rotate=-90,scale=\labelScale] {$-2$};  

\draw (1,500) node[scale=\labelSCALE] {\textbf{a}};
\draw (330,510) node[scale=\labelSCALE] {\textbf{b}};
\path[-] (468.41,125.16) -- (520.53,119.2) node[midway,sloped,scale=\labelSCALE] {\textbf{c}};

\path[-] (169.31,115.1) -- (223.15,130.62) node[midway,sloped,scale=\labelScale] {$C_\sigma(z)+\dC_\sigma(z)$};      
      
\path[-] (153.85,199.28) -- (214.97,213.43) node[midway,sloped,scale=\labelScale,white] {$C_\sigma(z)$};       
     
\path[-] (135.64,296.75-2) -- (205.73,308.65-2) node[midway,sloped,scale=\labelScale] {$C_\sigma(z)+\sigma z^{-1}\dC_\sigma\left(z^{-1}\right)$};

\path[-] (111.41000366,431.23) -- (194.23,437.66) node[midway,sloped,scale=\labelScale] {$C_\sigma(z)+\ddC_\sigma(z)$};  

     
\path[-] (412.11,192.55) -- (451.79,203.61) node[midway,sloped,scale=\labelScale] {$-\Phi(z)-1$};  
      
\path[-] (424.36,262.11) -- (467.41,272.03) node[midway,sloped,scale=\labelScale,white] {$\Phi(z)$};   
     
\path[-] (439.87,345.34) -- (486.25,353.09) node[midway,sloped,scale=\labelScale] {$-\Phi(z)+1$};

\path[-] (461.4,465.56) -- (513.06,469.26) node[midway,sloped,scale=\labelScale] {$\Phi(z)+2$};

\path[-] (305.97,127.56+15) -- (353.52,121.99+15) node[midway,sloped,scale=\labelSCALE] {$\zeta^{-1}\left(\mathbb{S}^1\right)$};
    
\path[-] (370.79+3,59.69+10) -- (419.07+3,53.12+10) node[midway,sloped,scale=\labelSCALE] {$\zeta\left(\mathbb{S}^1\right)$};

\path[-] (494.37-5,62.41+10) -- (444.34-5,68.82+10) node[midway,sloped,scale=\labelSCALE] {$\mathbb{S}^1$};
\end{tikzpicture}}
	\caption{\textbf{Domain coloring plots} of functions $C_\sigma(z)$ (\textbf{a}) and $\Phi(z)$ (\textbf{b}), and  $\zeta(z)$ (\textbf{c}). \textbf{a} and \textbf{b} extended analytically to an $\infty$ number of Riemann sheets (only 4 shown), connected through the branch cuts $\zeta^{\pm1}\left(\mathbb{S}^1\right)$ that are represented vertically. These are also branch cuts of $\zeta(z)$, which are represented in \textbf{c} together with $\mathbb{S}^1$ as thick black lines. The plots have a complex frequency of $\omega=(1-i) \left(\sqrt{5}-2\right)/2$, parameters $\Delta=0$ and $g=\sigma=1$, and a plot range of $\lvert \operatorname{Re }z\rvert\leq (1+\sqrt{5})^2/4$ and $\lvert \operatorname{Im }z\rvert\leq (1+\sqrt{5})/2$.}
	\label{fig:Towers}
	\end{center}
\end{figure}
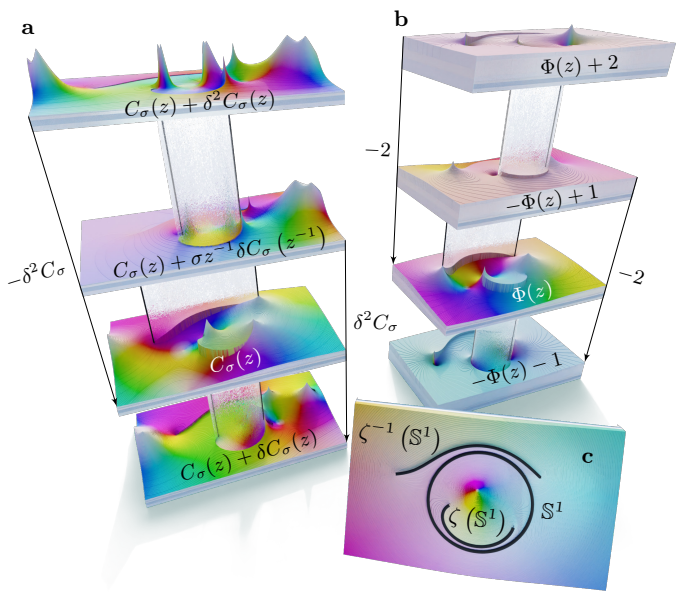

\begin{table*}[t]
\centering
\caption{Set of formulas for the evaluation of $C_\sigma(z)$ through~\eqref{eq:Csoln}.}
    \label{tab:closed formulas}
\begin{ruledtabular}
\begin{tabular}{cccc} 
\multicolumn{4}{c}{Increments of $C_\sigma$}
\\
 \multicolumn{2}{c}{$
    \dC_\sigma(z)= \frac{\left(2g  \tilde{C}_{1,1}\sigma-1\right)\frac{z+\sigma}{z}\frac{\sqrt{\domega-\sigma}}{\sqrt{\domega+\sigma}}    
    }{ g^{3}\left(\left[ g^{-2}\left(\domega-\Delta\right)-\sigma\right]^2-\frac{\domega-\sigma}{\domega+\sigma}\right)\left( g^{-2}\left(\frac{z+1/z}{2}+\Delta\right)+\frac{2\sigma z}{z-\sigma}\right)}$} 
    &
 \multicolumn{2}{c}{$
    \ddC_\sigma(z)= \frac{-2\sigma\left(2g \tilde C_{1,1}\sigma-1\right)
\frac{\sqrt{\domega-\sigma}}{\sqrt{\domega+\sigma}}    
    \frac{(z+\sigma)^2}{z(z-\sigma)}
    }{ g^3\left(\left[ g^{-2}\left(\domega-\Delta\right)-\sigma\right]^2-\frac{\domega-\sigma}{\domega+\sigma}\right)\left(\left[ g^{-2}\left(\frac{z+1/z}{2}+\Delta\right)+\sigma\right]^2-\frac{(z+\sigma)^2}{(z-\sigma)^2}\right)}$} \vspace{2pt}
    \\\hline\hline
    \multicolumn{4}{c}{The $\Phi$ function}
    \\
    \multicolumn{4}{c}{
    $\hspace{-2cm}\Phi(z) = \frac {-K(
       k^{-1})\operatorname{sgnRe}\omega } {2 \pi \sqrt{k}} \left(\frac {z^{-1} - 
        z} {1 - \zeta(z)} + \frac {z + 1} {z - 
        1} (1 + \omega) + 
     2\frac {z - 1} {z + 1} \zeta^{-1}(-1) - \frac {z^2 - 
        6 z + 1} {z^2 - 1} \zeta^{-1}(z)\right)$}
        \\
    \multicolumn{4}{c}{
    $\hspace{2cm}- \frac {2 \operatorname{sgnRe}\omega} {\pi i k}\left (K(k^{-1}) E(x;k) - k^2 E(k^{-1}) F(x;k) - (1 - k^2 ) K(k^{-1}) F(x;k) \right)$}\vspace{2pt}
    \\\hline
    \multicolumn{4}{c}{Elliptic integrals}
    \\
    \multicolumn{2}{c}{incomplete}
    &
    \multicolumn{2}{c}{complete}
    \\
    1$^{\text{st}}$ kind
    &
    2$^{\text{nd}}$ kind
    &
    1$^{\text{st}}$ kind
    &
    2$^{\text{nd}}$ kind
    \\
    \multicolumn{1}{c}{$F(x;k)\coloneqq\int _{0}^{x}{\frac {\dif t}{\sqrt{1-t^{2}}\sqrt{1-k^{2}t^{2}}}}$}
    &
    \multicolumn{1}{c}{$E(x;k)\coloneqq\int _{0}^{x}{\frac {\sqrt {1-k^{2}t^{2}}}{\sqrt {1-t^{2}}}}\,\dif t$}
    &
    \multicolumn{1}{c}{\hspace*{1.25cm} $K(k)\coloneqq F(1;k)$\hspace*{1.25cm}}
    &
    \multicolumn{1}{c}{$E(k)\coloneqq E(1;k)$}\vspace{2pt}
    \\\hline
    \multicolumn{2}{c}{Eccentricity}
    &
    \multicolumn{2}{c}{Amplitude}
    \\
    \multicolumn{2}{c}{$k=\frac{\omega^2-2+\omega\sqrt{\omega-2}\sqrt{\omega+2}}{2}$}
    &
    \multicolumn{2}{c}{$x= \frac{k^{-1/2}\operatorname{sgnIm}\omega}{
   \sqrt{-k - \zeta^{-1}(-1)}} \frac{z + 1}{z - 1} \sqrt{\frac{
   \zeta^{-1}(z) - \zeta(1)}{
   \zeta^{-1}(z) - \zeta(-1)}}
   \sqrt{\frac{\zeta^{-1}(-1)-\zeta(-1)}{\zeta^{-1}(z)-\zeta(-1)} \, \frac{\zeta^{-1}(z)-
      \zeta^{-1}(1)}{\zeta^{-1}(-1)-
      \zeta^{-1}(1)}}$}
    \\
    \multicolumn{2}{c}{$\operatorname{sgnIm}\omega=\left\{\begin{array}{l}
        +1  \text{ if Im}\omega>0\lor (\operatorname{Im}\omega=0\land\operatorname{Re}\omega<2) \\
        -1 \text{ otherwise}
    \end{array}\right.$}
    &
    \multicolumn{2}{c}{$\operatorname{sgnRe}\omega=\left\{\begin{array}{l}
        +1  \text{ if Re }\omega\geq0 \\
        -1 \text{ otherwise}
    \end{array}\right.$}\vspace{2pt}
    \\\hline\hline
    \multicolumn{4}{c}{The rational function $\mero_\sigma$}
    \\
    \multicolumn{4}{c}{
    $\hspace{-4cm}\mero_\sigma(z)=
    \frac{\sigma}{4}
    \left(
    \dC_\sigma(z)+\frac{\sigma}{z}\dC_\sigma\left(\frac{1}{z}\right)
    \right)\frac{\sqrt{\domega+\sigma}}{\sqrt{\domega-\sigma}}
    \left[
     g^{-2}\left(\domega-\Delta\right)-\sigma
    \right]
    $
    }
    \\
    \multicolumn{4}{c}{
    $\hspace{4cm}+(2g\tilde C_{1,1}\sigma-1)\sum_{i=1}^3 \alpha_\sigma(z_{\sigma i})
   \left(\sigma+z^{-1}\right)z_{\sigma i}\left(
   \frac{1}{1+2\domega\, z_{\sigma i}+ z_{\sigma i}^2}
   -
   \frac{1}{(1-z_{\sigma i} z)(1-z_{\sigma i}z^{-1})}
   \right)$
    }\vspace{2pt}
    \\
    \multicolumn{2}{c}{$(z_{\sigma i}^2+2\Delta z_{\sigma i}+1)(z_{\sigma i}-\sigma)=4 g^2 z_{\sigma i}$}
    & 
    \multicolumn{2}{c}{
    $\alpha_\sigma(z)\coloneqq 
   \frac{
   \left(2\sigma(2\Phi(z)+1)\frac{\sqrt{\domega-\sigma}}{\sqrt{\domega+\sigma}}-\left[ g^{-2}\left(\domega-\Delta\right)-\sigma\right]\right)
   \left(1-z^{-2}\right)
   }
   {
   2 g^3
   \left(\left[
    g^{-2}\left(\domega-\Delta\right)-\sigma\right]^2
   -\frac{\domega-\sigma}{\domega+\sigma}\right)
   \left( g^{-2}\left(1-z^{-2}\right)
   +\frac{4}{(\sigma-z)^2}\right)
   }$
   }\vspace{2pt}
    \\\hline\hline
    \multicolumn{4}{c}{Special field amplitudes}
    \\
    \multicolumn{2}{c}{
    $2g\tilde C_{1,1}=\frac{\sum_{\sigma,i}\sigma\alpha_\sigma(z_{\sigma i})}{\sum_{\sigma,i}\alpha_\sigma(z_{\sigma i})}$}
    &
    \multicolumn{2}{c}{
    $2g\tilde C_{1,2}=\frac{\omega-2\Delta}{2 g}\frac{\left(\sum_{\sigma,i}\sigma\alpha_\sigma(z_{\sigma i})\right)^2-\left(\sum_{\sigma,i}\alpha_\sigma(z_{\sigma i})\right)^2}{\sum_{\sigma,i}\alpha_\sigma(z_{\sigma i})} -1$}
\end{tabular}
\end{ruledtabular}
    \end{table*}

In this section, we investigate the analytic properties of $C_\sigma$ as a function of $z$, while omitting $\omega$ as a variable to avoid confusion. The analytic continuation of $C_\sigma(z)$ can be investigated by modifying the integration contour of \eqref{eq:secularEqn} as $z$ is displaced across the branch cuts (see Sec.~\ref{app:increments of C} for details). Crossing $\zeta\left(\mathbb{S}^1\right)$ results in an increment $\dC_\sigma(z)$ in the function, which is again lost if the same branch cut is subsequently crossed in the same direction. By inversion symmetry, the situation is analogous when crossing $\zeta^{-1}\left(\mathbb{S}^1\right)$, while the increment becomes $\sigma z^{-1} \dC_{\sigma}(z^{-1})$.\\

The analytically continued function is strikingly similar to the original: whereas the simple poles of $C_\sigma$ inherited by $\tilde a_\sigma$ in~\eqref{eq:secularEqn} relocate, the branch cuts (which are independent of $g$, $\Delta$ and $\sigma$, see~\eqref{eq:zeta}) reappear over and over in the same locations as illustrated in Fig.~\ref{fig:Towers}. Crossing one branch cut after the other leads to an increment of $\pm\ddC_\sigma$, where
\begin{equation}\label{eq:ddC}    \ddC_\sigma(z)\coloneqq \dC_\sigma(z)-\sigma z^{-1}\dC_\sigma\left(z^{-1}\right).
\end{equation}
This behavior can be written succinctly as
\begin{equation}\label{eq:Csoln}
    C_\sigma(z)=\frac{1}{2}\ddC_\sigma(z)\left(\Phi(z)+\frac{1}{2}\right)-\frac{1}{2}\dC_\sigma(z)+\frac{1}{2}\mero_\sigma(z),
\end{equation}
where we interpret this as a deconstruction of the analytic structure of $C_\sigma(z)$: $\Phi(z)$ contains the monodromy group of the function (it has the same branch cuts, see Fig.~\ref{fig:Towers}\textbf{b}) but unlike $C_\sigma$ it is independent of $g$, $\Delta$, $\sigma$, $\tilde{C}_{1,1}$ or $\tilde{C}_{1,2}$; then $\ddC_\sigma(z)$ and $\dC_\sigma(z)$ are algebraic functions that fix double and single leaps (respectively) in the Riemann sheets of the function; and $\mero_\sigma(z)$ is a rational function that fixes the form of $C_\sigma$ within one Riemann sheet. Closed formulas for these functions are given in Table~\ref{tab:closed formulas} and proof of these can be found in Appendices~\ref{app:increments of C}, \ref{app:the function Phi}, and \ref{app:constructive proof}.

\section{Spectrum and decay}\label{sec:spectrum}
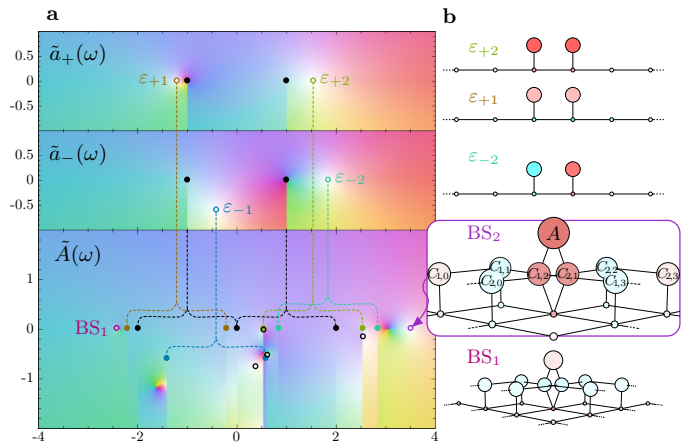
\begin{figure}[t]
    \begin{center}
\resizebox{0.5\textwidth}{!}{\begin{tikzpicture}[x=.5pt,y=.5pt]

\def\dy{15.7}
\def\dx{62.65}
\def\coordscale{0.5}
\def\coordScale{0.69}
\def\labelScale{0.75}
\def\Cscale{0.5}
\def\cradius{0.75}

\coordinate (P1) at (0.33,235.53);
\coordinate (O) at (0.35,0.34);

\coordinate (s11) at (87.71,219.77);
\coordinate (s12) at (174.05,219.77);

\coordinate (s21) at (112.64,138.06);
\coordinate (s22) at (183.43,157.06);

\coordinate (S1) at (49.63,63.05);
\coordinate (S2) at (81.3,44.06);
\coordinate (S3) at (63.07,63.05);
\coordinate (S4) at (56.3,63.05);
\coordinate (S5) at (119.05,63.05);
\coordinate (S6) at (125.75,63.05);
\coordinate (S7) at (143.94,44.06);
\coordinate (S8) at (142.56,63.05);
\coordinate (S9) at (152.08,63.05);
\coordinate (S10) at (188.56,63.05);
\coordinate (S11) at (205.34,63.05);
\coordinate (S12) at (235.7,63.05);
\coordinate (S13) at (214.78,63.05);
\coordinate (S14) at (137.56,38.82);
\coordinate (S15) at (144.98,46.09);
\coordinate (S16) at (142.65,62);
\coordinate (S17) at (205.55,57.75);

\def\keytension{0.5}
\def\lkey{(0,0) (3,7) (0.5*\dx-5,7) (0.5*\dx,12)}
\def\rkey{(0.5*\dx,12) (0.5*\dx+5,7) (\dx-3,7) (\dx,0)}
\def\keystyle{very thin, dash pattern=on 1pt off 0.5pt}

\coordinate (tS4) at (56.3,63.05+4);
\coordinate (tS8) at (142.56,63.05+4);
\coordinate (tS9) at (152.08,63.05+8);


\definecolor{colors11}{rgb}{0.6, 0.43, 0.04}
\definecolor{colors12}{rgb}{0.49, 0.7, 0}

\definecolor{colorBC11}{rgb}{0,0,0}
\definecolor{colorBC12}{rgb}{0,0,0}

\definecolor{colors21}{rgb}{0, 0.49, 0.7}
\definecolor{colors22}{rgb}{0.16, 0.8, 0.61}

\definecolor{colorBC21}{rgb}{0,0,0}
\definecolor{colorBC22}{rgb}{0,0,0}

\definecolor{colorS1}{rgb}{0.7, 0., 0.49}
\definecolor{colorS12}{rgb}{0.61, 0.16, 0.8}

\node[inner sep=0pt] at (125.71,125.37)
    {\includegraphics[width=125.71pt]{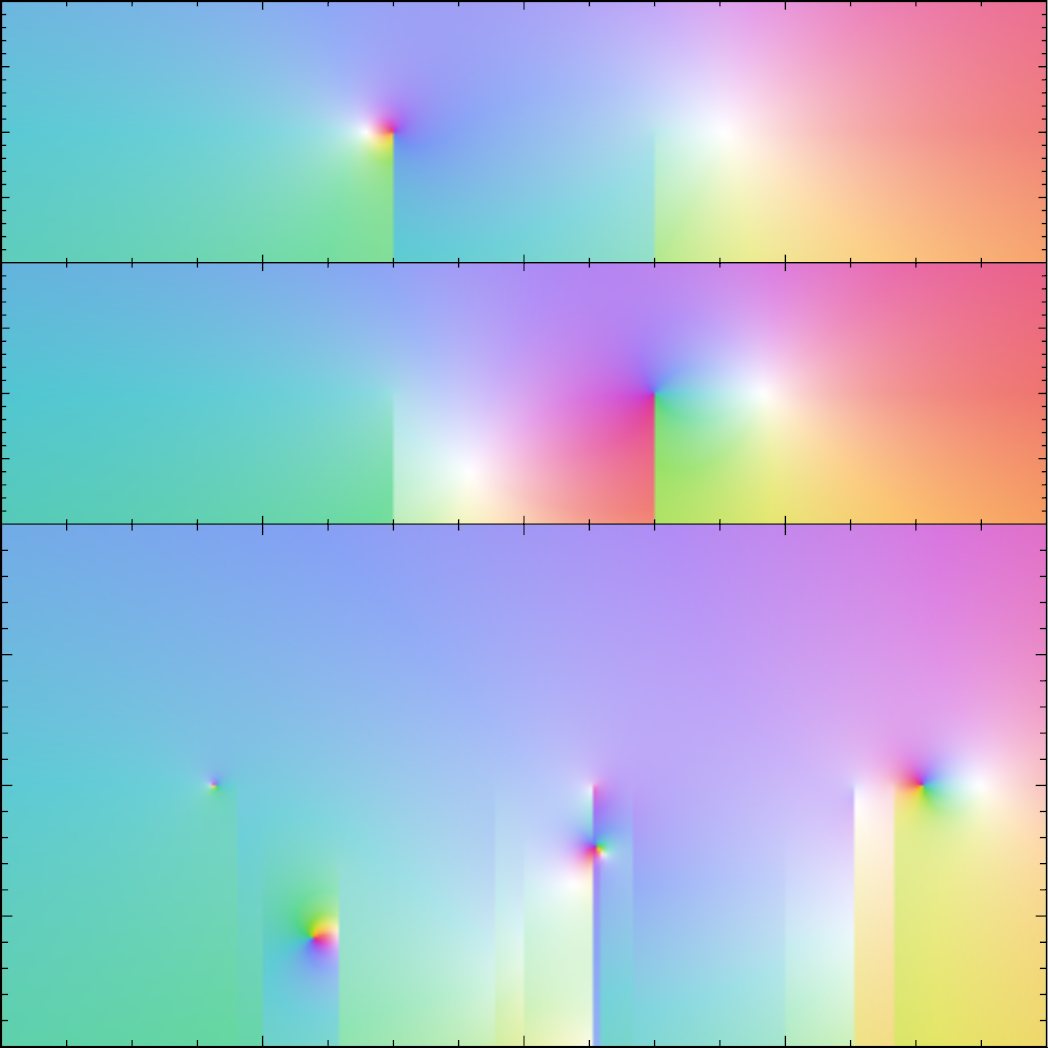}};
    
\node[scale=\labelScale] at ([xshift=0.2*\dx, yshift=7.5*\dy]O) {$\tilde a_{+}(\omega)$};
\node[scale=\labelScale] at ([xshift=0.2*\dx, yshift=5.5*\dy]O) {$\tilde a_{-}(\omega)$};
\node[scale=\labelScale] at ([xshift=0.2*\dx, yshift=3.5*\dy]O) {$\tilde A(\omega)$};

\node[scale=\labelScale,right] at ([yshift=8.3*\dy]O) {\textbf{a}};
\node[scale=\labelScale,right] at ([xshift=2*\dx,yshift=8.3*\dy]O) {\textbf{b}};

\foreach \pos/\num in {{0/0.5},{1/0},{2/-0.5},{4/0.5},{5/0},{6/-0.5},{9/1},{11/0},{13/-1}}
{\draw ([xshift=-0.5mm]P1)++(0,-\pos*\dy) node[left, scale=\coordscale,inner sep=0] {\num};}

\foreach \pos/\num in {{0/-4},{1/-2},{2/0},{3/2},{4/4}}
{\draw ([yshift=-0.5mm]O)++(\pos*\dx,0) node[below, scale=\coordscale,inner sep=0] {\num};}

\draw [colors11,apply style/.expand once=\keystyle] (tS4)++(0.5*\dx,12)--(s11);
\draw [colors11,apply style/.expand once=\keystyle] (S4)--(tS4);
\draw [colors11,apply style/.expand once=\keystyle] ([xshift=0.5*\dx]S4)--([xshift=0.5*\dx]tS4);
\draw [colors11, shift=(tS4),apply style/.expand once=\keystyle] plot [smooth, tension=\keytension] coordinates \lkey;
\draw [colors11, shift=(tS4),apply style/.expand once=\keystyle] plot [smooth, tension=\keytension] coordinates \rkey;

\draw [colors12,apply style/.expand once=\keystyle] (tS8)++(0.5*\dx,12)--(s12);
\draw [colors12,apply style/.expand once=\keystyle] (S8)--(tS8);
\draw [colors12,apply style/.expand once=\keystyle] ([xshift=0.5*\dx]S8)--([xshift=0.5*\dx]tS8);
\draw [colors12, shift=(tS8),apply style/.expand once=\keystyle] plot [smooth, tension=\keytension] coordinates \lkey;
\draw [colors12, shift=(tS8),apply style/.expand once=\keystyle] plot [smooth, tension=\keytension] coordinates \rkey;

\draw [colorBC21,shift=(S3),apply style/.expand once=\keystyle] (0.5*\dx,12)--(0.5*\dx,4*\dy);
\draw [colorBC21,shift=(S3),apply style/.expand once=\keystyle] plot [smooth, tension=\keytension] coordinates \lkey;
\draw [colorBC21,shift=(S3),apply style/.expand once=\keystyle] plot [smooth, tension=\keytension] coordinates \rkey;

\draw [colorBC22,shift=(S6),apply style/.expand once=\keystyle] (0.5*\dx,12)--(0.5*\dx,4*\dy);
\draw [colorBC22,shift=(S6),apply style/.expand once=\keystyle] plot [smooth, tension=\keytension] coordinates \lkey;
\draw [colorBC22,shift=(S6),apply style/.expand once=\keystyle] plot [smooth, tension=\keytension] coordinates \rkey;

\draw [colors21,shift=(S2),apply style/.expand once=\keystyle] (0.5*\dx,12)--(0.5*\dx,6*\dy);
\draw [colors21,shift=(S2),apply style/.expand once=\keystyle] plot [smooth, tension=\keytension] coordinates \lkey;
\draw [colors21,shift=(S2),apply style/.expand once=\keystyle] plot [smooth, tension=\keytension] coordinates \rkey;

\draw [colors22,apply style/.expand once=\keystyle] (tS9)++(0.5*\dx,12)--(s22);
\draw [colors22,apply style/.expand once=\keystyle] (S9)--(tS9);
\draw [colors22,apply style/.expand once=\keystyle] ([xshift=0.5*\dx]S9)--([xshift=0.5*\dx]tS9);
\draw [colors22,shift=(tS9),apply style/.expand once=\keystyle] plot [smooth, tension=\keytension] coordinates \lkey;
\draw [colors22,shift=(tS9),apply style/.expand once=\keystyle] plot [smooth, tension=\keytension] coordinates \rkey;

    \draw [colors11,fill=white] (s11) circle[radius=\cradius pt];
  \draw [colors12,fill=white] (s12)  circle[radius=\cradius pt];  
  \draw [colorBC11,fill=colorBC11] (P1)++(1.5*\dx,-\dy) circle[radius=\cradius pt];    
  \draw [colorBC12,fill=colorBC11] (P1)++(2.5*\dx,-\dy) circle[radius=\cradius pt];
  
    \draw [colors21,fill=white] (s21) circle[radius=\cradius pt];
  \draw [colors22,fill=white] (s22)  circle[radius=\cradius pt];  
  \draw [colorBC21,fill=colorBC21] (P1)++(1.5*\dx,-5*\dy) circle[radius=\cradius pt];    
  \draw [colorBC22,fill=colorBC22] (P1)++(2.5*\dx,-5*\dy) circle[radius=\cradius pt];
  
    \draw [colorS1] (S1) circle[radius=\cradius pt]; 
    \draw [colors21, fill=colors21] (S2) circle[radius=\cradius pt]; 
    \draw [colorBC21, fill=colorBC21] (S3) circle[radius=\cradius pt]; 
    \draw [colors11, fill=colors11] (S4) circle[radius=\cradius pt]; 
    \draw [colors11, fill=colors11] (S5) circle[radius=\cradius pt]; 
    \draw [colorBC21, fill=colorBC21] (S6) circle[radius=\cradius pt]; 
    \draw [colors21, fill=colors21] (S7) circle[radius=\cradius pt]; 
    \draw [colors12, fill=colors12] (S8) circle[radius=\cradius pt]; 
    \draw [colors22, fill=colors22] (S9) circle[radius=\cradius pt]; 
    \draw [colorBC22, fill=colorBC22] (S10) circle[radius=\cradius pt];
    \draw [colors12, fill=colors12] (S11) circle[radius=\cradius pt];
    \draw [colorS12] (S12) circle[radius=\cradius pt];
    \draw [colors22, fill=colors22] (S13) circle[radius=\cradius pt];
    \draw (S14) circle[radius=\cradius pt];
    \draw (S15) circle[radius=\cradius pt];
    \draw (S16) circle[radius=\cradius pt];
    \draw (S17) circle[radius=\cradius pt];

\node[] at ([xshift=2.6*\dx,yshift=6.5*\dy]O)
    {\includegraphics[width=70pt]{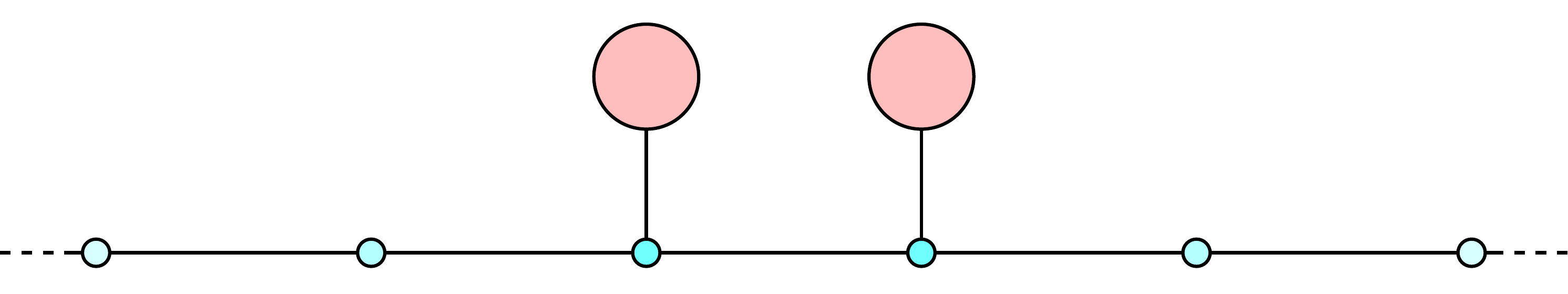}};
 
\node[scale=\coordScale,colors11] at ([xshift=2.25*\dx,yshift=6.6*\dy]O) {$\varepsilon_{+1}$};
\node[scale=\coordScale,colors11,left] at (s11) {$\varepsilon_{+1}$};
    
\node[] at ([xshift=2.6*\dx,yshift=7.5*\dy]O)
    {\includegraphics[width=70pt]{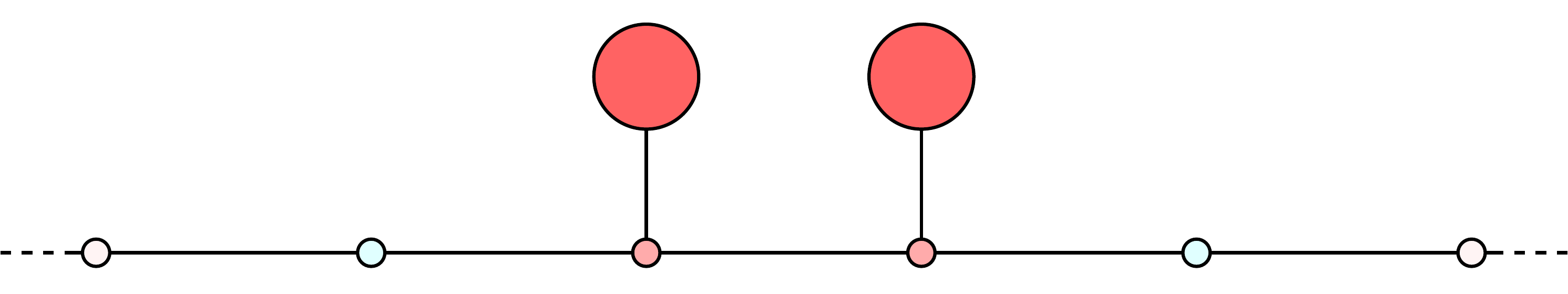}};
 
\node[scale=\coordScale,colors12] at ([xshift=2.25*\dx,yshift=7.6*\dy]O) {$\varepsilon_{+2}$};
\node[scale=\coordScale,colors12,right] at (s12) {$\varepsilon_{+2}$};

\node[] at ([xshift=2.6*\dx,yshift=5*\dy]O)
    {\includegraphics[width=70pt]{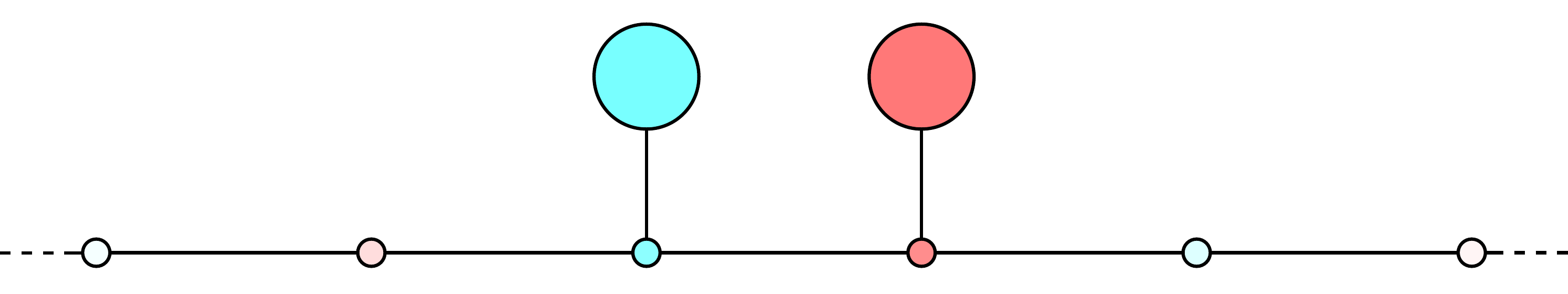}};
 
\node[scale=\coordScale,colors22] at ([xshift=2.25*\dx,yshift=5.4*\dy]O) {$\varepsilon_{-2}$};
\node[scale=\coordScale,colors22,right] at (s22) {$\varepsilon_{-2}$};

\node[scale=\coordScale,colors21,right] at (s21) {$\varepsilon_{-1}$};
      
\node[scale=\coordScale,colorS1] at ([xshift=2.25*\dx,yshift=1.4*\dy]O) {BS$_1$};
\node[] at ([xshift=2.6*\dx,yshift=0.8*\dy]O)
    {\includegraphics[width=70pt]{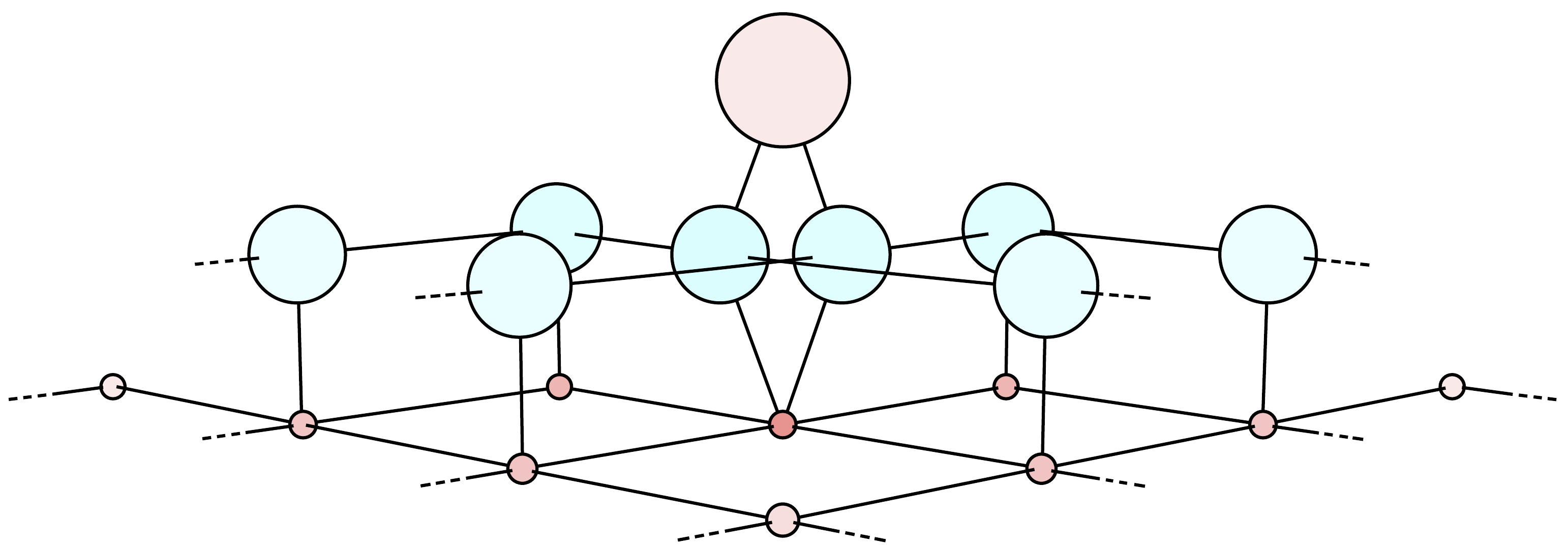}};
    
    \node[scale=\coordScale,colorS1,left] at (S1) {BS$_{1}$};

\node[inner xsep=0, inner ysep=-2pt, rectangle, draw, rounded corners,colorS12,fill=white] at ([xshift=2.6*\dx,yshift=3*\dy]O)
    {\includegraphics[width=80pt]{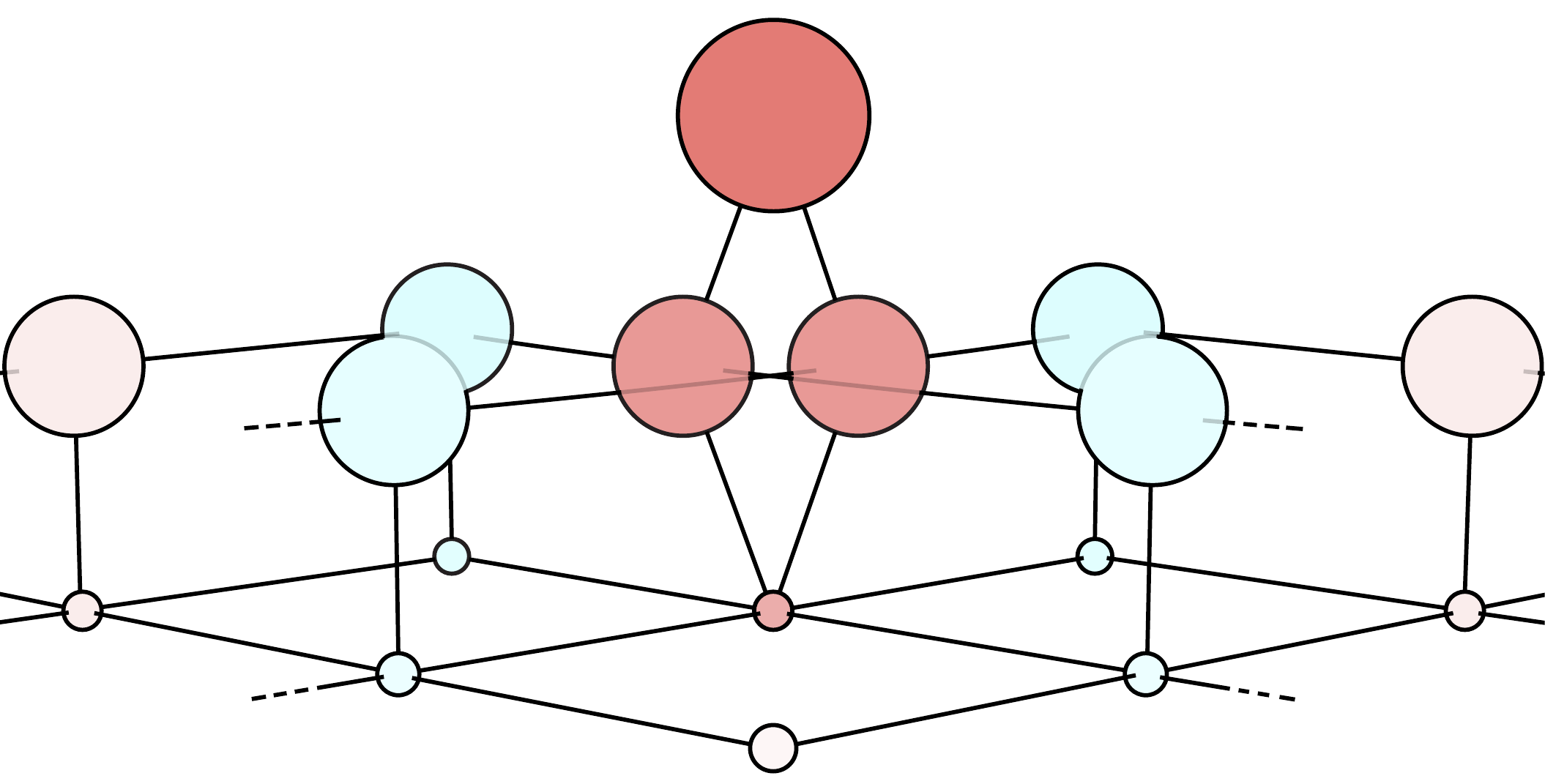}};
\draw [colorS12, -latex] plot [smooth, tension=1] coordinates {([xshift=2.6*\dx-40pt-0.em,yshift=3*\dy]O) ([xshift=2.45*\dx-25pt-0.2em-5pt,yshift=3*\dy-3pt]O) ([xshift=5.3pt,yshift=7pt]S12)(S12)};

\node[scale=\coordScale,colorS12] at ([xshift=2.25*\dx,yshift=3.9*\dy]O) {BS$_2$};

\node[scale=\coordScale] at ([xshift=2.6*\dx,yshift=3.9*\dy]O) {$A$};

\node[scale=\Cscale] at ([xshift=2.6*\dx-0.58*\dx,yshift=3.075*\dy]O) {$C_{\!1\!,0}$};
\node[scale=\Cscale] at ([xshift=2.6*\dx-0.267*\dx,yshift=3.2*\dy]O) {$C_{\!1\!,1}$};
\node[scale=\Cscale] at ([xshift=2.6*\dx-0.075*\dx,yshift=3.075*\dy]O) {$C_{\!1\!,2}$};
\node[scale=\Cscale] at ([xshift=2.6*\dx+0.315*\dx,yshift=2.93*\dy]O) {$C_{\!1\!,3}$};

\node[scale=\Cscale] at ([xshift=2.6*\dx+0.58*\dx,yshift=3.075*\dy]O) {$C_{\!2\!,3}$};
\node[scale=\Cscale] at ([xshift=2.6*\dx+0.273*\dx,yshift=3.2*\dy]O) {$C_{\!2\!,2}$};
\node[scale=\Cscale] at ([xshift=2.6*\dx+0.075*\dx,yshift=3.075*\dy]O) {$C_{\!2\!,1}$};
\node[scale=\Cscale] at ([xshift=2.6*\dx-0.31*\dx,yshift=2.93*\dy]O) {$C_{\!2\!,0}$};

\end{tikzpicture}}
\caption{\textbf{Spectrum and decay properties} of the system for $\Delta=g=1$. \textbf{a.} Simple poles (open circles) $\varepsilon_{\sigma i}$ of the transformed symmetric ($\sigma=+1$) and antisymmetric ($\sigma=-1$) emitter amplitudes $\tilde a_\sigma(\omega)$ split into branch points (solid circles) $\varepsilon_{\sigma i}\pm1$ of the doubly-excited amplitude $\tilde A(\omega)$. Black open circles correspond to collective exponential decay sources. \textbf{b.} Corresponding bound-state amplitudes in Wannier space for a single excitation ($\varepsilon_{+1},\varepsilon_{+2}$, and $\varepsilon_{-1}$) and two excitations (BS$_1$ and BS$_2$). Hue encodes the phase (red for positive, cyan for negative) and saturation encodes the absolute value.}
\label{fig:Spectra}
	\end{center}
\end{figure}

Combining the expression for $\tilde{A}$ in Eqn.~\eqref{eq:tildeA,BfromC} with expressions for $\tilde{C}_{1,2}$ (and nested definitions of $z_{\sigma i}$ and $\alpha_\sigma$) in Tab.~\ref{tab:closed formulas} yields
\begin{equation}\label{eq:Atilde}
    \tilde A(\omega)
    =    \frac{\left(\sum_{\sigma,i}\sigma\alpha_\sigma(\omega,z_{\sigma i})\right)^2-\left(\sum_{\sigma,i}\alpha_\sigma(\omega,z_{\sigma i})\right)^2}{2g\sum_{\sigma,i}\alpha_\sigma(\omega,z_{\sigma i})}
\end{equation}
for the amplitude of both emitters being simultaneously excited, which we use to study the spectrum and decay properties of the system. While the spectrum is determined by the complex singularities of $\tilde A(\omega)$ (all located on the real line), the decay properties are given by the analytic continuation in the complex $\omega$-plane \cite{Lanuza2022}. More specifically, the branch cuts of $\tilde A(\omega)$ are segments along the real line representing different continua in the spectrum;  redrawing them vertically reveals other singularities (as in Fig.~\ref{fig:Spectra}\textbf{a}) that characterize the decay dynamics.\\

The spectrum of the system is easily understood physically. There are simple real poles that represent bound states with both excitations located around the emitters. These poles satisfy the equation
\begin{equation}\label{eq:BSenergies}
\sum_{\sigma,i}\alpha_\sigma(\omega,z_{\sigma i})=0,
\end{equation}
and they can be located away from or on the branch cuts (in which case they represent bound states in the continuum and, consequently, they are hard to distinguish with simple numerical approaches, such as a brute-force diagonalization of the truncated Hamiltonian). The largest of these cuts is $[-2,2]$ and it represents two excitations emitted into the waveguide. Additionally, a cut is formed at 
$[\varepsilon_{\sigma i}-1,\varepsilon_{\sigma i}+1]$
(for some $i\in\{1,2,3\}$) when one excitation forms a bound state of parity $\sigma$ and dimensionless energy $\varepsilon_{\sigma i}=-\left(z_{\sigma i}+z_{\sigma i}^{-1}\right)/2$ (see Sec.~\ref{sec:single excitation}) and another excitation is delocalized over the waveguide. We note that Eqn.~\eqref{eq:BSenergies} is exact, in contrast to previous approaches used to compute multiexcitation bound state energies, such as variational~\cite{Calajo2016} or perturbative~\cite{Shi2018} methods.\\

To access the decay properties, one needs to analytically continue $\tilde{A}(\omega)$ beyond these branch cuts. For most of the calculations, this can be done by reverting the signs of the square roots appropriately. The situation is complicated by $\Phi(\omega,z)$ when $\omega$ crosses $[-2,2]$ (see Fig.~\ref{fig:AnalyticExtension}): the two branch cuts in $z$-space, $\zeta^{\pm1}\left(\omega,\mathbb{S}^1\right)$, come into contact when $\omega\in[-2,2]$ and, as they separate beyond this continuum, a new branch cut crosses from one to the other. This topological change in analytic structure is important for understanding the collective decay of our system.\\

The analysis reveals several sources of decay:\\

\textbf{\textit{i.}} Solutions to the analytic continuation of~\eqref{eq:BSenergies} that do not describe bound states, contain the collective exponential decay rates. Mathematically, they are simple poles $p$ in the continuation of $\tilde A(\omega)$ and, while their frequency is $\operatorname{Re}p$, the amplitude decay rates are given by $-\operatorname{Im}p$. In some cases, this decay rate might exceed the Markovian superradiant prediction $2\Gamma/2$ (for instance, for $\Delta=-0.8$ and $g=\Gamma=0.3$ the dominant pole is $-1.5077(1)-0.31956(1) i$). However, even in such cases and in contrast to \cite{Dinc2019,Sinha2020}, the total decay is generally slower than the Markovian prediction (App.~\ref{app:Markovian limit}) due to other non-Markovian effects such as the bound population at late times, or the dynamics starting out as Rabi oscillations between the QEs and the sites below them at early times \cite{Stewart2020}.\\

\textbf{\textit{ii.}} When $\varepsilon_{\sigma i}\in\mathbb{C}$ is a pole that contributes to the decay dynamics of a single excitation (Sec.~\ref{sec:single excitation}), then $\varepsilon_{\sigma i} \pm1$ is a branch point of the two-excitation problem, with frequency $\operatorname{Re}\{\varepsilon_{\sigma i} \pm1\}$ and a mixture of exponential decay with rate $-\operatorname{Im}\varepsilon_{\sigma i}$ (which can be $0$) and an algebraic decay with an order inherited from the corresponding single-excitation band edge (generally 3/2).\\

\textbf{\textit{iii.}} In between branch points $\omega=\pm2$ at the borders of $[-2,2]$, there is an additional branch point at the center ($\omega=0$). All three points are a source of algebraic decay, generally of order 3 (such as the expected for quantum emitters in 2D \cite{GTudela2019}). From a mathematical perspective, these singularities originate from $\Phi(\omega,z)$ and are thus the hardest to analyze (see the last part of App.~\ref{app:the function Phi}).\\

Indeed, \textbf{\textit{ii}} and \textbf{\textit{iii}} are what one would expect for two independently emitted excitations. However, the algebraic decay orders themselves can also present signs of collective decay. For instance, for $\Delta=0$ and $g=1$ (a critical coupling between Markovian decay and bound-state induced oscillations), one might expect $\omega=\pm (1+\sqrt{2})$ to be the dominant decay sources at very late times ($2Jt\gg10$) with an algebraic order of $1/2$. However, these sources actually have an order of $3/2$, making the branch points $\omega=\pm2$ the slowest decay sources instead. Additionally, assuming independently emitted excitations would imply those sources produce algebraic decay of order 1, but there is a $\log^2(t)$ correction to the asymptotic decay that they induce,
\begin{equation}
\sim\frac{e^{\mp2 i t}}{t \log^2 (t)}.
\end{equation}

We note that such a logarithmic correction will be difficult to access in measurements and simulations because the bound state contributions overshadow the algebraic decay at exponentially late times. For any other parameter values the logarithmic corrections are also present at $\omega=\pm2$ and $\omega=0$, although not generally at leading order.\\

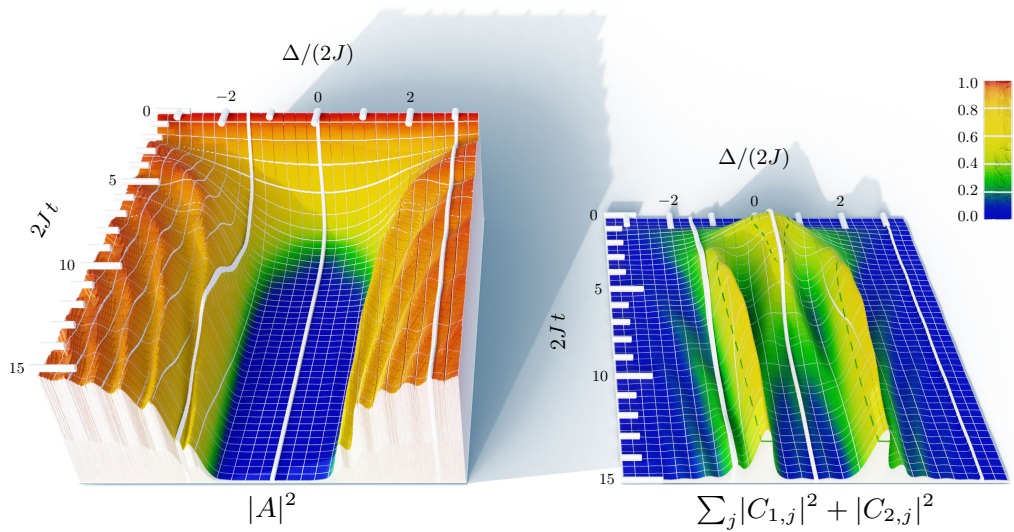
\begin{figure*}[t]
	\begin{center}
        \resizebox{0.75\textwidth}{!}{\begin{tikzpicture}[x=.5pt,y=.5pt]

\def\labelScale{2.5}
\def\labelSCALE{3.5}
\def\margin{50}
\def\legendwidth{51}

\coordinate (A) at (710.2,31);
\coordinate (C) at (2291.1,29.3);

\coordinate (Center) at (1442.5,718);
\coordinate (topleft) at (0,1436);

\coordinate (At15) at (2.2,363.6);
\coordinate (At0) at (369,1113.9);
\coordinate (atlabeL) at (89.7,876.3);
\coordinate (Atlabel) at (2,718);

\coordinate (ADm2) at (574.8,1123);
\coordinate (ADp2) at (1111.8,1123);

\coordinate (Ct15) at (1723,40.8);
\coordinate (Ct0) at (1672.7,811.3);
\coordinate (ctlabeL) at (1541.5,343.6);
\coordinate (Ctlabel) at (1538.4,598.5);

\coordinate (CDm2) at (1862.2,819);
\coordinate (CDp2) at (2362.4,819);

\coordinate (legend) at (2820,1000);
\coordinate (Crightbottomcorner) at (2851.3,31.6);

\node[inner sep=0pt] at (Center)
    {\includegraphics[width=1442.5pt]{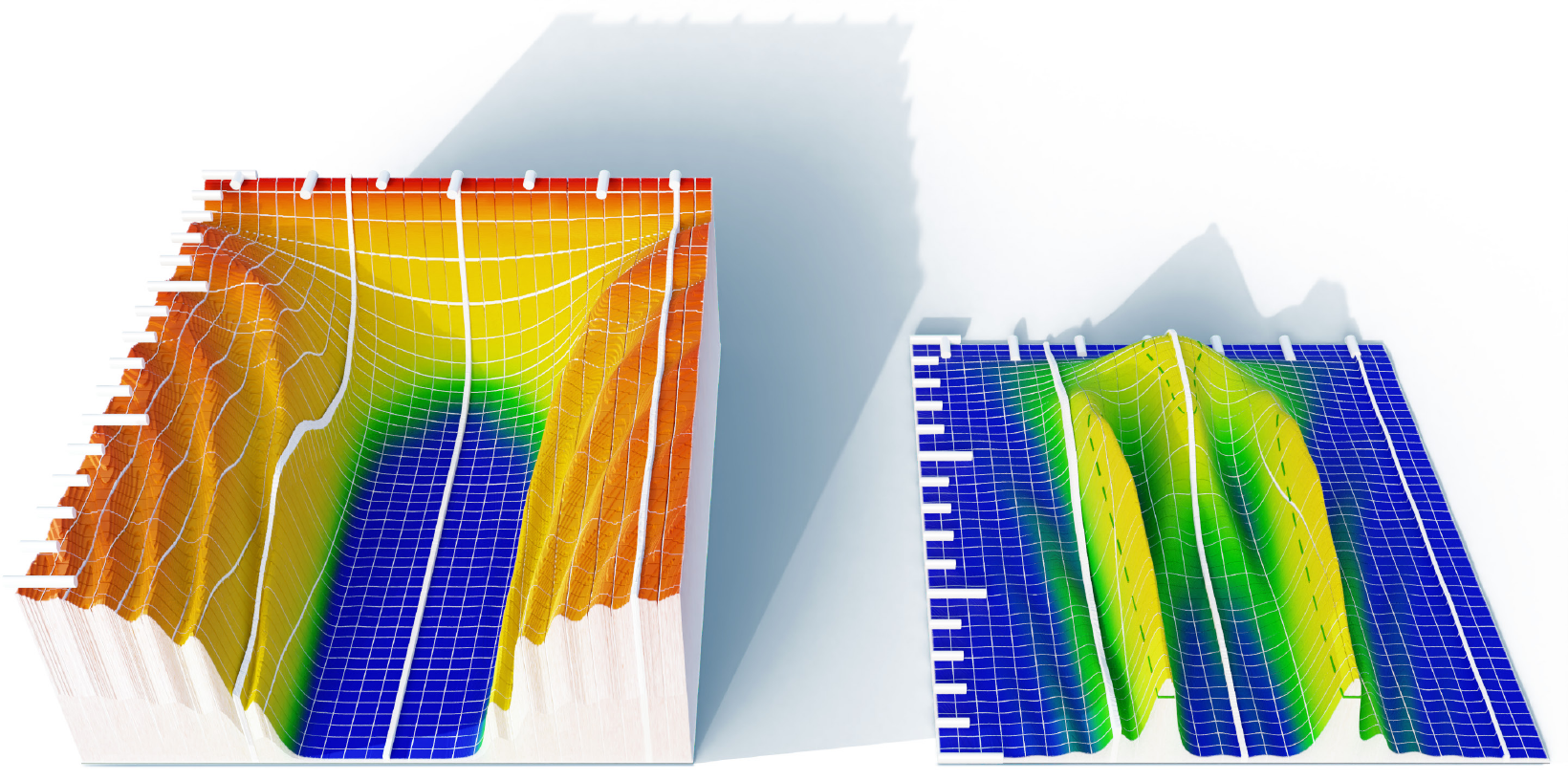}};

\foreach \pos/\num in {{0/-2},{0.5/0},{1/2}}
{\path[-] (ADm2) -- node[pos=\pos, above,scale=\labelScale] {$\num$} (ADp2);}
\foreach \pos/\num in {{0/0},{0.27/5},{0.6/10},{1/15}}
{\path[-] ([xshift=-2mm] At0) -- node[pos=\pos, left, inner sep=0, scale=\labelScale] {$\num$} ([xshift=-2mm] At15);}

\foreach \pos/\num in {{0/-2},{0.5/0},{1/2}}
{\path[-] (CDm2) -- node[pos=\pos, above,scale=\labelScale] {$\num$} (CDp2);}
\foreach \pos/\num in {{0/0},{0.28/5},{0.61/10},{1/15}}
{\path[-] (Ct0) -- node[pos=\pos, left,scale=\labelScale] {$\num$} (Ct15);}

\path[-] ([yshift=\margin] ADm2) -- node[sloped, midway, above,scale=\labelSCALE] {$\Delta/(2J)$} ([yshift=\margin] ADp2);
      
\path[-] ([yshift=\margin] CDm2) -- node[sloped, midway, above,scale=\labelSCALE] {$\Delta/(2J)$} ([yshift=\margin] CDp2);

\path[-] (Atlabel) -- node[sloped, midway,scale=\labelSCALE] {$2J\,t$} (atlabeL);
      
\path[-] (Ctlabel) -- node[sloped, midway, scale=\labelSCALE, rotate=180] {$2J\,t$} (ctlabeL);

\node[scale=1.3*\labelSCALE,below] at (A) {$\lvert A\rvert^2$};
\node[scale=1.3*\labelSCALE,below] at (C) {$\sum_j \lvert C_{1,j}\rvert^2+\lvert C_{2,j}\rvert^2$};

\node[inner sep=0pt] at (legend)
    {\includegraphics[width=\legendwidth pt]{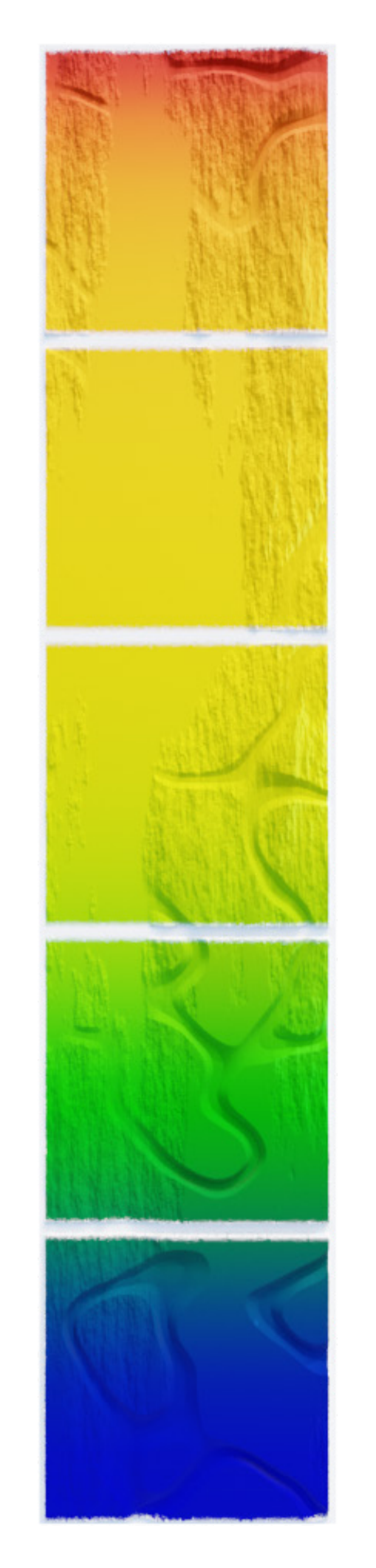}};
    
    \foreach \pos in {{0.0},{0.2},{0.4},{0.6},{0.8},{1.0}}
{\path[-] ([yshift=-1.9*\legendwidth pt,xshift=-0.4*\legendwidth pt] legend) -- node[pos=\pos, left,scale=\labelScale] {$\pos$} ([yshift=1.9*\legendwidth pt,xshift=-0.4*\legendwidth pt] legend);}
\end{tikzpicture}}
	\caption{\textbf{Dynamical evolution} of the probability $\lvert A(t)\rvert^2$ of having both emitters simultaneously excited and the probability $\sum_{j=-\infty}^\infty\lvert C_{1,j}(t)\rvert^2+\lvert C_{2,j}(t)\rvert^2$ of having only one excited for a coupling strength $g=1/3$. Three white lines in each plot delineate examples of dynamics dominated by Markovian decay (at $\Delta=0$), algebraic decay ($\Delta=-3J$), and a bound state ($\Delta=6J$). The dashed lines on the right plot are level lines of probability 0.5, the theoretical maximum if the excitations were independent.}
	\label{fig:Waterfall}
	\end{center}
\end{figure*}

Exponential, algebraic decay, and bound states always participate in the dynamics, but which one dominates the emission process depends on the interplay between $g$ and $\Delta$. Qualitatively, for weak couplings $g$ (see Fig.~\ref{fig:Waterfall}), the decay is mostly Markovian for deep in-band detunings $\Delta$, mostly algebraic when $\Delta$ is resonant with the band edges and suppressed for detunings far outside the band, when none of the two excitations may significantly leave its emitter (this means they form a two-excitation bound state of energy $\simeq2\Delta$). As one increases the coupling, the Markovian decay for in-band detunings becomes algebraic with pronounced oscillations that eventually become bound-state oscillations \cite{Stewart2020} for couplings that surpass the band width. The dynamics is not sensitive to detuning changes smaller than the coupling.\\

A clear signature of non-Markovian collective behavior can be found in the probability of having one excitation in the emitters and the other in the waveguide (the right plot of Fig.~\ref{fig:Waterfall}). If we use $p$ to denote the probability for the first excitation to be held by one of the emitters, then assuming independence between the two excitations would result in a probability $2p(1-p)$ of having just one emitter excited. Since $2p(1-p)\leq 1/2$ regardless of the value of $p$, 0.5 is a fundamental limitation for this probability, were the excitations independent. The collective nature of the decay allows it to break this limitation, as seen in the plot. As shown in Appendix~\ref{app:Markovian limit}, the violation is also a sign of non-Markovianity,  because the Markovian prediction for this probability peaks at 0.5 for $\Delta=0$ when the decay channels are not collectively enhanced or suppressed.\\

\section{Bound states}\label{sec:bound states}
The shape of two-excitation bound states can most easily be computed by reusing the equations that describe the transformed field amplitudes in Sections~\ref{sec:secular equations} and \ref{sec:analytic structure}. These equations (except those for $\tilde C_{1,1}$ and $\tilde C_{1,2}$ in Table~\ref{tab:closed formulas}) are affine, with a homogeneous part proportional to the field amplitudes and an inhomogeneous part. Removing the tildes and the inhomogeneous parts (e.g. through the substitutions $(2g\tilde C_{1,1}\sigma-1)\to 2g C_{1,1}\sigma$ in Table~\ref{tab:closed formulas}) yields equations that describe the field amplitudes of the bound states instead.\\

There are two ways to justify dropping the tildes and the inhomogeneous part to compute the bound states.
First, the eigenstate equation is formally identical to the transformed Schrödinger equation~\eqref{eq:tSchrEqn}, except for the tildes and a ``-1" that represents the initial conditions and breaks the linearity of the equations. Second, the transformed field amplitudes are divergent at the bound state energies, so the inhomogeneous terms are negligible in comparison. Scaling the amplitudes $A$ (from Eqn.~\eqref{eq:Atilde}), $C_{1,1}$ and $C_{1,2}$ (from Table~\ref{tab:closed formulas})  with the vanishing factor $\sum_{\sigma,i}\alpha_\sigma(z_{\sigma i})$ fixes the divergent terms and yields the correct (not yet normalized) multi-excitation bound-state amplitudes. This argument can be made more rigorous by using the residue theorem in the inverse transform (\ref{eq:inverseLaplace}) and separating the bound states using a harmonic decomposition of the solution. Examples of bound states are presented in Fig.~\ref{fig:Spectra}b.\\

\section{Conclusion}\label{sec:Conclusion}
In this work, we have made an original use of symmetries and analytic methods to solve the problem of two excited QEs coupled to a single-band waveguide. This is a minimal scenario for the study of collective non-Markovian decay, and the solution could be generalized to many other cases whose exact single-excitation dynamics are known \cite{Vacchini2010, Tufarelli2014, Rzazewski1982, Bay1997, Lombardo2014, Lanuza2022, Lambropoulos2000, GTudela2017, Facchi2019, Calajo2016, Burgess2022, Bello2019, Soro2023}. The most immediate generalizations are to revisit the multi-photon scattering problem \cite{Shi2015, Laakso2014, Roulet2016, Roy2017} by means of a different initial state; to change the distance between the emitters and investigate delay-induced entangled dark states \cite{Alvarez2023}; or to consider multiband waveguides, which can also be done analytically and efficiently by using infinite products \textit{\`a la Euler} \cite{Lanuza2022}.\\

Non-Markovian collective decay is commonly related to either strong coupling or retardation delays between emitters. This characterization oversimplifies the full complexity of the problem, which has infinitely many degrees of freedom. We instead treat the quantum emitters and the radiation modes as a collective in which every constituent partakes in the decay process since the spectral decay components of the solution cannot be traced back to individual origins.\\

The closed expression for the transformed field amplitudes in terms of elementary functions exposes a spectrum with a wide variety of decay types. Whereas the amount, frequencies, or decay rates of bound and exponentially-decaying states are independent of the single excitation spectrum, this is not the case for the algebraic decay components. The algebraic decay is generally caused by edges of continua involving free states, where the two excitations can be considered to be independent. A peculiar case of algebraic decay occurs at the center of the band, which is not singular for one excitation and does not lead to collective decay in the Markovian limit, and yet it is a source of collective non-Markovian decay with logarithmic modifications. These modifications might be the result of interference between superposed algebraic decays, as they also affect the edges of the continuum representing two free particles.\\

Despite their stark difference in functional dependence, logarithmic modifications to the decay would be extremely hard to measure, as they would require interrogating the emitters for exponentially long times with exponentially high precision. In contrast, finding more than a $50\%$ chance of having exactly one emitter excited should be feasible in state-of-the-art experiments while also being a conceptually clearer indicator of collective non-Markovian decay. Our results could be tested in a variety of platforms, including atoms
near photonic crystals \cite{Goban2015} 
 or optical fibers \cite{Liedl2023}, semiconductor quantum dots \cite{Lodahl2015}, matter-wave emitters \cite{Krinner2018}, quantum acoustic systems \cite{Chu2017}, or superconducting circuits \cite{Gu2017}.\\

We expect that the connections we established between the one- and two-excitation sectors can be extrapolated to arbitrary excitation numbers and that the solution presented here can also be used for studies of the fundamental connection between superradiance, synchronization \cite{Xu2014, Weiner2017}, and entanglement \cite{Gonzalez-Ballestero2013, Cascio2019, Lohof2023}.\\

\section*{Acknowledgements}\label{sec:Acknowledgements}
We thank Y. Kim for a critical reading of the manuscript. This work was supported by the U.S. National Science Foundation through grants PHY-1912546 and PHY-2208050.\\

\appendix
\renewcommand{\appendixname}{APPENDIX}
 \section{\MakeUppercase{Two excitations in the Markovian limit}}\label{app:Markovian limit}
  \begin{figure}[b]
	\begin{center}
	\includegraphics[width=1\columnwidth]{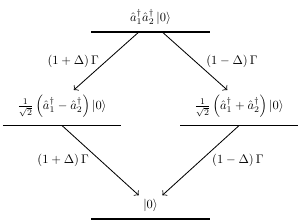}
	\caption{Decay channels and rates for the superradiant cascade of two excitations leaving two emitters through the coupling with a single sinusoidal band. \cite{Mlynek2014}}
	\label{fig:DecayChannels}
	\end{center}
\end{figure}

In this section, we present the Markovian approach \cite{Breuer2002} to this system for completeness. This section is somewhat independent of the rest of the paper because it will not be used to find the general solution.\\

 When each of the QEs is holding an excitation there is no phase relation between the two excitations, despite the initial state seeming symmetric. This allows for the system to decay into either the symmetric or anti-symmetric decay channels presented in Sec.~\ref{sec:single excitation}\textbf{\textit{ii}} (see Fig.~\ref{fig:DecayChannels}).\\

 The density matrix $\rho=\operatorname{Tr}_B \ket\psi\bra\psi$ resulting from partial-tracing over the waveguide subsystem $B$ follows a dynamics that in the Markovian limit are described by the master equation \cite{Breuer2002}\\
\begin{equation}
\begin{aligned}
\dot \rho
&=
-i \left[\sum_{j=1}^2 \Delta \hat a_j^\dagger\hat a_j,\rho\right]
\\&
+\sum_{\nu=1}^4\frac{\Gamma_\nu}{2}\left(2 \hat{\mathcal{O}}_\nu\rho\hat{\mathcal{O}}_\nu^\dagger-\rho \hat{\mathcal{O}}_\nu^\dagger \hat{\mathcal{O}}_\nu-\hat{\mathcal{O}}_\nu^\dagger \hat{\mathcal{O}}_\nu\rho\right),
\end{aligned}
\end{equation}\\
where $\nu$ runs over the decay channels of Fig.~\ref{fig:DecayChannels}, e.g. $\Gamma_1=(1+\Delta)\Gamma$ and $\hat{\mathcal{O}}_1=\sqrt{1/2}\left(\hat{a}_1^\dagger-\hat{a}_2^\dagger\right)\ket{0}\bra{0}\hat{a}_1\hat{a}_2$. This results in a $4\times4$ density matrix that is diagonal in the implied parity-explicit basis and describes exponential decay of the total population of the emitters,\\
\begin{equation}\label{eq:MarkovianDecay}
    \begin{aligned}
    \operatorname{Tr}&\left(\hat {a}^\dagger_1\hat {a}_1+\hat {a}^\dagger_2\hat {a}_2\right)\rho
    =\\&
    \frac{1-\Delta}{1+\Delta}e^{-\left(1-\Delta\right)\Gamma t}
    +\frac{1+\Delta}{1-\Delta}e^{-\left(1+\Delta\right)\Gamma t}-\frac{4\Delta^2}{1-\Delta^2}e^{-2\Gamma t}.
\end{aligned}
\end{equation}\\

We note that this decreases monotonically, so although they cooperate, two distant quantum emitters are not enough to produce a superradiant burst \cite{Masson2020}.\\

The decay channels illustrated in Fig.~\ref{fig:DecayChannels} become identical at $\Delta=0$, and~\eqref{eq:MarkovianDecay} reduces to the decay of two independent emitters. This signals a suppression of the collective effects, similar to the single-excitation case discussed in Section~\ref{sec:single excitation}.\\

Another relevant observable (see Section~\ref{sec:spectrum}) is the probability of finding exactly one excitation in the emitters,
\begin{equation}
    \begin{aligned}
    \operatorname{Tr}&\left(\hat {a}^\dagger_1\hat {a}_1+\hat {a}^\dagger_2\hat {a}_2-2\hat {a}^\dagger_1\hat {a}^\dagger_2\hat {a}_1\hat {a}_2\right)\rho
    =\\&
    \frac{1-\Delta }{1+\Delta}e^{-(1-\Delta)\Gamma  t}+\frac{1+\Delta}{1-\Delta }e^{-(1+\Delta)\Gamma t}-2\frac{1+\Delta ^2}{1-\Delta ^2}e^{-2 \Gamma  t}.
\end{aligned}
\end{equation}
The global maximum of this function is 0.5 at $\Delta=0$ and $\Gamma t=\log 2$, so any evidence of a higher value would signal physics beyond the Markovian approximation.\\

 \section{\MakeUppercase{Dynamics of the transformed  field amplitudes}}\label{app:dynamics of the transformed amplitudes}
For the exact non-Markovian solution we make use of the \Schr equation, which in terms of the transformed Bloch basis~\eqref{eq:tpsi} and the initial conditions $A(0)=1$ and $B_{p,q}(0)=C_{j,q}(0)=0$ becomes\\
\begin{equation}\label{eq:tSchrEqn}
\left\lbrace
\begin{array}{l}
    \omega\tilde A-1=2\Delta\tilde A+g\sum\limits_q \left(e^{iqx_1}\tilde C_{2,q}+e^{iqx_2}\tilde C_{1,q}\right)\\
    \omega\tilde B_{p,q}=(\omega_p+\omega_q)\tilde B_{p,q}\\
    \qquad +g\sum\limits_{j=1}^2 \left(e^{-ipx_j}\tilde C_{j,q}+e^{-iqx_j}\tilde C_{j,p}\right)\qquad\text{if }p<q\\
    \omega\tilde B_{q,q}=2\omega_q\tilde B_{q,q}+\sqrt{2}g\sum\limits_{j=1}^2 e^{-iqx_j}\tilde C_{j,q}\\
    \omega\tilde C_{j,q}=(\Delta+\omega_q)\tilde C_{j,q}+g e^{iqx_j}\tilde A\\
    \quad+g\left(\sum\limits_{p< q} e^{ipx_j}\tilde B_{p,q}+\sum\limits_{q< p} e^{ipx_j}\tilde B_{q,p}
    +\sqrt{2}e^{iqx_j}\tilde B_{q,q}\right),
\end{array}\right.
\end{equation}
where $x_1=-d/2$ and $x_2=+d/2$ are the positions of the quantum emitters on the waveguide. Solving for the $\tilde C_{j,q}$, we get Eqns.~\eqref{eq:tildeA,BfromC} and the additional equation
\begin{widetext}
\begin{equation}\label{eq:tC}
    \left(\omega-\Delta-\omega_q\right)\tilde C_{j,q}=\frac{ge^{iqx_j}}{\omega-2\Delta}+g^2\sum_p\frac{e^{i(px_1+qx_j)} \tilde C_{2,p}+e^{i(px_2+qx_j)} \tilde C_{1,p}}{\omega -2\Delta}+g^2\sum_{p}\sum_{j'=1}^2\frac{e^{ip(x_j-x_{j'})}\tilde C_{j',q}+e^{i(px_j-qx_{j'})}\tilde C_{j',p}}{\omega-\omega_p-\omega_q}.
\end{equation}
We can use this equation to prove that $\tilde C_{1,q}(\omega)=\tilde C^*_{2,q}(\omega)$ for all $\omega\in \mathbb{R}$. Beyond informing us about the form of the bound states (see Fig.~\ref{fig:Spectra}b), this allows us to obtain $\tilde C_{2,q}$ from $\tilde C_{1,q}$ through analytic continuation. For the proof, notice that conjugating this equation while rewriting $j\to\neg j$ and $j'\to\neg j'$, the equation becomes
\begin{equation}
    \left(\omega-\Delta-\omega_q\right)\tilde C^*_{\neg j,q}=\frac{ge^{iqx_j}}{\omega-2\Delta}+g^2\sum_p\frac{e^{i(px_1+qx_j)} \tilde C^*_{1,p}+e^{i(px_2+qx_j)} \tilde C^*_{2,p}}{\omega -2\Delta}+g^2\sum_{p}\sum_{j'=1}^2\frac{e^{ip(x_j-x_{j'})}\tilde C^*_{\neg j',q}+e^{i(px_j-qx_{j'})}\tilde C^*_{\neg j',p}}{\omega-\omega_p-\omega_q},
\end{equation}
which is identical to the original writing~\eqref{eq:tC} after exchanging $\tilde C_{1,q}\leftrightarrow\tilde C^*_{2,q}$. Since the dynamics are unequivocally determined by the \Schr equation and initial conditions, \eqref{eq:tC} has a unique solution and therefore $\tilde C_{1,q}=\tilde C^*_{2,q}$. We note that this also implies through \eqref{eq:tildeA,BfromC} that $\tilde A, \tilde B_{p,q}\in \mathbb{R}$ for $\omega\in \mathbb{R}$.\\

Using this result, Eqn.~\eqref{eq:tC} with $j=1$ and arbitrary $\omega\in\mathbb{C}$ can also be written as
\begin{equation}
   (\domega-\Delta) C(\omega,z)=\frac{g/z}{\omega-2\Delta}+\frac{g^2/z}{2\pi i}\oint\left\lbrace
    \frac{2C(\omega,z')}{\omega-2\Delta}+2\frac{zC(\omega,z)+C(\omega,1/z)/z'+(z+z')C(\omega,z')}{1+2\domega\,z'+z'^2}
    \right\rbrace\dif z'
\end{equation}
in terms of the function $C(\omega,z)$ defined in~\eqref{eq:CLaurent}. After splitting $C$ into symmetric and antisymmetric parts as in~\eqref{eq:Csym}, this equation simplifies into~\eqref{eq:secularEqn}.
\end{widetext}

 \section{\MakeUppercase{Increments of $C_\sigma$}}\label{app:increments of C}
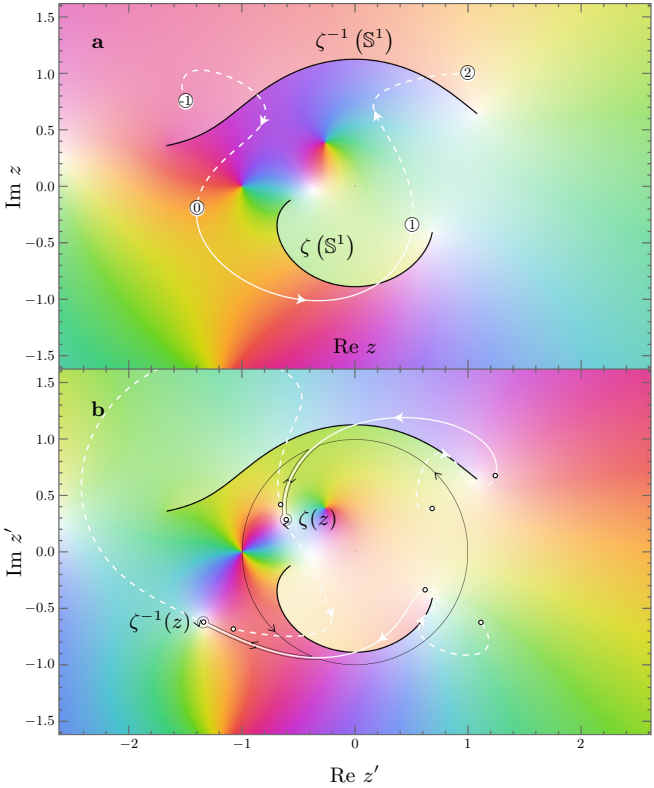
\begin{figure}[b]
    \begin{center}
\resizebox{0.485\textwidth}{!}{\begin{tikzpicture}[x=.5pt,y=.5pt]

\def\labelScale{1}
\def\labelSCALE{1.5}
\def\margin{36}
\def\legendwidth{51}

\coordinate (Center) at (379.5,469);
\coordinate (topleft) at (0,938);

\coordinate (a) at (50,938-50);
\coordinate (b) at (50,469-50);

\coordinate (ReMint) at (90.05,1.35);
\coordinate (RePint) at (668.71,1.35);
\coordinate (ImMint) at (0.97,18.41);
\coordinate (ImPint) at (0.97,452.18);

\coordinate (ReMC) at (90.05,469.28);
\coordinate (RePC) at (668.71,469.28);
\coordinate (ImMC) at (0.97,486.39);
\coordinate (ImPC) at (0.97,920.15);

\coordinate (zetainv S1) at (379.5,890);
\coordinate (zeta S1) at (345,625);

\coordinate (-1) at (162.63,812.25);
\coordinate (0) at (177.06,675.27);
\coordinate (1) at (452.84,652.96);
\coordinate (2) at (524.07,847.91);

\coordinate (Re z) at (379.38,469.28+30);
\coordinate (Im z) at (0.97-60,703.27);
\coordinate (Re zz) at (379.38,1.35-50);
\coordinate (Im zz) at (0.97-60,235.295);

\coordinate (zeta0) at (559.15,332.98);
\coordinate (zetainv0) at (469.32,186.52);
\coordinate (zeta-1) at (478.25,290.81);
\coordinate (zetainv-1) at (540.79,144.73);
\coordinate (zeta2) at (223.42,136.23);
\coordinate (zetainv2) at (284.05,295.86);
\coordinate (zeta1) at (291.28,276.56);
\coordinate (zetainv1) at (185.43,145.13);
\node[inner sep=0pt] at (Center)
    {\includegraphics[width=379.5pt]{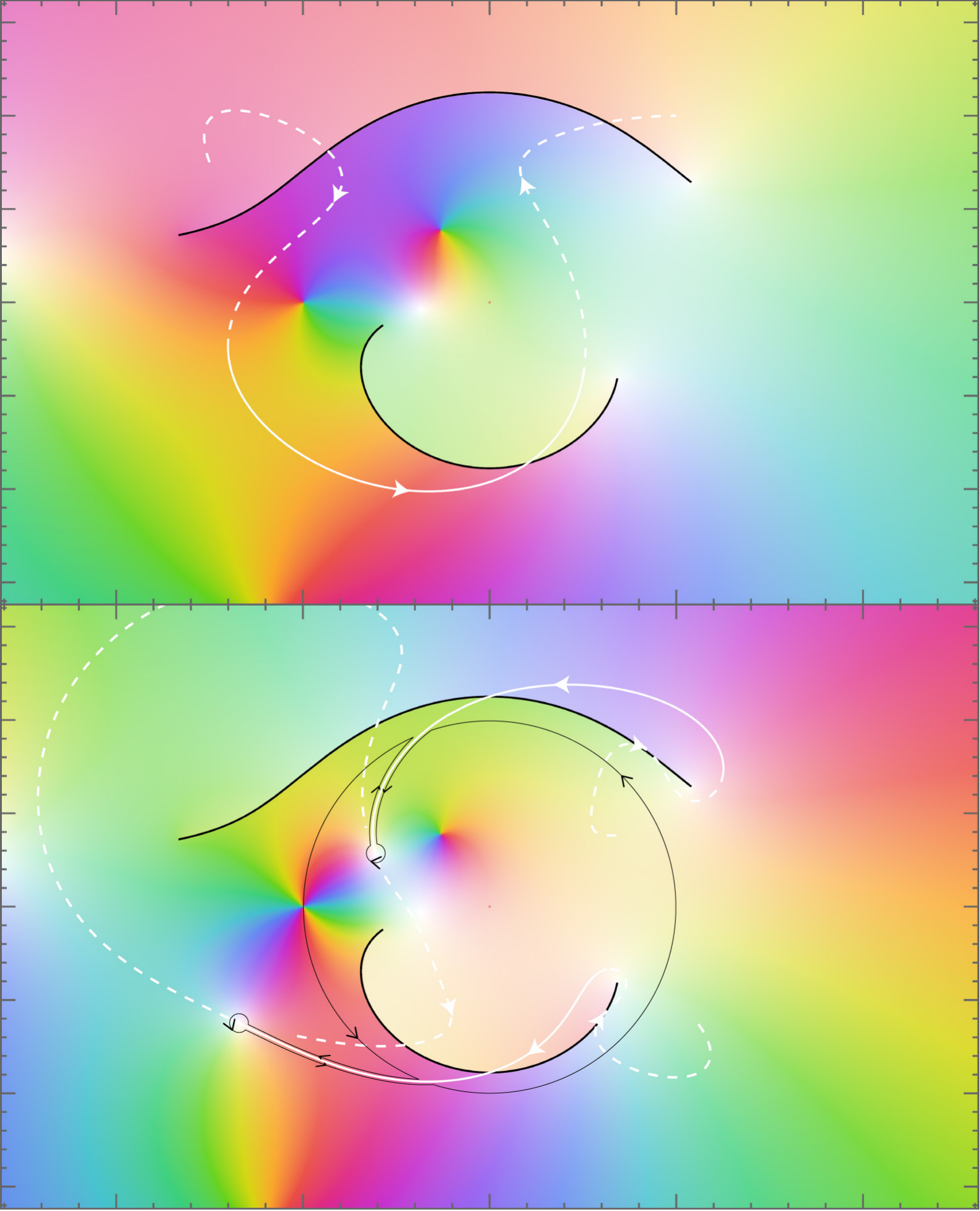}};

\foreach \pos/\num in {{0/-1.5},{1/-1.0},{2/-0.5},{3/0.0},{4/0.5},{5/1.0},{6/1.5}}
{\path[-] (ImMC) -- node[pos=0.16666*\pos, left,scale=\labelScale] {$\num$} (ImPC);}

\foreach \pos/\num in {{0/-2},{0.25/-1},{0.5/0},{0.75/1},{1/2}}
{\path[-] (ReMint) -- node[pos=\pos, below,scale=\labelScale] {$\num$} (RePint);}
\foreach \pos/\num in {{0/-1.5},{1/-1.0},{2/-0.5},{3/0.0},{4/0.5},{5/1.0},{6/1.5}}
{\path[-] (ImMint) -- node[pos=0.16666*\pos, left,scale=\labelScale] {$\num$} (ImPint);}

\node[scale=\labelSCALE] at (a) {\textbf{a}};
\node[scale=\labelSCALE] at (b) {\textbf{b}};

\foreach \n in {{-1},{0},{1},{2}}
{\node[circle,fill=white,inner sep=-0.5pt, scale=2] at (\n) {$\circled{\n}$};}

\node[scale=\labelSCALE] at (zetainv S1) {$\zeta^{-1}\left(\mathbb{S}^1\right)$};
\node[scale=\labelSCALE] at (zeta S1) {$\zeta\left(\mathbb{S}^1\right)$};

\node[scale=\labelSCALE,left] at ([xshift=-3pt]zetainv1) {$\zeta^{-1}(z)$};
\node[scale=\labelSCALE,right] at ([xshift=3pt]zeta1) {$\zeta(z)$};

\node[scale=\labelSCALE, rotate=90] at (Im z) {Im $z$};
\node[scale=\labelSCALE] at (Re z) {Re $z$};
\node[scale=\labelSCALE, rotate=90] at (Im zz) {Im $z'$};
\node[scale=\labelSCALE] at (Re zz) {Re $z'$};

\foreach \n in {{-1},{0},{1},{2}}
{\node[circle,draw,fill=white,inner sep=-1pt] at (zeta\n) {};
\node[circle,draw,fill=white,inner sep=-1pt] at (zetainv\n) {};}
\end{tikzpicture}}
\caption{\textbf{a} Domain coloring plot of $C^\circled{0}_\sigma (z)$ for the same parameters as in Fig.~\ref{fig:Towers}. In white, a path for $z$ that crosses the branch cuts $\zeta^{\pm1}\left(\mathbb{S}^1\right)$ (in black) is suggested. The circled numbers enumerate the analytic continuations $C^\circled{-1}_\sigma (z)$, $C^\circled{0}_\sigma (z)$, $C^\circled{1}_\sigma (z)$, and $C^\circled{2}_\sigma (z)$ that the path traverses. \textbf{b} Corresponding domain coloring plot of Eqn.~\eqref{eq:secularEqn}'s integrand at $z=0.5 - 0.4 i$. The integration contour (originally $\mathbb{S}^1$, in black) has been adapted to avoid the mobile poles $\zeta^{\pm1}(z)$, whose trajectory as $z$ follows the white path in \textbf{a} is marked with white lines and black circles.}
\label{fig:path deformation}
	\end{center}
\end{figure}

As mentioned in Sec.~\ref{sec:analytic structure}, Eqn.~\eqref{eq:secularEqn} imposes a very particular analytic structure on the function $C_\sigma(z)$. For starters, knowing the value of the function for values of $z\in \mathbb{S}^1$ (corresponding to real quasimomenta) allows to determine the analytic continuation $C^\circled{0}_\sigma (z)$ of the function to all $z\in\mathbb{C}\setminus\left(\zeta\left(\mathbb{S}^1\right)\cup\zeta^{-1}\left(\mathbb{S}^1\right)\right)$, through direct integration of the RHS in~\eqref{eq:secularEqn}.\\

Continuing the function to $C^\circled{1}_\sigma (z)$ beyond the $\zeta\left(\mathbb{S}^1\right)$ branch cut (in Fig 2a, this corresponds to crossing from the second lowest to the lowest sheet) is possible by deforming the integration contour, as shown in Fig.~\ref{fig:path deformation}. Subsequently applying the residue theorem results in
\begin{equation}\label{eq:C1}
\begin{aligned}
  &C^\circled{1}_\sigma (z)=\frac{\fun_{-,\sigma}(z)}{\fun_{+,\sigma}(z)}C^\circled{0}_\sigma (z)\\&
  -2\fun_{-,\sigma}(z)\frac{\zeta(z)+\sigma}{\zeta(z)-\zeta^{-1}(z)}\times\left\{
  \begin{array}{lc}
       C^\circled{0}_\sigma (\zeta(z))& \text{if } \lvert z\rvert<1 \\
       C^\circled{1}_\sigma(\zeta(z))&  \text{if } \lvert z\rvert>1 ,
  \end{array}
  \right.
\end{aligned}
\end{equation}
where for briefness we introduced
\begin{equation}
    \fun_{\pm,\sigma}(z)\coloneqq
\frac{\frac{1}{z}+\sigma}{g^{-2}\left(\domega-\Delta\right)\pm\sigma\frac{\sqrt{\domega-\sigma}}{\sqrt{\domega+\sigma}}-\sigma}.
\end{equation}
We note that crossing the same branch cut again in either direction brings you back to $C^\circled{0}_\sigma (z)$, as the modification to the integration contour is reverted around the poles. This is no longer the case (i.e. $C^\circled{2}_\sigma (z)\neq C^\circled{0}_\sigma (z)$) if one subsequently crosses the other branch cut instead, but unexpectedly both increments
\begin{equation}\label{eq:dC}
    \dC_\sigma(z)=C^\circled{1}_\sigma (z)-C^\circled{0}_\sigma (z)
\end{equation}
and
\begin{equation}
    \ddC_\sigma(z)=C^\circled{0}_\sigma (z)-C^\circled{2}_\sigma (z)
\end{equation}
become meromorphic functions when multiplied by $\sqrt{\domega+\sigma}/\sqrt{\domega-\sigma}$.\\

The proof for $\dC_\sigma$ is elaborated (for $\ddC_\sigma$ it then follows from~\eqref{eq:ddC}). We start by noticing that, by construction, the only possible branch cuts of $\dC_\sigma$ are $\zeta\left(\mathbb{S}^1\right)$ and $\zeta^{-1}\left(\mathbb{S}^1\right)$. This is how the function extends when we cross them:
\begin{equation*}
    \begin{array}{c}
    \dC_\sigma =C^\circled{1}_\sigma-C^\circled{0}_\sigma
    \xrightarrow[]{\zeta\left(\mathbb{S}^1\right)}C^\circled{0}_\sigma -C^\circled{1}_\sigma=-\dC_\sigma
\\
    \dC_\sigma =C^\circled{1}_\sigma-C^\circled{0}_\sigma
    \xrightarrow[]{\zeta^{-1}\left(\mathbb{S}^1\right)}
    C^\circled{2}_\sigma-C^\circled{-1}_\sigma.
    \end{array}
\end{equation*}
The last function can also be equated to $-\dC_\sigma$. To do so, one should write $C^\circled{2}_\sigma(z)$ and $C^\circled{-1}_\sigma(z)$ in terms of $C^\circled{0}_\sigma (z)$ and $C^\circled{0}_\sigma (\zeta(z))$, which is a tedious calculation but shows that indeed the equality holds iff
\begin{equation}
\begin{aligned}
4&\frac{\zeta (z)+\sigma }{\zeta (z)-\zeta^{-1}(z)}\frac{z+\sigma }{z-z^{-1}}
\\
&+\left(\fun_{+,\sigma}^{-1}(\zeta(z))-\fun_{-,\sigma}^{-1}(\zeta(z))\right)\left(\fun_{+,\sigma}^{-1}(z)-\fun_{-,\sigma}^{-1}(z)\right)=0,
\end{aligned}
\end{equation}
which, in turn, holds because
\begin{equation}
    \fun_{+,\sigma}^{-1}(z)-\fun_{-,\sigma}^{-1}(z)=\frac{2 \sigma }{\sigma +z^{-1}}\frac{\sqrt{\domega-\sigma }}{\sqrt{\domega+\sigma }}.
\end{equation}
This concludes that $\dC_\sigma(z){\sqrt{\domega+\sigma}}/{\sqrt{\domega-\sigma}}$ is meromorphic.\\

In fact, using these formulas and briefly denoting $\operatorname{sl}(z)\coloneqq \text{sign}\log\lvert z\rvert$ for all $z\notin\mathbb{S}^1\cup\{0\}$, one can rewrite the function as
\begin{equation}\label{eq:zetaSymmForC}
\begin{aligned}
&\dC_\sigma(z)\frac{\fun_{+,\sigma}(\zeta(z))}{\fun_{-,\sigma}(z)\fun_{-\operatorname{sl}(z),\sigma}(\zeta(z))}=
\\&
\frac{2 \sigma }{\sigma +z^{-1}}\frac{\sqrt{\domega-\sigma }}{\sqrt{\domega+\sigma }}C_\sigma(z)
-2\frac{\zeta(z)+\sigma}{\zeta(z)-\zeta^{-1}(z)}C_\sigma(\zeta(z))
\end{aligned}
\end{equation}
which can be regarded as the way in which $C_\sigma(z)$ transforms under the $z\leftrightarrow\zeta(z)$ symmetry (in the sense specified in App.~\ref{app:soft symmetry}). We use this formula to extract the symmetries of $\dC_\sigma$,
\begin{equation}
\begin{array}{c}
    \dC_\sigma\left(\frac{1}{z}\right)=\sigma z\frac{\fun_{+\operatorname{sl}(z),\sigma}(\zeta(z))}{\fun_{-\operatorname{sl}(z),\sigma}(\zeta(z))}\dC_\sigma(z)
\\
\dC_\sigma\left(\zeta(z)\right)=\operatorname{sl}(z) \frac{\sqrt{\domega+\sigma }}{\sqrt{\domega-\sigma }}

\frac{\fun_{+\operatorname{sl}(z),\sigma}(\zeta(z))}{\fun_{-,\sigma}(z)}\frac{z+\sigma}{z-\sigma}\dC_\sigma(z).
\end{array}
\end{equation}
With these symmetries, we build up the projectors
\begin{equation}
\begin{array}{c}
    P_{\operatorname{inv}}\lbrace f_\sigma\rbrace(z)=\frac{1}{2}\left(f_\sigma(z)+\frac{\sigma}{z}\frac{\fun_{-\operatorname{sl}(z),\sigma}(\zeta(z))}{\fun_{+\operatorname{sl}(z),\sigma}(\zeta(z))}f_\sigma\left(z^{-1}\right)\right)
\\
    P_{\zeta}\lbrace f_\sigma\rbrace(z)=\frac{1}{2}\Big(f_\sigma(z)+\hspace{4.2cm}
    \\
    \hspace{1.5cm}+\operatorname{sl}(z)  \frac{\sqrt{\domega-\sigma }}{\sqrt{\domega+\sigma }}
    \frac{\fun_{-,\sigma}(z)}{\fun_{+\operatorname{sl}(z),\sigma}(\zeta(z))}\frac{z-\sigma}{z+\sigma}    f_\sigma\left(\zeta(z)\right)\Big)
\end{array}
\end{equation}
that take functions of the form meromorphic$\times$ $ {\sqrt{\domega-\sigma}}/{\sqrt{\domega+\sigma}}$ and imbues in them the corresponding symmetry. $P_{\zeta}\circ P_{\operatorname{inv}}$ spans a linear subspace of functions so limited that only one has the correct asymptotic dependence for $\dC_\sigma(z\to\infty)$, which can be found by using~\eqref{eq:secularEqn} to extract
\begin{equation}\begin{aligned}
 &C_\sigma(z)=
  \sigma z^{-1}\frac{g \left(1+2g\tilde C_{1,2}\right)}
    {\omega-2\Delta}
  +
  \\&
  g\frac{1-2\sigma  (\omega -\Delta )+2g\tilde C_{1,1} (\omega-2\Delta)+2g\tilde C_{1,2} (1-\sigma  \omega )} {z^{2}(\omega-2\Delta)}
  \\&
 \hspace{5cm}+O\left(z^{-3}\right)
    \end{aligned}
\end{equation}
and combine it with~\eqref{eq:zetaSymmForC} to obtain
\begin{equation}
\begin{aligned}
  \dC_\sigma(z)&=
  8g^3 z^{-3}(2g\tilde C_{1,1}\sigma-1)+O\left(z^{-4}\right)
  \\&
  =
  8g^3 \sigma z^{2}(2g\tilde C_{1,1}\sigma-1)+O\left(z^{3}\right).
  \end{aligned}
\end{equation}
The only function matching the symmetries and asymptotic dependence is listed as $\dC_\sigma(z)$ in Table~\ref{tab:closed formulas}.

 \section{\MakeUppercase{The function $\Phi$}}\label{app:the function Phi}
The core idea behind our solution is that since the overall form of the analytic structure of $C_\sigma(z)$ (see Fig.~\ref{fig:Towers}) is independent of the system parameters, we can define a simpler parameter-free version $\Phi(z)$ of the function hosting the same overall structure.\\

As a starting point, one can tweak Eqn.~\eqref{eq:secularEqn} to simplify it while keeping the essential analytic structure of the solution unaltered. There are many choices for this; we propose
\begin{equation}\label{eq:PhiDefn}
    \Phi(z)=\tfrac{z+1}{z-1}\left(\frac{1}{4}+\oint\tfrac{z'-1/z'}{1+2\domega z'+z'^2}\frac{\sqrt{\domega'+1}}{\sqrt{\domega'-1}}\frac{ \Phi(z')\dif z'}{2\pi i}\right),
\end{equation}
where $\domega'\coloneqq\omega+(z'+z'^{-1})/2$ and the integration takes place around a positively oriented circle of radius $1^-$ and center in $0$.\\

Some properties that follow from the equation are the following. i) Symmetries:
\begin{equation}\label{eq:Phisyms}
   \left\lbrace \begin{array}{l}
   \Phi(z)=-\Phi(1/z)  \\
    \Phi(z)=\operatorname{sl}(z)(\Phi(\zeta(z))+1/2)
   \end{array}
   \right.
\end{equation}

ii) Asymptotic expansion:
\begin{equation}
    \Phi(z)=\frac{1}{4}\frac{z+1}{z-1}+O\left(\lvert z\rvert^{-1}\right)
\end{equation}

iii) $\Phi(-1)=0$ and $(z-1)\Phi(z)\in \mathcal{H}\big(\mathbb{C}\setminus \big(\zeta\left(\mathbb{S}^1\right)\cup\zeta^{-1}\left(\mathbb{S}^1\right)\big)\big)$. This can be used to regularize the integral equation of $(z-1)\Phi(z)$ in the unit circumference. This allows for a Picard iteration scheme to both prove existence and uniqueness and to compute the numerical solution of the equation, albeit not efficiently.\\

iv) Analytic extension (see Fig.~\ref{fig:Towers}b). It can be achieved by carefully changing the shape of the integration contour, completely analogous to the study of the analytic extensions of $C_\sigma(z)$ presented in App.~\ref{app:increments of C}.\\

v) An algorithm to compute $\Phi$ is the following. Due to the analytic structure of this function, it can be written as
\begin{equation}\label{eq:PhiFit}
    \Phi(z)=-\frac{1}{2}+\frac{1+\sum_{n=1}^\infty a_n z^n}{4(1-z)}\sqrt{\frac{\domega-1}{\domega+1}}\quad\forall\lvert z\rvert<\frac{1}{\lvert\zeta(\pm1)\rvert}.
\end{equation}
Applying \eqref{eq:Phisyms}, this has to be equal to
\begin{equation}
     \Phi_\zeta(z)=\frac{1+\sum_{n=1}^\infty a_n \zeta(z)^n}{4(1-\zeta(z))}\frac{z+1}{z-1}\quad\forall z\in \mathbb{C},
\end{equation}
where the convergence is improved to the whole complex plane thanks to the fact that always $\lvert \zeta(z)\rvert\leq1$. Now, one can show that any function with the generic analytic shape of \eqref{eq:PhiFit} that also has property \eqref{eq:Phisyms} satisfies equation \eqref{eq:PhiDefn}. In other words, enforcing
\begin{equation}\label{eq:PhiZeta}
     \Phi(z)=\Phi_\zeta(z)\qquad\forall\lvert z\rvert<\operatorname{min}\lbrace\lvert\zeta(1)\rvert^{-1},\lvert\zeta(-1)\rvert^{-1}\rbrace
\end{equation}
can be used to fit the real coefficients $\lbrace a_n\rbrace_{n=1}^\infty$ unequivocally, and subsequently evaluate $\Phi$ in any point of the complex plane through the expression for $\Phi_\zeta(z)$ in~\eqref{eq:PhiZeta}. Since it is essentially solving a least-square problem, this method is efficient in practice if one chooses to enforce this along equispaced points on the circumference $\mathbb{S}^1$ excluding the singular $z=1$ (this choice makes the branch cut of the $a_n(\omega)$ to be at $\omega\in[-2,2]$). And the accuracy of the algorithm gets compromised the closer one gets to this branch cut.\\

vi) The analytic structure displayed in Fig.~\ref{fig:Towers}b simplifies greatly by taking the derivative with respect to $z$, to the point that it can be expressed algebraically as
\begin{equation}\label{eq:Phi'}
\begin{aligned}
    &\quad\Phi'(z)\propto 
    \\&
    \frac{(z-z_0)(z-z^{-1}_0)(z-\zeta(z_0))(z-\zeta^{-1}(z_0))}{(\domega^2-1)z^2(z-1)^2}\frac{\sqrt{\domega-1}}{\sqrt{\domega+1}},
\end{aligned}
\end{equation}
where $z_0$ is one of the four symmetric points where $\Phi'(z_0)=0$.\\

vii) Reintegrating the expression above means that $\Phi(z)$ can be expressed in terms of elliptic integrals \cite{Abramowitz1972}. In the Legendre normal form, the incomplete elliptic integral of the first and second kind we use are presented in Table~\ref{tab:closed formulas} together with the integrated expression for $\Phi(z)$. Our choice of branch cuts for these functions matches the branch cuts of their respective integrands (where we also choose arg$\sqrt{z}\in(-\pi/2,\pi/2]$ ).\\

While integrating, it is useful to notice that $z_0\equiv z_0(\omega)$ can be extracted from
\begin{equation}
\begin{aligned}
    \omega +&\frac{z_0+z_0^{-1}}{2}
    =
    \\&
    \frac{\omega}{2}-\frac{\omega +2}{2}\sqrt{1-\frac{2\omega}{\omega+1+ k\cdot\left(1-\frac{E\left(k^{-1}\right)}{K\left(k^{-1}\right)}\right)}},
\end{aligned}
\end{equation} 

whereas taking the opposite sign in the large square root yields $\omega+(\zeta(z_0)+\zeta^{-1}(z_0))/2$ instead. Also one can express the residue
\begin{equation}
    \operatorname{Res}\lbrace \Phi(z),z=1\rbrace=\frac{\sqrt{\omega(2+\omega)k^{-1}}}{\pi}K\left(k^{-1}\right).
\end{equation}

The second line in the closed formula of $\Phi(z)$ at Tab.~\ref{tab:closed formulas} is very particular and, although there is no value of $\omega$ or $z$ such that $x(z)=1$, if we enforce this, the second line becomes $\pm1$ due to Legendre's relation. In other words, excluding algebraic functions (see App.~\ref{app:soft symmetry}), $\Phi(z)$ is a modification of Legendre's relation for incomplete elliptic integrals.\\

viii) If we include the dependencies in $\omega$ explicitly, we have the additional symmetry\\
\begin{equation}\label{eq:omegaSym}
\begin{aligned}
    \Phi& (\omega ,z)=\Phi (-\omega ,-z) +
    \\&
    \frac{\operatorname{sgnRe}\omega\ \ 2 \omega  \left(z^2+\omega  z+1\right) K\left(k^{-1}(\omega)\right)}{\pi  \left(z^2-1\right) \sqrt{k(\omega )} \sqrt{\domega+1} \sqrt{\domega-1}}.
\end{aligned}
\end{equation}\\

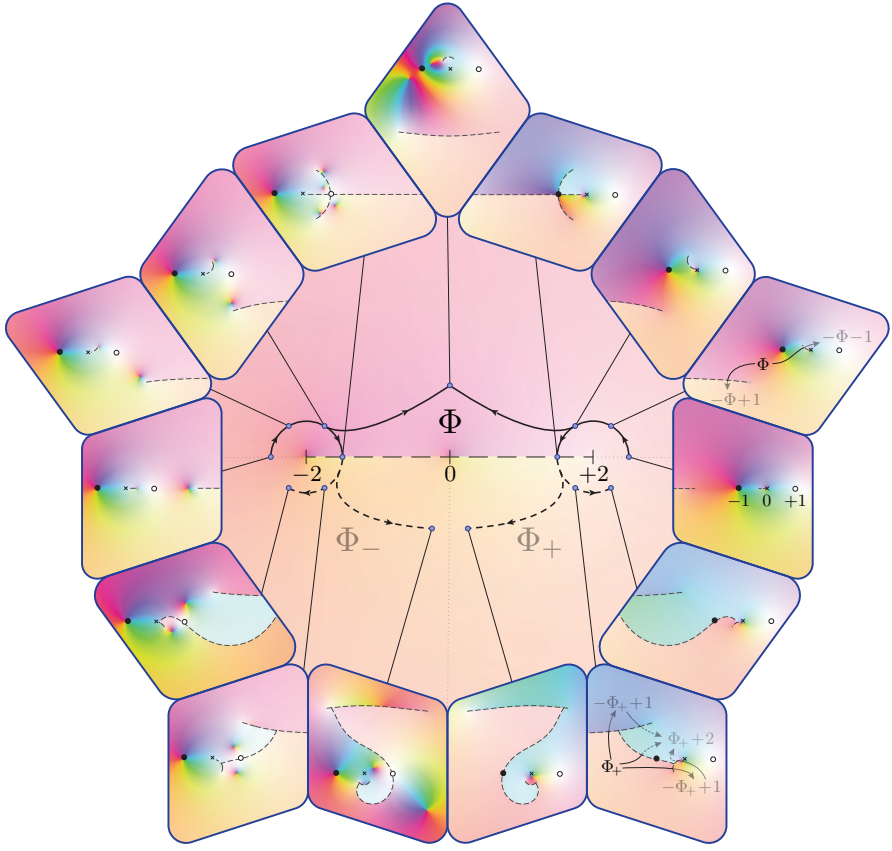
\begin{figure*}[t]
	\begin{center}
        \resizebox{0.66\textwidth}{!}{\begin{tikzpicture}[x=.5pt,y=.5pt]
\def\labelscale{0.6}
\def\labelScale{0.8}
\def\labelSCALE{1.2}

\coordinate (-1) at (461.54,224.63);
\coordinate (+1) at (497.06,224.63);

\coordinate (-2) at (190.88,244.41-0.5);
\coordinate (+2) at (370.72,244.41-0.5);

\coordinate (-miniphi-1) at (529.8,316.8+3);
\coordinate (miniphi) at (476.97,302.93);
\coordinate (-miniphi+1) at (450.68+9,279.4+1);

\coordinate (miniphiplus+2) at (417.66+14,67.1);
\coordinate (-miniphiplus+1) at (437.04-5,36.93);
\coordinate (-miniphiplus+01) at (389.05,89.2);
\coordinate (miniphiplus) at (382.34,49.22);

\coordinate (phiplus) at (337.17,190.6);
\coordinate (phi) at (280.66,266.55);
\coordinate (phiminus) at (223.82,190.59);

\node[inner sep=0pt] at (278.96,265.5)
    {\includegraphics[width=278.96pt]{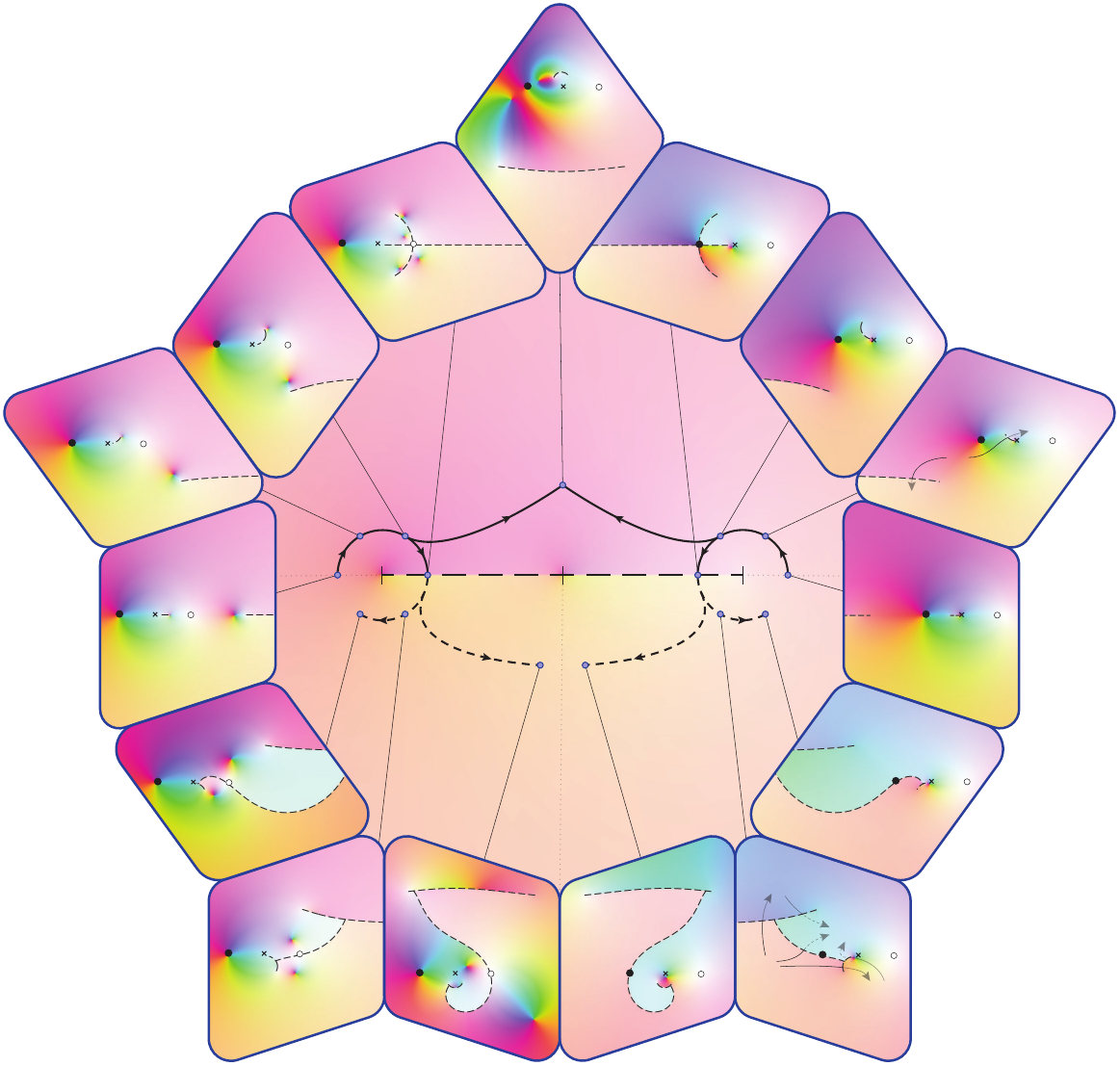}};

\node[below,scale=\labelScale] at (-2) {$-2$};
\path[-latex'] (-2) -- (+2) node[midway,below,scale=\labelScale] {$0$};
\node[below,scale=\labelScale] at (+2) {$+2$};

\node[scale=\labelSCALE,opacity=0.4] at (phiplus) {$\Phi_+$};
\node[scale=\labelSCALE] at (phi) {$\Phi$};
\node[scale=\labelSCALE,opacity=0.4] at (phiminus) {$\Phi_-$};

\node[below,scale=\labelscale] at (-1) {$-1$};
\path[-latex'] (-1) -- (+1) node[midway,below,scale=\labelscale] {$0$};
\node[below,scale=\labelscale] at (+1) {$+1$};

\node[scale=\labelscale,opacity=0.4] at (-miniphi+1) {$-\Phi\!+\!1$};
\node[scale=\labelscale] at (miniphi) {$\Phi$};
\node[scale=\labelscale,opacity=0.4] at (-miniphi-1) {$-\Phi\!-\!1$};

\node[scale=\labelscale,opacity=0.4] at (-miniphiplus+1) {$-\Phi_{\!+}\!+\!1$};
\node[scale=\labelscale,opacity=0.4] at (-miniphiplus+01) {$-\Phi_{\!+}\!+\!1$};
\node[scale=\labelscale] at (miniphiplus) {$\Phi_{\!+}$};
\node[scale=\labelscale,opacity=0.3] at (miniphiplus+2) {$\Phi_{\!+}\!+\!2$};

\end{tikzpicture}}	\caption{\textbf{Analytic continuations of} $\Phi(\omega,z)$. The central image is a domain coloring plot of $\operatorname{Res}\{\Phi(\omega,z),z=1\}$ in the complex $\omega$-plane. Around it, we show domain coloring plots (this time in the complex $z$-plane) of $\Phi(i,z)$ and $\Phi(\pm2+e^{i\pi n/3}/2,z)$ for $n\in\{0,1,2,3\}$, together with the analytic continuations $\Phi_\pm(-i\pm1/4,z)$ and $\Phi_\pm(\pm2+e^{i\pi n/3}/2,z)$ for $n\in\{4,5\}$. At values $\omega=2\pm e^{i\pi /3}/2$, we also indicate how the function analytically continues in $z\in\mathbb{C}$. For simplicity, instead of complex axes we use the $\omega$-values 0 and $\pm2$ and the $z$-values of -1 (black dots), 0 (small crosses), and +1 (white dots) as a reference system for the complex coordinates.}
	\label{fig:AnalyticExtension}
 \end{center}
\end{figure*}

ix) There is a two-particle continuum at $\omega\in[-2,2]$, which represents a branch cut in the $\omega$ plane. We can use the closed form of $\Phi(\omega,z)$ to analytically continue it beyond the branch cut, which will be needed to compute the super- and subradiant states of the system. For that purpose, we introduce $\xi\equiv\xi(\omega,z)$ as
\begin{equation}
\begin{aligned}
    \xi&=\frac{\zeta (z)-\zeta (-1) }{1-\zeta (-1) \zeta (z)}\sqrt{x\left(0^+\right)^2 \frac{\zeta (z)-\zeta^{-1} (-1)}{\zeta (-1)-\zeta (z)}} 
\\&\times
\sqrt{x\left(0^+\right)^{-2} \frac{\zeta^{-1} (z)-\zeta (-1)}{\zeta^{-1} (-1)-\zeta^{-1} (z)}}
\end{aligned}
\end{equation}
where $x\left(0^+\right)$ denotes the limit of the amplitude $x$ (see Tab.~\ref{tab:closed formulas}) when $z\to0$. It can be shown that $\xi\in\lbrace-1,1\rbrace$ is a sign dependent on $\omega$ and $z$. Consequently, the extension $\Phi_+(\omega,z)$ of $\Phi(\omega,z)$ when $\omega$ crosses the positive part of the continuum $(0,2)$ can then be written as
\begin{widetext}
 \begin{equation}\label{eq:Phi+}
    \begin{aligned}
        \Phi_+(z) = &\frac {-K(
       k)\sqrt{k}\operatorname{sgnRe}\omega } {2 \pi} \left(\frac {z^{-1} - 
        z} {1 - \zeta(z)} + \frac {z + 1} {z - 
        1} (1 + \omega) + 
     2\frac {z - 1} {z + 1} \zeta(-1) - \frac {z^2 - 
        6 z + 1} {z^2 - 1} \zeta^{-1}(z)\right) 
        \\&
        +\xi\frac {2k \operatorname{sgnRe}\omega} {\pi i}\Big(K(k) E\left(x^{-1};k^{-1}\right) - k^{-2} E(k) F\left(x^{-1};k^{-1}\right) - (1 - k^{-2} ) K(k) F\left(x^{-1};k^{-1}\right) \Big).
       \end{aligned}
    \end{equation}
\end{widetext}
It should be pointed out that $\Phi_+(\omega,z)$ is a continuation in the space of functions that still have a $z$ dependence to be evaluated. To know the analytic extension of the function evaluated at a particular $z_0\in\mathbb C$, one additionally has to take into account all the instances in which a branch cut of $\Phi(z)$ or $\Phi_+(z)$ crosses the point $z_0$ and correct for them accordingly. This is needed e.g. to plot $\tilde A(\omega)$ in Fig.~\ref{fig:Spectra}a.\\

We caution that expression~\eqref{eq:Phi+} is only valid for Re~$\omega>0$ and Im~$\omega<0$ if $(0,2)$ is crossed from above to below (as illustrated in Fig.~\ref{fig:AnalyticExtension}), or for Re~$\omega>0$ and Im~$\omega>0$ if it is crossed from below. The corresponding continuation $\Phi_-$  for $\omega$ crossing $(-2,0)\subset\mathbb{C}$ can be computed as
\begin{equation}
\begin{aligned}
   \Phi_-&(\omega ,z)=\Phi_+(-\omega ,-z) +
   \\&
   \frac{\operatorname{sgnRe}\omega\ \ 2 \omega \sqrt{k(\omega )} \left(z^2+\omega  z+1\right) K\left(k(\omega)\right)}{\pi  \left(z^2-1\right)  \sqrt{\domega+1} \sqrt{\domega-1}}
\end{aligned}
\end{equation}
using the symmetry \eqref{eq:omegaSym}. Interestingly, $\Phi_+$ and $\Phi_-$ have the same $z\leftrightarrow z^{-1}$ and $z\leftrightarrow \zeta(z)$ symmetries as $\Phi$ (see (i)), but they cannot be obtained through the algorithm described in (v) because their analytic structure is fundamentally different. They have an additional branch cut joining the previous two cuts ($\zeta^{\pm1}\left(\mathbb{S}^1\right)$, see Fig.~\ref{fig:AnalyticExtension}). The appearance of this branch cut is topological and is the essence underlying the difficulty in computing superradiant decays in the system at hand.\\

x) An additional surprise is the center of the continuum $\omega=0$, which is a branch point in the middle of the branch cut. This again can be appreciated in  Fig.~\ref{fig:AnalyticExtension}, where two paths with different winding around this center lead to very different functions, although the paths end at similar frequencies. Obtaining an asymptotic expansion around $\omega=0$ is very difficult, even with computer assistance and despite having a closed formula for the $\Phi$ function and its analytic extensions. The reason is the intricate $\omega$-dependence of the function, which makes the problem intractable by ordinary methods. However, we can prove that
\begin{widetext}
\begin{equation}
\begin{aligned}
    \operatorname{sl}(z)\Phi(\omega,z)=&\frac{1}{4}+\frac{i}{\pi}\log\left(\operatorname{sl}(z)\frac{z-1}{z+1}\right)+\frac{i}{\pi}\frac{z(z^2+1)}{(z^2-1)^2}\omega\log\omega+\frac{4i+(\pi-6i\log2)(z+z^{-1})}{2\pi(z-z^{-1})^2}\omega
    \\&
    \frac{4-\left(z+z^{-1}\right)+\left(z+z^{-1}\right)^2-\frac{1}{4} \left(z+z^{-1}\right)^3}{\pi  i \left(z-z^{-1}\right)^4}\omega^2\log\omega+O(\omega^2)
\end{aligned}
\end{equation}
by comparison  to the logarithmic $z$-derivative of \eqref{eq:Phi'}, which splits in simple summands. As we can see from the $\log\omega$ terms, $\Phi$ is responsible for the logarithmic corrections to the decay. Analogously one can find the behavior for $\omega\simeq-2$,
\begin{equation}\label{eq:Phiaround-2}
\begin{aligned}
    \Phi(\omega,z)=&\frac{2}{\pi}\arctan\left(
    (1+\sqrt{2})\frac{\sqrt{3-2\sqrt{2}-z}}{\sqrt{3+2\sqrt{2}-z}}
    \right)-\frac{1}{2}
    +
    \frac{z(1+z)}{\sqrt{3-2\sqrt{2}-z}\sqrt{3+2\sqrt{2}-z}}\frac{(\omega+2)}{2\pi}
    \\&
    +\frac{z(z+1)\sqrt{3-2\sqrt{2}-z}\sqrt{3+2\sqrt{2}-z}}{16\pi(z-1)^4}\left(\frac{1}{2}+i\pi+4\log2-\frac{8z(1-4z+z^2)}{(1-6z+z^2)^2}-\log(\omega+2)\right)(\omega+2)^2
    \\&
    +O\left(\log(\omega+2)(\omega+2)^3\right).
\end{aligned}
\end{equation}
Using~\eqref{eq:Phiaround-2} together with the symmetry~\eqref{eq:omegaSym}, we find as well the behaviour for $\omega\simeq+2$
\begin{equation}
\begin{aligned}
    \Phi(\omega,z)=&\frac{2 z \log \left(\frac{\omega -2}{16}\right)}{\pi  (1-z) \sqrt{z+2 \sqrt{2}+3} \sqrt{z-2 \sqrt{2}+3}}+\frac{2}{\pi}\arctan \left(\left(\sqrt{2}+1\right)\frac{ \sqrt{z-2 \sqrt{2}+3}}{\sqrt{z+2 \sqrt{2}+3}}\right)-\frac{1}{2}
    \\&
    -\frac{\frac{(z+1)^2 \log \left(\frac{\omega -2}{16}\right)}{z^2+6 z+1}-\frac{4 z}{(z+1)^2}}{\sqrt{z+2 \sqrt{2}+3} \sqrt{z-2 \sqrt{2}+3}}\frac{z(\omega -2)}{2 \pi (z-1)}
    \\&
    +\left(\frac{\left(z^4+28 z^3+38 z^2+28 z+1\right) \log \left(\frac{\omega -2}{16}\right)}{\left(z+2 \sqrt{2}+3\right)^{5/2} \left(z-2 \sqrt{2}+3\right)^{5/2}}+\frac{2 \left(z^4+4 z^3-18 z^2+4 z+1\right)}{(z+1)^4 \sqrt{z+2 \sqrt{2}+3} \sqrt{z-2 \sqrt{2}+3}}\right)\frac{z(\omega -2)^2}{32 \pi  (z-1)}
    \\&
    +O\left(\log(\omega+2)(\omega+2)^3\right).
\end{aligned}
\end{equation}
\end{widetext}

 \section{\MakeUppercase{Constructive proof of the connection between $\Phi$ and $C_\sigma$}}\label{app:constructive proof}
By design, $C_\sigma(z)$ and $\frac{1}{2}\ddC_\sigma(z)\Phi(z)$ have a similar analytic structure with the same increment every two Riemann sheets (see Fig.~\ref{fig:Towers}). As a result, $C_\sigma(z)-\frac{1}{2}\ddC_\sigma(z)\Phi(z)$ can be thought of as having only two Riemann sheets. The sum of these two sheets gives $\mero_\sigma(z)\coloneqq 2 C_\sigma(z)-\ddC_\sigma(z)(\Phi(z)+1/2)+\dC_\sigma(z)$ (equivalent to Eqn.~\eqref{eq:Csoln}), which is a meromorphic function by construction.\\

The functions generating $\mero_\sigma(z)$ all obey certain symmetry rules for the inversion $z\leftrightarrow 1/z$ and for $z\leftrightarrow\zeta(z)$, which in terms of $\mero_\sigma$ read as
\begin{equation}\label{eq:minvsym}
    \mero_\sigma\left(\frac{1}{z}\right)= \mero_\sigma(z)\sigma z
\end{equation}
and
\begin{equation}
    \mero_\sigma\left(\zeta(z)\right)= -\frac{1+\sigma/\zeta(z)}{1+\sigma/z}\mero_\sigma(z)+\gamma_\sigma(z),
\end{equation}
where
\begin{equation}
\gamma_\sigma(z)=\tfrac{\ddC_\sigma(z)-2\dC_\sigma(z)}{\zeta(z)+\sigma}\left(
\tfrac{\domega-\sigma}{1+\sigma/z}+\tfrac{\zeta(z)-\zeta^{-1}(z)}{2\fun_{-,\sigma}(z)}
\right).
\end{equation}

These symmetries can be imposed in any ansatz $f_\sigma(z)$ for $\mero_\sigma(z)$ through the affine projectors (in the space of meromorphic functions)
\begin{equation}
    \left\lbrace
\begin{array}{l}
    \mathcal{P}_{inv}\lbrace f_\sigma\rbrace(z)\coloneqq\left(f_\sigma(z)+\frac{\sigma}{z}f_\sigma\left(\frac{1}{z}\right)\right)/2   \\
   \mathcal{P}_\zeta \lbrace f_\sigma\rbrace(z)\coloneqq \frac{f_\sigma(z)}{2}-\frac{1}{2}\frac{1+\sigma/z}{1+\sigma/\zeta(z)}\left(f_\sigma(\zeta(z))-\gamma_\sigma(z)\right).
\end{array}
    \right.
\end{equation}
Furthermore, joining the asymptotic expansions for the components of $\mero_\sigma(z)$, one obtains
\begin{equation}\label{eq:mInfinity}
\begin{aligned}
\mero_\sigma&(z)=2\sigma g\frac{1+ 2g\tilde C_{1,2}}{(\omega-2\Delta)z}
+\\&
2 g\tfrac{1+ 2g\tilde C_{1,1}(\omega-2\Delta)-2\sigma(\omega-\Delta)+ 2g\tilde C_{1,2}(1-\sigma\omega)}{(\omega-2\Delta)z^2}
+O\left(\frac{1}{z^3}\right).
\end{aligned}
\end{equation}
This allows expressing $\mero_\sigma(z)$ as a rational function resulting of the sum of all of its 12 poles, which are simple and match those of $\ddC_\sigma(z)$ (see Tab.~\ref{tab:closed formulas}). However, thanks to the multiple symmetries of $\mero_\sigma$, not all of them need to be computed. It is sufficient to compute the three roots $\lbrace z_{\sigma i}\rbrace_{i=1}^3$ of the polynomial $(z^2+2\Delta z+1)(z-\sigma)-4g^2z$, since the remaining ones are given by $1/z_{\sigma i}$, $\zeta(z_{\sigma i})$ and $1/\zeta(z_{\sigma i})$. Thus one can take $f_\sigma(z)$ to be a sum of three simple poles located at $z_{\sigma i}$ and, when assigned with the appropriate residues, $\mero_\sigma=\mathcal{P}_\zeta\lbrace \mathcal{P}_{inv}\lbrace f_\sigma\rbrace\rbrace$ (equivalent to the Eqn. for $\mero_\sigma$ in Tab.~\ref{tab:closed formulas}), because the projectors force the correct asymptotic behavior and the rest of the poles on the function. The particular choice of these $z_{\sigma i}$ is intentional, as they correspond with poles of $\ddC_\sigma(z)$ but not of $C_\sigma(z)$, $\dC_\sigma(z)$ or $\Phi(z)$, which helps to calculate their residues (leading to the expression for $\alpha_\sigma(z)$ in Tab.~\ref{tab:closed formulas}).\\

Comparing the leading order of $\mero_\sigma(z\to\infty)$ as given by Tab.~\ref{tab:closed formulas} and~\eqref{eq:mInfinity} we get
\begin{equation}
    g\sigma\frac{1+2g\tilde C_{1,2}}{\omega-2\Delta}=(2g\tilde C_{1,1}-\sigma)\sum_{i=1}^3 \alpha_\sigma\left(z_{\sigma i}\right),
\end{equation}
which can be interpreted as two linear equations ($\sigma=\pm1$) with two unknowns ($\tilde C_{1,2}$ and $\tilde C_{1,1}$). The solutions are the expressions for $2g\tilde C_{1,1}$ and $2g\tilde C_{1,2}$ in Tab.~\ref{tab:closed formulas}.\\

 \section{\MakeUppercase{A soft form of symmetry}}\label{app:soft symmetry}
In this section, we specify the mathematical context in which we think of transformation $z\to\zeta(z)$ as a symmetry. Along the section, we again treat $\omega$ as an implicit constant and only consider $z$-dependencies. For simplicity, we restrict the $z$-domain to $\mathbb{C}\setminus\left(\mathbb{S}^1\cup\zeta\left(\mathbb{S}^1\right)\cup\zeta^{-1}\left(\mathbb{S}^1\right)\right)$. Let us denote by
\begin{equation}
    \mathcal{R}=\left\{\frac{P(z)}{Q(z)}\bigg\rvert P,Q\in \mathbb{C}[z]\right\}
\end{equation}
the set of rational complex functions and by
\begin{equation}
\mathcal{A}=\mathcal{R}\left\langle1,\zeta,\operatorname{sl},\operatorname{sl}\zeta\right\rangle
\end{equation}
an algebra (under the usual operations) of functions of the form
\begin{equation}
    \alpha(z)=\left\{\begin{array}{cc}
        r_1(z)+r_2(z)\frac{\sqrt{\domega-1}}{\sqrt{\domega+1}} &  \text{if }\lvert z\rvert<1\\
         r_3(z)+r_4(z)\frac{\sqrt{\domega-1}}{\sqrt{\domega+1}} & \text{if }\lvert z\rvert>1
    \end{array}\right.
\end{equation}
where $r_1,r_2,r_3,r_4\in\mathcal{R}$. This algebra is closed under the symmetry group discussed in Sec.~\ref{sec:symmetries}. We note that the functions in $\mathcal{A}$ need to be defined by parts to accommodate for the symmetry $z\to\zeta(\zeta(z))=z^{-1}(1+\operatorname{sl}(z))/2+z(1-\operatorname{sl}(z))/2$.\\

Since functions in $\mathcal{A}$ have simple expressions that can be found \textit{a posteriori}, we can `set them aside' through an equivalence relation
\begin{equation}
    f_1\sim f_2\Leftrightarrow\exists \alpha,\alpha^{-1}\in\mathcal{A}\mid f_1-\alpha f_2\in \mathcal{A}.
\end{equation}
This relation greatly simplifies the problem by, for instance, equiparating the solution of the secular equation~\eqref{eq:secularEqn} to an incomplete Legendre's relation
\begin{equation}
\begin{aligned}
    &C_\sigma(z)\sim\Phi(z)\sim K(k^{-1}) E(x(z);k) 
    \\
    - k^2 E&(k^{-1}) F(x(z);k) - (1 - k^2 ) K(k^{-1}) F(x(z);k)
\end{aligned}
\end{equation}
(see Eqn.~\eqref{eq:Csoln} and Tab.~\ref{tab:closed formulas}). With this, we can use $\Phi$ as a canonical representative for the solution, $[\Phi(z)]=[C_\sigma(z)]$, and use symmetries~\eqref{eq:Phisyms} to write
\begin{equation}
    [C_\sigma(z)]=[C_\sigma(\zeta(z))].
\end{equation}
This characterizes $z\to\zeta(z)$ not as a symmetry of $C_\sigma(z)$ in the space of functions, but in the quotient space. Such a condition is a lot less restrictive on the requirements of invariance that symmetries should satisfy.


\begin{thebibliography}{55}%
\makeatletter
\providecommand \@ifxundefined [1]{%
 \@ifx{#1\undefined}
}%
\providecommand \@ifnum [1]{%
 \ifnum #1\expandafter \@firstoftwo
 \else \expandafter \@secondoftwo
 \fi
}%
\providecommand \@ifx [1]{%
 \ifx #1\expandafter \@firstoftwo
 \else \expandafter \@secondoftwo
 \fi
}%
\providecommand \natexlab [1]{#1}%
\providecommand \enquote  [1]{``#1''}%
\providecommand \bibnamefont  [1]{#1}%
\providecommand \bibfnamefont [1]{#1}%
\providecommand \citenamefont [1]{#1}%
\providecommand \href@noop [0]{\@secondoftwo}%
\providecommand \href [0]{\begingroup \@sanitize@url \@href}%
\providecommand \@href[1]{\@@startlink{#1}\@@href}%
\providecommand \@@href[1]{\endgroup#1\@@endlink}%
\providecommand \@sanitize@url [0]{\catcode `\\12\catcode `\$12\catcode `\&12\catcode `\#12\catcode `\^12\catcode `\_12\catcode `\%12\relax}%
\providecommand \@@startlink[1]{}%
\providecommand \@@endlink[0]{}%
\providecommand \url  [0]{\begingroup\@sanitize@url \@url }%
\providecommand \@url [1]{\endgroup\@href {#1}{\urlprefix }}%
\providecommand \urlprefix  [0]{URL }%
\providecommand \Eprint [0]{\href }%
\providecommand \doibase [0]{https://doi.org/}%
\providecommand \selectlanguage [0]{\@gobble}%
\providecommand \bibinfo  [0]{\@secondoftwo}%
\providecommand \bibfield  [0]{\@secondoftwo}%
\providecommand \translation [1]{[#1]}%
\providecommand \BibitemOpen [0]{}%
\providecommand \bibitemStop [0]{}%
\providecommand \bibitemNoStop [0]{.\EOS\space}%
\providecommand \EOS [0]{\spacefactor3000\relax}%
\providecommand \BibitemShut  [1]{\csname bibitem#1\endcsname}%
\let\auto@bib@innerbib\@empty
\bibitem [{\citenamefont {Dicke}(1954)}]{Dicke1954}%
  \BibitemOpen
  \bibfield  {author} {\bibinfo {author} {\bibfnamefont {R.~H.}\ \bibnamefont {Dicke}},\ }\bibfield  {title} {\bibinfo {title} {Coherence in spontaneous radiation processes},\ }\href {https://doi.org/10.1103/PhysRev.93.99} {\bibfield  {journal} {\bibinfo  {journal} {Phys. Rev.}\ }\textbf {\bibinfo {volume} {93}},\ \bibinfo {pages} {99} (\bibinfo {year} {1954})}\BibitemShut {NoStop}%
\bibitem [{\citenamefont {Dirac}(1927)}]{Dirac1927}%
  \BibitemOpen
  \bibfield  {author} {\bibinfo {author} {\bibfnamefont {P.~A.~M.}\ \bibnamefont {Dirac}},\ }\bibfield  {title} {\bibinfo {title} {The quantum theory of the emission and absorption of radiation},\ }\href {http://www.jstor.org/stable/94746} {\bibfield  {journal} {\bibinfo  {journal} {Proc. R. Soc. Lond.}\ }\textbf {\bibinfo {volume} {114}},\ \bibinfo {pages} {243} (\bibinfo {year} {1927})}\BibitemShut {NoStop}%
\bibitem [{\citenamefont {Rzazewski}\ \emph {et~al.}(1982)\citenamefont {Rzazewski}, \citenamefont {Lewenstein},\ and\ \citenamefont {Eberly}}]{Rzazewski1982}%
  \BibitemOpen
  \bibfield  {author} {\bibinfo {author} {\bibfnamefont {K.}~\bibnamefont {Rzazewski}}, \bibinfo {author} {\bibfnamefont {M.}~\bibnamefont {Lewenstein}},\ and\ \bibinfo {author} {\bibfnamefont {J.}~\bibnamefont {Eberly}},\ }\bibfield  {title} {\bibinfo {title} {Threshold effects in strong-field photodetachment},\ }\href {https://doi.org/10.1088/0022-3700/15/18/004} {\bibfield  {journal} {\bibinfo  {journal} {J. Phys. B At. Mol. Opt. Phys.}\ }\textbf {\bibinfo {volume} {15}},\ \bibinfo {pages} {L661} (\bibinfo {year} {1982})}\BibitemShut {NoStop}%
\bibitem [{\citenamefont {Bay}\ \emph {et~al.}(1997)\citenamefont {Bay}, \citenamefont {Lambropoulos},\ and\ \citenamefont {M\o{}lmer}}]{Bay1997}%
  \BibitemOpen
  \bibfield  {author} {\bibinfo {author} {\bibfnamefont {S.}~\bibnamefont {Bay}}, \bibinfo {author} {\bibfnamefont {P.}~\bibnamefont {Lambropoulos}},\ and\ \bibinfo {author} {\bibfnamefont {K.}~\bibnamefont {M\o{}lmer}},\ }\bibfield  {title} {\bibinfo {title} {Atom-atom interaction in strongly modified reservoirs},\ }\href {https://doi.org/10.1103/PhysRevA.55.1485} {\bibfield  {journal} {\bibinfo  {journal} {Phys. Rev. A}\ }\textbf {\bibinfo {volume} {55}},\ \bibinfo {pages} {1485} (\bibinfo {year} {1997})}\BibitemShut {NoStop}%
\bibitem [{\citenamefont {Bykov}(1975)}]{Bykov1975}%
  \BibitemOpen
  \bibfield  {author} {\bibinfo {author} {\bibfnamefont {V.~P.}\ \bibnamefont {Bykov}},\ }\bibfield  {title} {\bibinfo {title} {Spontaneous emission from a medium with a band spectrum},\ }\href {https://doi.org/10.1070/QE1975v004n07ABEH009654} {\bibfield  {journal} {\bibinfo  {journal} {Sov. J. Quantum Electron.}\ }\textbf {\bibinfo {volume} {4}},\ \bibinfo {pages} {861} (\bibinfo {year} {1975})}\BibitemShut {NoStop}%
\bibitem [{\citenamefont {John}\ and\ \citenamefont {Wang}(1990)}]{John1990}%
  \BibitemOpen
  \bibfield  {author} {\bibinfo {author} {\bibfnamefont {S.}~\bibnamefont {John}}\ and\ \bibinfo {author} {\bibfnamefont {J.}~\bibnamefont {Wang}},\ }\bibfield  {title} {\bibinfo {title} {Quantum electrodynamics near a photonic band gap: Photon bound states and dressed atoms},\ }\href {https://doi.org/10.1103/PhysRevLett.64.2418} {\bibfield  {journal} {\bibinfo  {journal} {Phys. Rev. Lett.}\ }\textbf {\bibinfo {volume} {64}},\ \bibinfo {pages} {2418} (\bibinfo {year} {1990})}\BibitemShut {NoStop}%
\bibitem [{\citenamefont {Lambropoulos}\ \emph {et~al.}(2000)\citenamefont {Lambropoulos}, \citenamefont {Nikolopoulos}, \citenamefont {Nielsen},\ and\ \citenamefont {Bay}}]{Lambropoulos2000}%
  \BibitemOpen
  \bibfield  {author} {\bibinfo {author} {\bibfnamefont {P.}~\bibnamefont {Lambropoulos}}, \bibinfo {author} {\bibfnamefont {G.~M.}\ \bibnamefont {Nikolopoulos}}, \bibinfo {author} {\bibfnamefont {T.~R.}\ \bibnamefont {Nielsen}},\ and\ \bibinfo {author} {\bibfnamefont {S.}~\bibnamefont {Bay}},\ }\bibfield  {title} {\bibinfo {title} {Fundamental quantum optics in structured reservoirs},\ }\href {https://doi.org/10.1088/0034-4885/63/4/201} {\bibfield  {journal} {\bibinfo  {journal} {Rep. Prog. Phys.}\ }\textbf {\bibinfo {volume} {63}},\ \bibinfo {pages} {455} (\bibinfo {year} {2000})}\BibitemShut {NoStop}%
\bibitem [{\citenamefont {Calaj\'o}\ \emph {et~al.}(2016)\citenamefont {Calaj\'o}, \citenamefont {Ciccarello}, \citenamefont {Chang},\ and\ \citenamefont {Rabl}}]{Calajo2016}%
  \BibitemOpen
  \bibfield  {author} {\bibinfo {author} {\bibfnamefont {G.}~\bibnamefont {Calaj\'o}}, \bibinfo {author} {\bibfnamefont {F.}~\bibnamefont {Ciccarello}}, \bibinfo {author} {\bibfnamefont {D.}~\bibnamefont {Chang}},\ and\ \bibinfo {author} {\bibfnamefont {P.}~\bibnamefont {Rabl}},\ }\bibfield  {title} {\bibinfo {title} {Atom-field dressed states in slow-light waveguide {QED}},\ }\href {https://doi.org/10.1103/PhysRevA.93.033833} {\bibfield  {journal} {\bibinfo  {journal} {Phys. Rev. A}\ }\textbf {\bibinfo {volume} {93}},\ \bibinfo {pages} {033833} (\bibinfo {year} {2016})}\BibitemShut {NoStop}%
\bibitem [{\citenamefont {Lanuza}\ \emph {et~al.}(2022)\citenamefont {Lanuza}, \citenamefont {Kwon}, \citenamefont {Kim},\ and\ \citenamefont {Schneble}}]{Lanuza2022}%
  \BibitemOpen
  \bibfield  {author} {\bibinfo {author} {\bibfnamefont {A.}~\bibnamefont {Lanuza}}, \bibinfo {author} {\bibfnamefont {J.}~\bibnamefont {Kwon}}, \bibinfo {author} {\bibfnamefont {Y.}~\bibnamefont {Kim}},\ and\ \bibinfo {author} {\bibfnamefont {D.}~\bibnamefont {Schneble}},\ }\bibfield  {title} {\bibinfo {title} {Multiband and array effects in matter-wave-based waveguide {QED}},\ }\href {https://doi.org/10.1103/PhysRevA.105.023703} {\bibfield  {journal} {\bibinfo  {journal} {Phys. Rev. A}\ }\textbf {\bibinfo {volume} {105}},\ \bibinfo {pages} {023703} (\bibinfo {year} {2022})}\BibitemShut {NoStop}%
\bibitem [{\citenamefont {Bello}\ \emph {et~al.}(2019)\citenamefont {Bello}, \citenamefont {Platero}, \citenamefont {Cirac},\ and\ \citenamefont {Gonz\'alez-Tudela}}]{Bello2019}%
  \BibitemOpen
  \bibfield  {author} {\bibinfo {author} {\bibfnamefont {M.}~\bibnamefont {Bello}}, \bibinfo {author} {\bibfnamefont {G.}~\bibnamefont {Platero}}, \bibinfo {author} {\bibfnamefont {J.~I.}\ \bibnamefont {Cirac}},\ and\ \bibinfo {author} {\bibfnamefont {A.}~\bibnamefont {Gonz\'alez-Tudela}},\ }\bibfield  {title} {\bibinfo {title} {Unconventional quantum optics in topological waveguide {QED}},\ }\href {https://doi.org/10.1126/sciadv.aaw0297} {\bibfield  {journal} {\bibinfo  {journal} {Sci. Adv.}\ }\textbf {\bibinfo {volume} {5}},\ \bibinfo {pages} {eaaw0297} (\bibinfo {year} {2019})}\BibitemShut {NoStop}%
\bibitem [{\citenamefont {Gonz\'alez-Tudela}\ and\ \citenamefont {Cirac}(2017)}]{GTudela2017}%
  \BibitemOpen
  \bibfield  {author} {\bibinfo {author} {\bibfnamefont {A.}~\bibnamefont {Gonz\'alez-Tudela}}\ and\ \bibinfo {author} {\bibfnamefont {J.~I.}\ \bibnamefont {Cirac}},\ }\bibfield  {title} {\bibinfo {title} {{M}arkovian and non-{M}arkovian dynamics of quantum emitters coupled to two-dimensional structured reservoirs},\ }\href {https://link.aps.org/doi/10.1103/PhysRevA.96.043811} {\bibfield  {journal} {\bibinfo  {journal} {Phys. Rev. A}\ }\textbf {\bibinfo {volume} {96}},\ \bibinfo {pages} {043811} (\bibinfo {year} {2017})}\BibitemShut {NoStop}%
\bibitem [{\citenamefont {Guo}\ \emph {et~al.}(2017)\citenamefont {Guo}, \citenamefont {Grimsmo}, \citenamefont {Kockum}, \citenamefont {Pletyukhov},\ and\ \citenamefont {Johansson}}]{Guo2017}%
  \BibitemOpen
  \bibfield  {author} {\bibinfo {author} {\bibfnamefont {L.~Z.}\ \bibnamefont {Guo}}, \bibinfo {author} {\bibfnamefont {A.}~\bibnamefont {Grimsmo}}, \bibinfo {author} {\bibfnamefont {A.~F.}\ \bibnamefont {Kockum}}, \bibinfo {author} {\bibfnamefont {M.}~\bibnamefont {Pletyukhov}},\ and\ \bibinfo {author} {\bibfnamefont {G.}~\bibnamefont {Johansson}},\ }\bibfield  {title} {\bibinfo {title} {Giant acoustic atom: A single quantum system with a deterministic time delay},\ }\href {https://doi.org/10.1103/PhysRevA.95.053821} {\bibfield  {journal} {\bibinfo  {journal} {Phys. Rev. A}\ }\textbf {\bibinfo {volume} {95}},\ \bibinfo {pages} {053821} (\bibinfo {year} {2017})}\BibitemShut {NoStop}%
\bibitem [{\citenamefont {Guo}\ \emph {et~al.}(2020)\citenamefont {Guo}, \citenamefont {Kockum}, \citenamefont {Marquardt},\ and\ \citenamefont {Johansson}}]{Guo2020}%
  \BibitemOpen
  \bibfield  {author} {\bibinfo {author} {\bibfnamefont {L.~Z.}\ \bibnamefont {Guo}}, \bibinfo {author} {\bibfnamefont {A.~F.}\ \bibnamefont {Kockum}}, \bibinfo {author} {\bibfnamefont {F.}~\bibnamefont {Marquardt}},\ and\ \bibinfo {author} {\bibfnamefont {G.}~\bibnamefont {Johansson}},\ }\bibfield  {title} {\bibinfo {title} {Oscillating bound states for a giant atom},\ }\href {https://doi.org/10.1103/PhysRevResearch.2.043014} {\bibfield  {journal} {\bibinfo  {journal} {Phys. Rev. Res.}\ }\textbf {\bibinfo {volume} {2}},\ \bibinfo {pages} {043014} (\bibinfo {year} {2020})}\BibitemShut {NoStop}%
\bibitem [{\citenamefont {Soro}\ \emph {et~al.}(2023)\citenamefont {Soro}, \citenamefont {Mu\~noz},\ and\ \citenamefont {Kockum}}]{Soro2023}%
  \BibitemOpen
  \bibfield  {author} {\bibinfo {author} {\bibfnamefont {A.}~\bibnamefont {Soro}}, \bibinfo {author} {\bibfnamefont {C.~S.}\ \bibnamefont {Mu\~noz}},\ and\ \bibinfo {author} {\bibfnamefont {A.~F.}\ \bibnamefont {Kockum}},\ }\bibfield  {title} {\bibinfo {title} {Interaction between giant atoms in a one-dimensional structured environment},\ }\href {https://doi.org/10.1103/PhysRevA.107.013710} {\bibfield  {journal} {\bibinfo  {journal} {Phys. Rev. A}\ }\textbf {\bibinfo {volume} {107}},\ \bibinfo {pages} {013710} (\bibinfo {year} {2023})}\BibitemShut {NoStop}%
\bibitem [{\citenamefont {Facchi}\ \emph {et~al.}(2019)\citenamefont {Facchi}, \citenamefont {Lonigro}, \citenamefont {Pascazio}, \citenamefont {Pepe},\ and\ \citenamefont {Pomarico}}]{Facchi2019}%
  \BibitemOpen
  \bibfield  {author} {\bibinfo {author} {\bibfnamefont {P.}~\bibnamefont {Facchi}}, \bibinfo {author} {\bibfnamefont {D.}~\bibnamefont {Lonigro}}, \bibinfo {author} {\bibfnamefont {S.}~\bibnamefont {Pascazio}}, \bibinfo {author} {\bibfnamefont {F.~V.}\ \bibnamefont {Pepe}},\ and\ \bibinfo {author} {\bibfnamefont {D.}~\bibnamefont {Pomarico}},\ }\bibfield  {title} {\bibinfo {title} {Bound states in the continuum for an array of quantum emitters},\ }\href {https://doi.org/10.1103/PhysRevA.100.023834} {\bibfield  {journal} {\bibinfo  {journal} {Phys. Rev. A}\ }\textbf {\bibinfo {volume} {100}},\ \bibinfo {pages} {023834} (\bibinfo {year} {2019})}\BibitemShut {NoStop}%
\bibitem [{\citenamefont {Burgess}\ and\ \citenamefont {Florescu}(2022)}]{Burgess2022}%
  \BibitemOpen
  \bibfield  {author} {\bibinfo {author} {\bibfnamefont {A.}~\bibnamefont {Burgess}}\ and\ \bibinfo {author} {\bibfnamefont {M.}~\bibnamefont {Florescu}},\ }\bibfield  {title} {\bibinfo {title} {Non-{M}arkovian dynamics of a single excitation within many-body dissipative systems},\ }\href {https://doi.org/10.1103/PhysRevA.105.062207} {\bibfield  {journal} {\bibinfo  {journal} {Phys. Rev. A}\ }\textbf {\bibinfo {volume} {105}},\ \bibinfo {pages} {062207} (\bibinfo {year} {2022})}\BibitemShut {NoStop}%
\bibitem [{\citenamefont {Calaj\'o}\ \emph {et~al.}(2019)\citenamefont {Calaj\'o}, \citenamefont {Fang}, \citenamefont {Baranger},\ and\ \citenamefont {Ciccarello}}]{Calajo2019}%
  \BibitemOpen
  \bibfield  {author} {\bibinfo {author} {\bibfnamefont {G.}~\bibnamefont {Calaj\'o}}, \bibinfo {author} {\bibfnamefont {Y.~L.}\ \bibnamefont {Fang}}, \bibinfo {author} {\bibfnamefont {H.~U.}\ \bibnamefont {Baranger}},\ and\ \bibinfo {author} {\bibfnamefont {F.}~\bibnamefont {Ciccarello}},\ }\bibfield  {title} {\bibinfo {title} {Exciting a bound state in the continuum through multiphoton scattering plus delayed quantum feedback},\ }\href {https://doi.org/10.1103/PhysRevLett.122.073601} {\bibfield  {journal} {\bibinfo  {journal} {Phys. Rev. Lett.}\ }\textbf {\bibinfo {volume} {122}},\ \bibinfo {pages} {073601} (\bibinfo {year} {2019})}\BibitemShut {NoStop}%
\bibitem [{\citenamefont {Sinha}\ \emph {et~al.}(2020)\citenamefont {Sinha}, \citenamefont {Meystre}, \citenamefont {Goldschmidt}, \citenamefont {Fatemi}, \citenamefont {Rolston},\ and\ \citenamefont {Solano}}]{Sinha2020}%
  \BibitemOpen
  \bibfield  {author} {\bibinfo {author} {\bibfnamefont {K.}~\bibnamefont {Sinha}}, \bibinfo {author} {\bibfnamefont {P.}~\bibnamefont {Meystre}}, \bibinfo {author} {\bibfnamefont {E.~A.}\ \bibnamefont {Goldschmidt}}, \bibinfo {author} {\bibfnamefont {F.~K.}\ \bibnamefont {Fatemi}}, \bibinfo {author} {\bibfnamefont {S.~L.}\ \bibnamefont {Rolston}},\ and\ \bibinfo {author} {\bibfnamefont {P.}~\bibnamefont {Solano}},\ }\bibfield  {title} {\bibinfo {title} {Non-{M}arkovian collective emission from macroscopically separated emitters},\ }\href {https://doi.org/10.1103/PhysRevLett.124.043603} {\bibfield  {journal} {\bibinfo  {journal} {Phys. Rev. Lett.}\ }\textbf {\bibinfo {volume} {124}},\ \bibinfo {pages} {043603} (\bibinfo {year} {2020})}\BibitemShut {NoStop}%
\bibitem [{\citenamefont {Del~\'Angel}\ \emph {et~al.}(2023)\citenamefont {Del~\'Angel}, \citenamefont {Solano},\ and\ \citenamefont {Barberis-Blostein}}]{delAngel2023}%
  \BibitemOpen
  \bibfield  {author} {\bibinfo {author} {\bibfnamefont {A.}~\bibnamefont {Del~\'Angel}}, \bibinfo {author} {\bibfnamefont {P.}~\bibnamefont {Solano}},\ and\ \bibinfo {author} {\bibfnamefont {P.}~\bibnamefont {Barberis-Blostein}},\ }\bibfield  {title} {\bibinfo {title} {Effects of environment correlations on the onset of collective decay in waveguide {QED}},\ }\href {https://doi.org/10.1103/PhysRevA.108.013703} {\bibfield  {journal} {\bibinfo  {journal} {Phys. Rev. A}\ }\textbf {\bibinfo {volume} {108}},\ \bibinfo {pages} {013703} (\bibinfo {year} {2023})}\BibitemShut {NoStop}%
\bibitem [{\citenamefont {Alvarez-Giron}\ \emph {et~al.}(2023)\citenamefont {Alvarez-Giron}, \citenamefont {Solano}, \citenamefont {Sinha},\ and\ \citenamefont {Barberis-Blostein}}]{Alvarez2023}%
  \BibitemOpen
  \bibfield  {author} {\bibinfo {author} {\bibfnamefont {W.}~\bibnamefont {Alvarez-Giron}}, \bibinfo {author} {\bibfnamefont {P.}~\bibnamefont {Solano}}, \bibinfo {author} {\bibfnamefont {K.}~\bibnamefont {Sinha}},\ and\ \bibinfo {author} {\bibfnamefont {P.}~\bibnamefont {Barberis-Blostein}},\ }{\bibinfo {title} {Delay-induced spontaneous dark state generation from two distant excited atoms}} (\bibinfo {year} {2023}),\ \Eprint {https://arxiv.org/abs/2303.06559} {arXiv:2303.06559 [quant-ph]} \BibitemShut {NoStop}%
\bibitem [{\citenamefont {Dinc}\ and\ \citenamefont {Bra\'nczyk}(2019)}]{Dinc2019}%
  \BibitemOpen
  \bibfield  {author} {\bibinfo {author} {\bibfnamefont {F.}~\bibnamefont {Dinc}}\ and\ \bibinfo {author} {\bibfnamefont {A.~M.}\ \bibnamefont {Bra\'nczyk}},\ }\bibfield  {title} {\bibinfo {title} {Non-{M}arkovian super-superradiance in a linear chain of up to 100 qubits},\ }\href {https://doi.org/10.1103/PhysRevResearch.1.032042} {\bibfield  {journal} {\bibinfo  {journal} {Phys. Rev. Res.}\ }\textbf {\bibinfo {volume} {1}},\ \bibinfo {pages} {032042(R)} (\bibinfo {year} {2019})}\BibitemShut {NoStop}%
\bibitem [{\citenamefont {Zheng}\ and\ \citenamefont {Baranger}(2013)}]{Zheng2013b}%
  \BibitemOpen
  \bibfield  {author} {\bibinfo {author} {\bibfnamefont {H.}~\bibnamefont {Zheng}}\ and\ \bibinfo {author} {\bibfnamefont {H.~U.}\ \bibnamefont {Baranger}},\ }\bibfield  {title} {\bibinfo {title} {Persistent quantum beats and long-distance entanglement from waveguide-mediated interactions},\ }\href {https://doi.org/10.1103/PhysRevLett.110.113601} {\bibfield  {journal} {\bibinfo  {journal} {Phys. Rev. Lett.}\ }\textbf {\bibinfo {volume} {110}},\ \bibinfo {pages} {113601} (\bibinfo {year} {2013})}\BibitemShut {NoStop}%
\bibitem [{\citenamefont {Kimble}(2008)}]{Kimble2008}%
  \BibitemOpen
  \bibfield  {author} {\bibinfo {author} {\bibfnamefont {H.~J.}\ \bibnamefont {Kimble}},\ }\bibfield  {title} {\bibinfo {title} {The quantum internet},\ }\href {https://doi.org/10.1038/nature07127} {\bibfield  {journal} {\bibinfo  {journal} {Nature}\ }\textbf {\bibinfo {volume} {453}},\ \bibinfo {pages} {1023} (\bibinfo {year} {2008})}\BibitemShut {NoStop}%
\bibitem [{\citenamefont {Sheremet}\ \emph {et~al.}(2023)\citenamefont {Sheremet}, \citenamefont {Petrov}, \citenamefont {Iorsh}, \citenamefont {Poshakinskiy},\ and\ \citenamefont {Poddubny}}]{Sheremet2023}%
  \BibitemOpen
  \bibfield  {author} {\bibinfo {author} {\bibfnamefont {A.~S.}\ \bibnamefont {Sheremet}}, \bibinfo {author} {\bibfnamefont {M.~I.}\ \bibnamefont {Petrov}}, \bibinfo {author} {\bibfnamefont {I.~V.}\ \bibnamefont {Iorsh}}, \bibinfo {author} {\bibfnamefont {A.~V.}\ \bibnamefont {Poshakinskiy}},\ and\ \bibinfo {author} {\bibfnamefont {A.~N.}\ \bibnamefont {Poddubny}},\ }\bibfield  {title} {\bibinfo {title} {Waveguide quantum electrodynamics: Collective radiance and photon-photon correlations},\ }\href {https://doi.org/10.1103/RevModPhys.95.015002} {\bibfield  {journal} {\bibinfo  {journal} {Rev. Mod. Phys.}\ }\textbf {\bibinfo {volume} {95}},\ \bibinfo {pages} {015002} (\bibinfo {year} {2023})}\BibitemShut {NoStop}%
\bibitem [{\citenamefont {Kim}\ \emph {et~al.}(2023)\citenamefont {Kim}, \citenamefont {Lanuza},\ and\ \citenamefont {Schneble}}]{Kim2024}%
  \BibitemOpen
  \bibfield  {author} {\bibinfo {author} {\bibfnamefont {Y.}~\bibnamefont {Kim}}, \bibinfo {author} {\bibfnamefont {A.}~\bibnamefont {Lanuza}},\ and\ \bibinfo {author} {\bibfnamefont {D.}~\bibnamefont {Schneble}},\ }{\bibinfo {title} {Super- and subradiant dynamics of quantum emitters mediated by atomic matter waves}} (\bibinfo {year} {2023}),\ \Eprint {https://arxiv.org/abs/2311.09474} {arXiv:2311.09474 [quant-ph]} \BibitemShut {NoStop}%
\bibitem [{\citenamefont {Shi}\ \emph {et~al.}(2015)\citenamefont {Shi}, \citenamefont {Chang},\ and\ \citenamefont {Cirac}}]{Shi2015}%
  \BibitemOpen
  \bibfield  {author} {\bibinfo {author} {\bibfnamefont {T.}~\bibnamefont {Shi}}, \bibinfo {author} {\bibfnamefont {D.~E.}\ \bibnamefont {Chang}},\ and\ \bibinfo {author} {\bibfnamefont {J.~I.}\ \bibnamefont {Cirac}},\ }\bibfield  {title} {\bibinfo {title} {Multiphoton-scattering theory and generalized master equations},\ }\href {https://doi.org/10.1103/PhysRevA.92.053834} {\bibfield  {journal} {\bibinfo  {journal} {Phys. Rev. A}\ }\textbf {\bibinfo {volume} {92}},\ \bibinfo {pages} {053834} (\bibinfo {year} {2015})}\BibitemShut {NoStop}%
\bibitem [{\citenamefont {Shi}\ \emph {et~al.}(2018)\citenamefont {Shi}, \citenamefont {Wu}, \citenamefont {Gonz\'alez-Tudela},\ and\ \citenamefont {Cirac}}]{Shi2018}%
  \BibitemOpen
  \bibfield  {author} {\bibinfo {author} {\bibfnamefont {T.}~\bibnamefont {Shi}}, \bibinfo {author} {\bibfnamefont {Y.-H.}\ \bibnamefont {Wu}}, \bibinfo {author} {\bibfnamefont {A.}~\bibnamefont {Gonz\'alez-Tudela}},\ and\ \bibinfo {author} {\bibfnamefont {J.~I.}\ \bibnamefont {Cirac}},\ }\bibfield  {title} {\bibinfo {title} {Effective many-body hamiltonians of qubit-photon bound states},\ }\href {https://doi.org/10.1088/1367-2630/aae4a9} {\bibfield  {journal} {\bibinfo  {journal} {New J. Phys.}\ }\textbf {\bibinfo {volume} {20}},\ \bibinfo {pages} {105005} (\bibinfo {year} {2018})}\BibitemShut {NoStop}%
\bibitem [{\citenamefont {Gonz\'alez-Guti\'errez}\ \emph {et~al.}(2021)\citenamefont {Gonz\'alez-Guti\'errez}, \citenamefont {Rom\'an-Roche},\ and\ \citenamefont {Zueco}}]{GGutierrez2021}%
  \BibitemOpen
  \bibfield  {author} {\bibinfo {author} {\bibfnamefont {C.~A.}\ \bibnamefont {Gonz\'alez-Guti\'errez}}, \bibinfo {author} {\bibfnamefont {J.}~\bibnamefont {Rom\'an-Roche}},\ and\ \bibinfo {author} {\bibfnamefont {D.}~\bibnamefont {Zueco}},\ }\bibfield  {title} {\bibinfo {title} {Distant emitters in ultrastrong waveguide {QED}: Ground-state properties and non-{M}arkovian dynamics},\ }\href {https://doi.org/10.1103/PhysRevA.104.053701} {\bibfield  {journal} {\bibinfo  {journal} {Phys. Rev. A}\ }\textbf {\bibinfo {volume} {104}},\ \bibinfo {pages} {053701} (\bibinfo {year} {2021})}\BibitemShut {NoStop}%
\bibitem [{\citenamefont {Cascio}\ \emph {et~al.}(2019)\citenamefont {Cascio}, \citenamefont {Halimeh}, \citenamefont {McCulloch}, \citenamefont {Recati},\ and\ \citenamefont {de~Vega}}]{Cascio2019}%
  \BibitemOpen
  \bibfield  {author} {\bibinfo {author} {\bibfnamefont {C.}~\bibnamefont {Cascio}}, \bibinfo {author} {\bibfnamefont {J.~C.}\ \bibnamefont {Halimeh}}, \bibinfo {author} {\bibfnamefont {I.~P.}\ \bibnamefont {McCulloch}}, \bibinfo {author} {\bibfnamefont {A.}~\bibnamefont {Recati}},\ and\ \bibinfo {author} {\bibfnamefont {I.}~\bibnamefont {de~Vega}},\ }\bibfield  {title} {\bibinfo {title} {Dynamics of multiple atoms in one-dimensional fields},\ }\href {https://doi.org/10.1103/PhysRevA.99.013845} {\bibfield  {journal} {\bibinfo  {journal} {Phys. Rev. A}\ }\textbf {\bibinfo {volume} {99}},\ \bibinfo {pages} {013845} (\bibinfo {year} {2019})}\BibitemShut {NoStop}%
\bibitem [{\citenamefont {de~Vega}\ and\ \citenamefont {Alonso}(2017)}]{deVega2017}%
  \BibitemOpen
  \bibfield  {author} {\bibinfo {author} {\bibfnamefont {I.}~\bibnamefont {de~Vega}}\ and\ \bibinfo {author} {\bibfnamefont {D.}~\bibnamefont {Alonso}},\ }\bibfield  {title} {\bibinfo {title} {Dynamics of non-{M}arkovian open quantum systems},\ }\href {https://doi.org/10.1103/RevModPhys.89.015001} {\bibfield  {journal} {\bibinfo  {journal} {Rev. Mod. Phys.}\ }\textbf {\bibinfo {volume} {89}},\ \bibinfo {pages} {015001} (\bibinfo {year} {2017})}\BibitemShut {NoStop}%
\bibitem [{\citenamefont {Laakso}\ and\ \citenamefont {Pletyukhov}(2014)}]{Laakso2014}%
  \BibitemOpen
  \bibfield  {author} {\bibinfo {author} {\bibfnamefont {M.}~\bibnamefont {Laakso}}\ and\ \bibinfo {author} {\bibfnamefont {M.}~\bibnamefont {Pletyukhov}},\ }\bibfield  {title} {\bibinfo {title} {Scattering of two photons from two distant qubits: Exact solution},\ }\href {https://doi.org/10.1103/PhysRevLett.113.183601} {\bibfield  {journal} {\bibinfo  {journal} {Phys. Rev. Lett.}\ }\textbf {\bibinfo {volume} {113}},\ \bibinfo {pages} {183601} (\bibinfo {year} {2014})}\BibitemShut {NoStop}%
\bibitem [{\citenamefont {Schneider}\ \emph {et~al.}(2016)\citenamefont {Schneider}, \citenamefont {Sproll}, \citenamefont {Stawiarski}, \citenamefont {Schmitteckert},\ and\ \citenamefont {Busch}}]{Schneider2016}%
  \BibitemOpen
  \bibfield  {author} {\bibinfo {author} {\bibfnamefont {M.~P.}\ \bibnamefont {Schneider}}, \bibinfo {author} {\bibfnamefont {T.}~\bibnamefont {Sproll}}, \bibinfo {author} {\bibfnamefont {C.}~\bibnamefont {Stawiarski}}, \bibinfo {author} {\bibfnamefont {P.}~\bibnamefont {Schmitteckert}},\ and\ \bibinfo {author} {\bibfnamefont {K.}~\bibnamefont {Busch}},\ }\bibfield  {title} {\bibinfo {title} {Green's-function formalism for waveguide {QED} applications},\ }\href {https://doi.org/10.1103/PhysRevA.93.013828} {\bibfield  {journal} {\bibinfo  {journal} {Phys. Rev. A}\ }\textbf {\bibinfo {volume} {93}},\ \bibinfo {pages} {013828} (\bibinfo {year} {2016})}\BibitemShut {NoStop}%
\bibitem [{\citenamefont {Roulet}\ and\ \citenamefont {Scarani}(2016)}]{Roulet2016}%
  \BibitemOpen
  \bibfield  {author} {\bibinfo {author} {\bibfnamefont {A.}~\bibnamefont {Roulet}}\ and\ \bibinfo {author} {\bibfnamefont {V.}~\bibnamefont {Scarani}},\ }\bibfield  {title} {\bibinfo {title} {Solving the scattering of n photons on a two-level atom without computation},\ }\href {https://doi.org/10.1088/1367-2630/18/9/093035} {\bibfield  {journal} {\bibinfo  {journal} {New J. Phys.}\ }\textbf {\bibinfo {volume} {18}},\ \bibinfo {pages} {093035} (\bibinfo {year} {2016})}\BibitemShut {NoStop}%
\bibitem [{\citenamefont {Masson}\ and\ \citenamefont {Asenjo-Garcia}(2022)}]{Masson2022}%
  \BibitemOpen
  \bibfield  {author} {\bibinfo {author} {\bibfnamefont {S.~J.}\ \bibnamefont {Masson}}\ and\ \bibinfo {author} {\bibfnamefont {A.}~\bibnamefont {Asenjo-Garcia}},\ }\bibfield  {title} {\bibinfo {title} {Universality of {D}icke superradiance in arrays of quantum emitters},\ }\href {https://doi.org/10.1038/s41467-022-29805-4} {\bibfield  {journal} {\bibinfo  {journal} {Nat. Commun.}\ }\textbf {\bibinfo {volume} {13}},\ \bibinfo {pages} {2285} (\bibinfo {year} {2022})}\BibitemShut {NoStop}%
\bibitem [{\citenamefont {Lombardo}\ \emph {et~al.}(2014)\citenamefont {Lombardo}, \citenamefont {Ciccarello},\ and\ \citenamefont {Palma}}]{Lombardo2014}%
  \BibitemOpen
  \bibfield  {author} {\bibinfo {author} {\bibfnamefont {F.}~\bibnamefont {Lombardo}}, \bibinfo {author} {\bibfnamefont {F.}~\bibnamefont {Ciccarello}},\ and\ \bibinfo {author} {\bibfnamefont {G.~M.}\ \bibnamefont {Palma}},\ }\bibfield  {title} {\bibinfo {title} {Photon localization versus population trapping in a coupled-cavity array},\ }\href {https://doi.org/10.1103/PhysRevA.89.053826} {\bibfield  {journal} {\bibinfo  {journal} {Phys. Rev. A}\ }\textbf {\bibinfo {volume} {89}},\ \bibinfo {pages} {053826} (\bibinfo {year} {2014})}\BibitemShut {NoStop}%
\bibitem [{\citenamefont {Lenggenhager}\ \emph {et~al.}(2022)\citenamefont {Lenggenhager}, \citenamefont {Stegmaier}, \citenamefont {Upreti}, \citenamefont {Hofmann}, \citenamefont {Helbig}, \citenamefont {Vollhardt}, \citenamefont {Greiter}, \citenamefont {Lee}, \citenamefont {Imhof}, \citenamefont {Brand}, \citenamefont {Kießling}, \citenamefont {Boettcher}, \citenamefont {Neupert}, \citenamefont {Thomale},\ and\ \citenamefont {Bzdušek}}]{Lenggenhager2022}%
  \BibitemOpen
  \bibfield  {author} {\bibinfo {author} {\bibfnamefont {P.~M.}\ \bibnamefont {Lenggenhager}}, \bibinfo {author} {\bibfnamefont {A.}~\bibnamefont {Stegmaier}}, \bibinfo {author} {\bibfnamefont {L.~K.}\ \bibnamefont {Upreti}}, \bibinfo {author} {\bibfnamefont {T.}~\bibnamefont {Hofmann}}, \bibinfo {author} {\bibfnamefont {T.}~\bibnamefont {Helbig}}, \bibinfo {author} {\bibfnamefont {A.}~\bibnamefont {Vollhardt}}, \bibinfo {author} {\bibfnamefont {M.}~\bibnamefont {Greiter}}, \bibinfo {author} {\bibfnamefont {C.~H.}\ \bibnamefont {Lee}}, \bibinfo {author} {\bibfnamefont {S.}~\bibnamefont {Imhof}}, \bibinfo {author} {\bibfnamefont {H.}~\bibnamefont {Brand}}, \bibinfo {author} {\bibfnamefont {T.}~\bibnamefont {Kießling}}, \bibinfo {author} {\bibfnamefont {I.}~\bibnamefont {Boettcher}}, \bibinfo {author} {\bibfnamefont {T.}~\bibnamefont {Neupert}}, \bibinfo {author} {\bibfnamefont {R.}~\bibnamefont {Thomale}},\ and\ \bibinfo {author} {\bibfnamefont {T.}~\bibnamefont {Bzdušek}},\ }\bibfield  {title} {\bibinfo
  {title} {Simulating hyperbolic space on a circuit board},\ }\href {https://doi.org/10.1038/s41467-022-32042-4} {\bibfield  {journal} {\bibinfo  {journal} {Nat. Commun.}\ }\textbf {\bibinfo {volume} {13}},\ \bibinfo {pages} {4373} (\bibinfo {year} {2022})}\BibitemShut {NoStop}%
\bibitem [{\citenamefont {Stewart}\ \emph {et~al.}(2020)\citenamefont {Stewart}, \citenamefont {Kwon}, \citenamefont {Lanuza},\ and\ \citenamefont {Schneble}}]{Stewart2020}%
  \BibitemOpen
  \bibfield  {author} {\bibinfo {author} {\bibfnamefont {M.}~\bibnamefont {Stewart}}, \bibinfo {author} {\bibfnamefont {J.}~\bibnamefont {Kwon}}, \bibinfo {author} {\bibfnamefont {A.}~\bibnamefont {Lanuza}},\ and\ \bibinfo {author} {\bibfnamefont {D.}~\bibnamefont {Schneble}},\ }\bibfield  {title} {\bibinfo {title} {Dynamics of matter-wave quantum emitters in a structured vacuum},\ }\href {https://doi.org/10.1103/PhysRevResearch.2.043307} {\bibfield  {journal} {\bibinfo  {journal} {Phys. Rev. Res.}\ }\textbf {\bibinfo {volume} {2}},\ \bibinfo {pages} {043307} (\bibinfo {year} {2020})}\BibitemShut {NoStop}%
\bibitem [{\citenamefont {Gonz\'alez-Tudela}\ \emph {et~al.}(2019)\citenamefont {Gonz\'alez-Tudela}, \citenamefont {Mu\~noz},\ and\ \citenamefont {Cirac}}]{GTudela2019}%
  \BibitemOpen
  \bibfield  {author} {\bibinfo {author} {\bibfnamefont {A.}~\bibnamefont {Gonz\'alez-Tudela}}, \bibinfo {author} {\bibfnamefont {C.~S.}\ \bibnamefont {Mu\~noz}},\ and\ \bibinfo {author} {\bibfnamefont {J.~I.}\ \bibnamefont {Cirac}},\ }\bibfield  {title} {\bibinfo {title} {Engineering and harnessing giant atoms in high-dimensional baths: A proposal for implementation with cold atoms},\ }\href {https://doi.org/10.1103/PhysRevLett.122.203603} {\bibfield  {journal} {\bibinfo  {journal} {Phys. Rev. Lett.}\ }\textbf {\bibinfo {volume} {122}},\ \bibinfo {pages} {203603} (\bibinfo {year} {2019})}\BibitemShut {NoStop}%
\bibitem [{\citenamefont {Vacchini}\ and\ \citenamefont {Breuer}(2010)}]{Vacchini2010}%
  \BibitemOpen
  \bibfield  {author} {\bibinfo {author} {\bibfnamefont {B.}~\bibnamefont {Vacchini}}\ and\ \bibinfo {author} {\bibfnamefont {H.-P.}\ \bibnamefont {Breuer}},\ }\bibfield  {title} {\bibinfo {title} {Exact master equations for the non-{M}arkovian decay of a qubit},\ }\href {https://doi.org/10.1103/PhysRevA.81.042103} {\bibfield  {journal} {\bibinfo  {journal} {Phys. Rev. A}\ }\textbf {\bibinfo {volume} {81}},\ \bibinfo {pages} {042103} (\bibinfo {year} {2010})}\BibitemShut {NoStop}%
\bibitem [{\citenamefont {Tufarelli}\ \emph {et~al.}(2014)\citenamefont {Tufarelli}, \citenamefont {Kim},\ and\ \citenamefont {Ciccarello}}]{Tufarelli2014}%
  \BibitemOpen
  \bibfield  {author} {\bibinfo {author} {\bibfnamefont {T.}~\bibnamefont {Tufarelli}}, \bibinfo {author} {\bibfnamefont {M.~S.}\ \bibnamefont {Kim}},\ and\ \bibinfo {author} {\bibfnamefont {F.}~\bibnamefont {Ciccarello}},\ }\bibfield  {title} {\bibinfo {title} {Non-{M}arkovianity of a quantum emitter in front of a mirror},\ }\href {https://doi.org/10.1103/PhysRevA.90.012113} {\bibfield  {journal} {\bibinfo  {journal} {Phys. Rev. A}\ }\textbf {\bibinfo {volume} {90}},\ \bibinfo {pages} {012113} (\bibinfo {year} {2014})}\BibitemShut {NoStop}%
\bibitem [{\citenamefont {Roy}\ \emph {et~al.}(2017)\citenamefont {Roy}, \citenamefont {Wilson},\ and\ \citenamefont {Firstenberg}}]{Roy2017}%
  \BibitemOpen
  \bibfield  {author} {\bibinfo {author} {\bibfnamefont {D.}~\bibnamefont {Roy}}, \bibinfo {author} {\bibfnamefont {C.~M.}\ \bibnamefont {Wilson}},\ and\ \bibinfo {author} {\bibfnamefont {O.}~\bibnamefont {Firstenberg}},\ }\bibfield  {title} {\bibinfo {title} {Colloquium: Strongly interacting photons in one-dimensional continuum},\ }\href {https://doi.org/10.1103/RevModPhys.89.021001} {\bibfield  {journal} {\bibinfo  {journal} {Rev. Mod. Phys.}\ }\textbf {\bibinfo {volume} {89}},\ \bibinfo {pages} {021001} (\bibinfo {year} {2017})}\BibitemShut {NoStop}%
\bibitem [{\citenamefont {Goban}\ \emph {et~al.}(2015)\citenamefont {Goban}, \citenamefont {Hung}, \citenamefont {Hood}, \citenamefont {Yu}, \citenamefont {Muniz}, \citenamefont {Painter},\ and\ \citenamefont {Kimble}}]{Goban2015}%
  \BibitemOpen
  \bibfield  {author} {\bibinfo {author} {\bibfnamefont {A.}~\bibnamefont {Goban}}, \bibinfo {author} {\bibfnamefont {C.-L.}\ \bibnamefont {Hung}}, \bibinfo {author} {\bibfnamefont {J.~D.}\ \bibnamefont {Hood}}, \bibinfo {author} {\bibfnamefont {S.-P.}\ \bibnamefont {Yu}}, \bibinfo {author} {\bibfnamefont {J.~A.}\ \bibnamefont {Muniz}}, \bibinfo {author} {\bibfnamefont {O.}~\bibnamefont {Painter}},\ and\ \bibinfo {author} {\bibfnamefont {H.~J.}\ \bibnamefont {Kimble}},\ }\bibfield  {title} {\bibinfo {title} {Superradiance for atoms trapped along a photonic crystal waveguide},\ }\href {https://doi.org/10.1103/PhysRevLett.115.063601} {\bibfield  {journal} {\bibinfo  {journal} {Phys. Rev. Lett.}\ }\textbf {\bibinfo {volume} {115}},\ \bibinfo {pages} {063601} (\bibinfo {year} {2015})}\BibitemShut {NoStop}%
\bibitem [{\citenamefont {Liedl}\ \emph {et~al.}(2023)\citenamefont {Liedl}, \citenamefont {Pucher}, \citenamefont {Tebbenjohanns}, \citenamefont {Schneeweiss},\ and\ \citenamefont {Rauschenbeutel}}]{Liedl2023}%
  \BibitemOpen
  \bibfield  {author} {\bibinfo {author} {\bibfnamefont {C.}~\bibnamefont {Liedl}}, \bibinfo {author} {\bibfnamefont {S.}~\bibnamefont {Pucher}}, \bibinfo {author} {\bibfnamefont {F.}~\bibnamefont {Tebbenjohanns}}, \bibinfo {author} {\bibfnamefont {P.}~\bibnamefont {Schneeweiss}},\ and\ \bibinfo {author} {\bibfnamefont {A.}~\bibnamefont {Rauschenbeutel}},\ }\bibfield  {title} {\bibinfo {title} {Collective radiation of a cascaded quantum system: From timed {D}icke states to inverted ensembles},\ }\href {https://doi.org/10.1103/PhysRevLett.130.163602} {\bibfield  {journal} {\bibinfo  {journal} {Phys. Rev. Lett.}\ }\textbf {\bibinfo {volume} {130}},\ \bibinfo {pages} {163602} (\bibinfo {year} {2023})}\BibitemShut {NoStop}%
\bibitem [{\citenamefont {Lodahl}\ \emph {et~al.}(2015)\citenamefont {Lodahl}, \citenamefont {Mahmoodian},\ and\ \citenamefont {Stobbe}}]{Lodahl2015}%
  \BibitemOpen
  \bibfield  {author} {\bibinfo {author} {\bibfnamefont {P.}~\bibnamefont {Lodahl}}, \bibinfo {author} {\bibfnamefont {S.}~\bibnamefont {Mahmoodian}},\ and\ \bibinfo {author} {\bibfnamefont {S.}~\bibnamefont {Stobbe}},\ }\bibfield  {title} {\bibinfo {title} {Interfacing single photons and single quantum dots with photonic nanostructures},\ }\href {https://doi.org/10.1103/RevModPhys.87.347} {\bibfield  {journal} {\bibinfo  {journal} {Rev. Mod. Phys.}\ }\textbf {\bibinfo {volume} {87}},\ \bibinfo {pages} {347} (\bibinfo {year} {2015})}\BibitemShut {NoStop}%
\bibitem [{\citenamefont {Krinner}\ \emph {et~al.}(2018)\citenamefont {Krinner}, \citenamefont {Stewart}, \citenamefont {Pazmi\~no}, \citenamefont {Kwon},\ and\ \citenamefont {Schneble}}]{Krinner2018}%
  \BibitemOpen
  \bibfield  {author} {\bibinfo {author} {\bibfnamefont {L.}~\bibnamefont {Krinner}}, \bibinfo {author} {\bibfnamefont {M.}~\bibnamefont {Stewart}}, \bibinfo {author} {\bibfnamefont {A.}~\bibnamefont {Pazmi\~no}}, \bibinfo {author} {\bibfnamefont {J.}~\bibnamefont {Kwon}},\ and\ \bibinfo {author} {\bibfnamefont {D.}~\bibnamefont {Schneble}},\ }\bibfield  {title} {\bibinfo {title} {Spontaneous emission of matter waves from a tunable open quantum system},\ }\href {https://doi.org/10.1038/s41586-018-0348-z} {\bibfield  {journal} {\bibinfo  {journal} {Nature}\ }\textbf {\bibinfo {volume} {559}},\ \bibinfo {pages} {589} (\bibinfo {year} {2018})}\BibitemShut {NoStop}%
\bibitem [{\citenamefont {Chu}\ \emph {et~al.}(2017)\citenamefont {Chu}, \citenamefont {Kharel}, \citenamefont {Renninger}, \citenamefont {Burkhart}, \citenamefont {Frunzio}, \citenamefont {Rakich},\ and\ \citenamefont {Schoelkopf}}]{Chu2017}%
  \BibitemOpen
  \bibfield  {author} {\bibinfo {author} {\bibfnamefont {Y.}~\bibnamefont {Chu}}, \bibinfo {author} {\bibfnamefont {P.}~\bibnamefont {Kharel}}, \bibinfo {author} {\bibfnamefont {W.~H.}\ \bibnamefont {Renninger}}, \bibinfo {author} {\bibfnamefont {L.~D.}\ \bibnamefont {Burkhart}}, \bibinfo {author} {\bibfnamefont {L.}~\bibnamefont {Frunzio}}, \bibinfo {author} {\bibfnamefont {P.~T.}\ \bibnamefont {Rakich}},\ and\ \bibinfo {author} {\bibfnamefont {R.~J.}\ \bibnamefont {Schoelkopf}},\ }\bibfield  {title} {\bibinfo {title} {Quantum acoustics with superconducting qubits},\ }\href {https://doi.org/doi:10.1126/science.aao1511} {\bibfield  {journal} {\bibinfo  {journal} {Science}\ }\textbf {\bibinfo {volume} {358}},\ \bibinfo {pages} {199} (\bibinfo {year} {2017})}\BibitemShut {NoStop}%
\bibitem [{\citenamefont {Gu}\ \emph {et~al.}(2017)\citenamefont {Gu}, \citenamefont {Kockum}, \citenamefont {Miranowicz}, \citenamefont {Liu},\ and\ \citenamefont {Nori}}]{Gu2017}%
  \BibitemOpen
  \bibfield  {author} {\bibinfo {author} {\bibfnamefont {X.}~\bibnamefont {Gu}}, \bibinfo {author} {\bibfnamefont {A.~F.}\ \bibnamefont {Kockum}}, \bibinfo {author} {\bibfnamefont {A.}~\bibnamefont {Miranowicz}}, \bibinfo {author} {\bibfnamefont {Y.~X.}\ \bibnamefont {Liu}},\ and\ \bibinfo {author} {\bibfnamefont {F.}~\bibnamefont {Nori}},\ }\bibfield  {title} {\bibinfo {title} {Microwave photonics with superconducting quantum circuits},\ }\href {https://doi.org/10.1016/j.physrep.2017.10.002} {\bibfield  {journal} {\bibinfo  {journal} {Phys. Rep.}\ }\textbf {\bibinfo {volume} {718}},\ \bibinfo {pages} {1} (\bibinfo {year} {2017})}\BibitemShut {NoStop}%
\bibitem [{\citenamefont {Xu}\ \emph {et~al.}(2014)\citenamefont {Xu}, \citenamefont {Tieri}, \citenamefont {Fine}, \citenamefont {Thompson},\ and\ \citenamefont {Holland}}]{Xu2014}%
  \BibitemOpen
  \bibfield  {author} {\bibinfo {author} {\bibfnamefont {M.}~\bibnamefont {Xu}}, \bibinfo {author} {\bibfnamefont {D.~A.}\ \bibnamefont {Tieri}}, \bibinfo {author} {\bibfnamefont {E.~C.}\ \bibnamefont {Fine}}, \bibinfo {author} {\bibfnamefont {J.~K.}\ \bibnamefont {Thompson}},\ and\ \bibinfo {author} {\bibfnamefont {M.~J.}\ \bibnamefont {Holland}},\ }\bibfield  {title} {\bibinfo {title} {Synchronization of two ensembles of atoms},\ }\href {https://doi.org/10.1103/PhysRevLett.113.154101} {\bibfield  {journal} {\bibinfo  {journal} {Phys. Rev. Lett.}\ }\textbf {\bibinfo {volume} {113}},\ \bibinfo {pages} {154101} (\bibinfo {year} {2014})}\BibitemShut {NoStop}%
\bibitem [{\citenamefont {Weiner}\ \emph {et~al.}(2017)\citenamefont {Weiner}, \citenamefont {Cox}, \citenamefont {Bohnet},\ and\ \citenamefont {Thompson}}]{Weiner2017}%
  \BibitemOpen
  \bibfield  {author} {\bibinfo {author} {\bibfnamefont {J.~M.}\ \bibnamefont {Weiner}}, \bibinfo {author} {\bibfnamefont {K.~C.}\ \bibnamefont {Cox}}, \bibinfo {author} {\bibfnamefont {J.~G.}\ \bibnamefont {Bohnet}},\ and\ \bibinfo {author} {\bibfnamefont {J.~K.}\ \bibnamefont {Thompson}},\ }\bibfield  {title} {\bibinfo {title} {Phase synchronization inside a superradiant laser},\ }\href {https://doi.org/10.1103/PhysRevA.95.033808} {\bibfield  {journal} {\bibinfo  {journal} {Phys. Rev. A}\ }\textbf {\bibinfo {volume} {95}},\ \bibinfo {pages} {033808} (\bibinfo {year} {2017})}\BibitemShut {NoStop}%
\bibitem [{\citenamefont {Gonzalez-Ballestero}\ \emph {et~al.}(2013)\citenamefont {Gonzalez-Ballestero}, \citenamefont {Garc\'ia-Vidal},\ and\ \citenamefont {Moreno}}]{Gonzalez-Ballestero2013}%
  \BibitemOpen
  \bibfield  {author} {\bibinfo {author} {\bibfnamefont {C.}~\bibnamefont {Gonzalez-Ballestero}}, \bibinfo {author} {\bibfnamefont {F.~J.}\ \bibnamefont {Garc\'ia-Vidal}},\ and\ \bibinfo {author} {\bibfnamefont {E.}~\bibnamefont {Moreno}},\ }\bibfield  {title} {\bibinfo {title} {Non-{M}arkovian effects in waveguide-mediated entanglement},\ }\href {https://doi.org/10.1088/1367-2630/15/7/073015} {\bibfield  {journal} {\bibinfo  {journal} {New J. Phys.}\ }\textbf {\bibinfo {volume} {15}},\ \bibinfo {pages} {073015} (\bibinfo {year} {2013})}\BibitemShut {NoStop}%
\bibitem [{\citenamefont {Lohof}\ \emph {et~al.}(2023)\citenamefont {Lohof}, \citenamefont {Schumayer}, \citenamefont {Hutchinson},\ and\ \citenamefont {Gies}}]{Lohof2023}%
  \BibitemOpen
  \bibfield  {author} {\bibinfo {author} {\bibfnamefont {F.}~\bibnamefont {Lohof}}, \bibinfo {author} {\bibfnamefont {D.}~\bibnamefont {Schumayer}}, \bibinfo {author} {\bibfnamefont {D.~A.~W.}\ \bibnamefont {Hutchinson}},\ and\ \bibinfo {author} {\bibfnamefont {C.}~\bibnamefont {Gies}},\ }\bibfield  {title} {\bibinfo {title} {Signatures of superradiance as a witness to multipartite entanglement},\ }\href {https://doi.org/10.1103/PhysRevLett.131.063601} {\bibfield  {journal} {\bibinfo  {journal} {Phys. Rev. Lett.}\ }\textbf {\bibinfo {volume} {131}},\ \bibinfo {pages} {063601} (\bibinfo {year} {2023})}\BibitemShut {NoStop}%
\bibitem [{\citenamefont {Mlynek}\ \emph {et~al.}(2014)\citenamefont {Mlynek}, \citenamefont {Abdumalikov}, \citenamefont {Eichler},\ and\ \citenamefont {Wallraff}}]{Mlynek2014}%
  \BibitemOpen
  \bibfield  {author} {\bibinfo {author} {\bibfnamefont {J.}~\bibnamefont {Mlynek}}, \bibinfo {author} {\bibfnamefont {A.}~\bibnamefont {Abdumalikov}}, \bibinfo {author} {\bibfnamefont {C.}~\bibnamefont {Eichler}},\ and\ \bibinfo {author} {\bibfnamefont {A.}~\bibnamefont {Wallraff}},\ }\bibfield  {title} {\bibinfo {title} {Observation of {D}icke superradiance for two artificial atoms in a cavity with high decay rate},\ }\href {https://doi.org/10.1038/ncomms6186} {\bibfield  {journal} {\bibinfo  {journal} {Nat. Commun.}\ }\textbf {\bibinfo {volume} {5}},\ \bibinfo {pages} {5186} (\bibinfo {year} {2014})}\BibitemShut {NoStop}%
\bibitem [{\citenamefont {Breuer}\ and\ \citenamefont {Petruccione}(2002)}]{Breuer2002}%
  \BibitemOpen
  \bibfield  {author} {\bibinfo {author} {\bibfnamefont {H.}~\bibnamefont {Breuer}}\ and\ \bibinfo {author} {\bibfnamefont {F.}~\bibnamefont {Petruccione}},\ }\emph {\bibinfo {title} {The Theory of Open Quantum Systems}}\ (\bibinfo  {publisher} {Oxford University Press},\ \bibinfo {year} {2002})\BibitemShut {NoStop}%
\bibitem [{\citenamefont {Masson}\ \emph {et~al.}(2020)\citenamefont {Masson}, \citenamefont {Ferrier-Barbut}, \citenamefont {Orozco}, \citenamefont {Browaeys},\ and\ \citenamefont {Asenjo-Garcia}}]{Masson2020}%
  \BibitemOpen
  \bibfield  {author} {\bibinfo {author} {\bibfnamefont {S.~J.}\ \bibnamefont {Masson}}, \bibinfo {author} {\bibfnamefont {I.}~\bibnamefont {Ferrier-Barbut}}, \bibinfo {author} {\bibfnamefont {L.~A.}\ \bibnamefont {Orozco}}, \bibinfo {author} {\bibfnamefont {A.}~\bibnamefont {Browaeys}},\ and\ \bibinfo {author} {\bibfnamefont {A.}~\bibnamefont {Asenjo-Garcia}},\ }\bibfield  {title} {\bibinfo {title} {Many-body signatures of collective decay in atomic chains},\ }\href {https://doi.org/10.1103/PhysRevLett.125.263601} {\bibfield  {journal} {\bibinfo  {journal} {Phys. Rev. Lett.}\ }\textbf {\bibinfo {volume} {125}},\ \bibinfo {pages} {263601} (\bibinfo {year} {2020})}\BibitemShut {NoStop}%
\bibitem [{\citenamefont {Abramowitz}\ and\ \citenamefont {Stegun}(1972)}]{Abramowitz1972}%
  \BibitemOpen
  \bibfield  {author} {\bibinfo {author} {\bibfnamefont {M.}~\bibnamefont {Abramowitz}}\ and\ \bibinfo {author} {\bibfnamefont {I.~A.}\ \bibnamefont {Stegun}},\ }\emph{\bibinfo {title} {Handbook of mathematical functions}}\ (\bibinfo  {publisher} {National Bureau of Standards},\ \bibinfo {address} {USA},\ \bibinfo {year} {1972})\ Chap.~\bibinfo {chapter} {17}, pp.\ \bibinfo {pages} {590--591}\BibitemShut {NoStop}%
\end{thebibliography}
\end{document}